\documentclass[prx,twocolumn,english,superscriptaddress,floatfix,longbibliography,nofootinbib]{revtex4-2}

\newcommand{\curlE}{\mathcal{E}}
\newcommand{\ii}{\text{i}}
\newcommand{\ms}{\text{MS}}

\newcommand\brpm{\mathbin{\vcenter{\hbox{\oalign{$\scriptstyle({+})$\cr
					\noalign{\kern-.3ex}
					\hfil$\scriptscriptstyle-$\hfil\cr}}}}}

\let\originalleft\left
\let\originalright\right
\renewcommand{\left}{\mathopen{}\mathclose\bgroup\originalleft}
\renewcommand{\right}{\aftergroup\egroup\originalright}

\usepackage[pdftex]{graphicx}
\usepackage{tikz}
\definecolor{mygreen}{rgb}{0.328125,0.6796875,0.1953125}
\definecolor{myblue}{rgb}{0.12156862745098039, 0.4666666666666667, 0.7058823529411765}
\usepackage{hyperref}
\hypersetup{
     colorlinks=true,
     linkcolor=myblue,
     filecolor=magenta,      
     urlcolor=myblue,
     pdftitle={Measurement-free fault-tolerant quantum error correction in near-term devices},
     citecolor = myblue
}

\usetikzlibrary{quantikz}
\usepackage{yquant}

\usepackage{amsmath}
\usepackage{mathtools}
\usepackage{amssymb}
\usepackage{xcolor}
\usepackage{nicefrac}
\usepackage{siunitx}
\usepackage{bbold}
\usepackage{braket}
\usepackage{hhline}

\usepackage[caption=false]{subfig}
\usepackage{ragged2e}
\DeclareCaptionJustification{justified}{\justifying}

\begin{document}

\title{Measurement-free fault-tolerant quantum error correction in near-term devices}

\author{Sascha Heu\ss en}
\email{sascha.heussen@rwth-aachen.de}
\affiliation{Institute for Quantum Information, RWTH Aachen University, 52056 Aachen, Germany}
\affiliation{Institute for Theoretical Nanoelectronics (PGI-2), Forschungszentrum J\"{u}lich, 52425 J\"{u}lich, Germany}

\author{David F. Locher}
\affiliation{Institute for Quantum Information, RWTH Aachen University, 52056 Aachen, Germany}
\affiliation{Institute for Theoretical Nanoelectronics (PGI-2), Forschungszentrum J\"{u}lich, 52425 J\"{u}lich, Germany}

\author{Markus M\"{u}ller}
\affiliation{Institute for Quantum Information, RWTH Aachen University, 52056 Aachen, Germany}
\affiliation{Institute for Theoretical Nanoelectronics (PGI-2), Forschungszentrum J\"{u}lich, 52425 J\"{u}lich, Germany}

\begin{abstract}

Logical qubits can be protected from decoherence by performing quantum error correction (QEC) cycles repeatedly. Algorithms for fault-tolerant QEC must be compiled to the specific hardware platform under consideration in order to practically realize a quantum memory that operates for in principle arbitrary long times. All circuit components must be assumed as noisy unless specific assumptions about the form of the noise are made. Modern QEC schemes are challenging to implement experimentally in physical architectures where in-sequence measurements and feed-forward of classical information cannot be reliably executed fast enough or even at all. Here we provide a novel scheme to perform QEC cycles without the need of measuring qubits that is fully fault-tolerant with respect to all components used in the circuit. Our scheme can be used for any low-distance CSS code since its only requirement towards the underlying code is a transversal CNOT gate. Similarly to Steane-type EC, we coherently copy errors to a logical auxiliary qubit but then apply a coherent feedback operation from the auxiliary system to the logical data qubit. The logical auxiliary qubit is prepared fault-tolerantly without measurements, too. We benchmark logical failure rates of the scheme in comparison to a flag-qubit based EC cycle. We map out a parameter region where our scheme is feasible and estimate physical error rates necessary to achieve the break-even point of beneficial QEC with our scheme. We outline how our scheme could be implemented in ion traps and with neutral atoms in a tweezer array. For recently demonstrated capabilities of atom shuttling and native multi-atom Rydberg gates, we achieve moderate circuit depths and beneficial performance of our scheme while not breaking fault tolerance. These results thereby enable practical fault-tolerant QEC in hardware architectures that do not support mid-circuit measurements. \end{abstract}

\maketitle

\section{Introduction}

Implementation of quantum error correction (QEC) routines into inevitably noisy physical hardware is conjectured to be indispensable in order to enable large scale universal quantum computation \cite{Campbell2017}. If errors on the quantum register that holds the logical information can be corrected faster than they occur, the threshold theorem guarantees that the computation can be continued for in principle arbitrary long times \cite{knill1996threshold, Aharonov2008}. Fault-tolerant (FT) quantum circuits come with a qubit, gate or time overhead compared to non-FT circuits but can lead to lower logical failure rates provided the error rates of physical components are below a break-even point \cite{preskill1998reliable}. For a QEC code capable of correcting $t$ errors, any possible combination of $t$ Pauli faults on all components\footnote{For a given component acting on $q$ qubits, a possible fault is every element of the $q$-qubit Pauli group $\langle X,Y,Z\rangle^{\otimes q}$.} of a quantum circuit can never lead to failure of the circuit in order for the protocol to be fully FT.

The ability to perform measurements is considered crucial to perform FT QEC. Several milestones towards error corrected quantum computation have been achieved using FT circuit designs \cite{linke2017fault, Takita2017, andersen2020repeated, Egan2021, Erhard2021, abobeih2022fault, postler2022demonstration, hilder2022fault, ryan2022implementing, marques2022logical}. 
Repeated QEC cycles were realized in both ion traps \cite{Ryan-Anderson2021} and superconducting transmons \cite{google2021exponential, krinner2022realizing, zhao2022realization}. Recently, it was shown experimentally that increasing the size of the QEC code can suppress the logical failure rate in a surface code experiment~\cite{google2023suppressing}.
Neutral-atom platforms are catching up quickly. The preparation of logical states of the Steane, surface and toric code has been demonstrated in an experiment with mobile atoms in optical tweezers \cite{bluvstein2022aquantum}.
Proposals for FT quantum computing  have been put forward that take into account specific aspects of this physical platform such as enhanced leakage errors \cite{cong2022hardware, wu2022erasure} and first experimental observations were made recently \cite{scholl2023erasure, ma2023highfidelity}.

\begin{figure*}\includegraphics[width=0.99\linewidth]{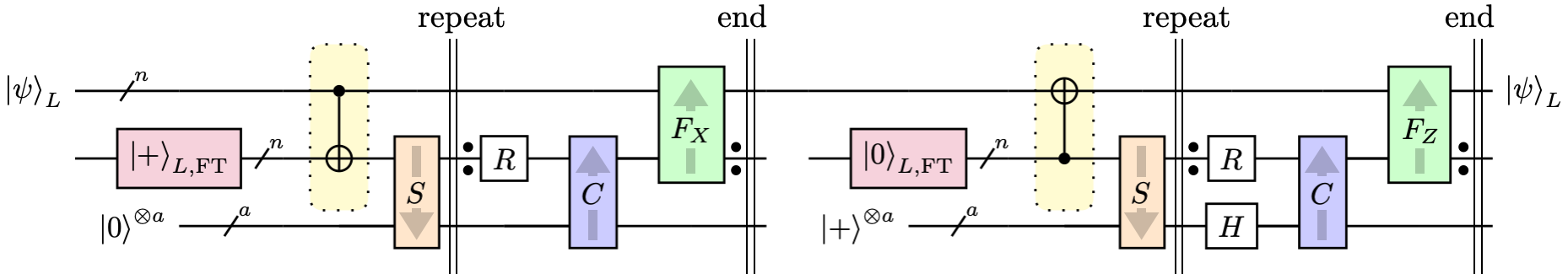}\caption{Logical building blocks of the measurement-free fault-tolerant quantum error correction cycle. Errors on the logical qubit $\ket{\psi}_L$ are copied to a logical auxiliary qubit initialized fault-tolerantly in the appropriate basis. Then the $a$-bit syndrome is mapped from the logical auxiliary qubit (block $S$) to the second auxiliary block of $a$ physical qubits. For each syndrome-correction pair, we repeatedly reset ($R$) the physical qubits to $\ket{0}$, copy ($C$) the syndrome and apply the matching feedback operation $F_\sigma$ of Pauli type $\sigma \in \{X, Z\}$. The repetition refers to all $2^a-1$ possible non-trivial syndromes. Instead of qubit reset, one may also supply fresh auxiliary qubits. The copy and feedback can be parallelized to avoid repetition and to reduce circuit depth and thereby the overall duration of the protocol if a larger number of physical auxiliary qubits is available (see Sec.~\ref{sec:resources}).}
	\label{circ:logical}
\end{figure*}

Although most quantum computing hardware platforms are able to perform measurements of physical qubits, each have their own limitations that hinder straightforward application of QEC. In superconducting transmons for example, error rates of measurements are typically larger than error rates of physical gates and measurement crosstalk can affect neighboring qubits \cite{google2023suppressing}.
In many hardware platforms, measurements are much slower than gate operations, leading to errors on qubits that are idling during measurement and feedback. In trapped-ion and neutral-atom platforms, this problem is exacerbated by the necessity of applying relatively slow laser recooling of ions after qubit detection \cite{pogorelov2021compact} or laser cooling during detection to avoid atom loss \cite{saffman2016quantum}.
In trapped ion systems, alternative routes explore using different ion species for sympathetic cooling \cite{blinov2002sympathetic, barrett2003sympathetic} or physical shuttling of ions into dedicated readout zones separated from the rest of the system to avoid heating of a large ion crystal \cite{kielpinski2002architecture, chiaverini2005surface, hilder2022fault, moses2023race}.
In neutral-atom platforms, mid-circuit measurements have only recently been demonstrated \cite{graham2023mid, norcia2023midcircuit, lis2023midcircuit}; however, they are still orders of magnitude slower than typical gate times on that platform.
Real-time feedback based on measurements remains experimentally challenging but has been demonstrated lately \cite{huie2023repetitive, singh2023midcircuit}. Future experiments that involve in-sequence logic might require technologies such as cavity-enhanced fluorescence imaging \cite{bochmann2010lossless, deist2022fast}, shuttling of atoms into dedicated readout-zones \cite{bluvstein2022aquantum}, or the use of multiple atom species \cite{singh2022dual}.

Because of these challenges, there has been continuous effort in finding QEC schemes that circumvent the need of measuring individual qubits to obtain information about potential errors while at the same time maintaining fault tolerance. Note that some dissipative element is still needed for QEC to remove entropy from the system; either by the ability to reset qubits or to have a sufficiently large reservoir of fresh qubits \cite{schindler2011experimental, barreiro2011open, schindler2013quantum}. Autonomous quantum error correction protocols make use of engineered dissipation in a non-FT way \cite{harrington2022engineered}. In bosonic codes it is common to engineer Lindbladians that have the code states as fixed points \cite{terhal2020towards, gertler2021protecting}. Measurement-free EC need not be substantially inferior than conventional QEC in principle \cite{cruikshank2017role, cruikshank2017high}. A non-FT measurement-free surface code implementation has been shown in Ref.~\cite{ercan2018measurement}. Techniques to devise the T-gate and the Toffoli gate fault-tolerantly without measurements were given in Ref.~\cite{boykin2010algorithms}. In Ref.~\cite{paz2010fault} a different measurement-free FT EC protocol was devised specific to the Bacon-Shor code while extendable to the class of Calderbank-Shor-Steane (CSS) codes. The authors proved a competitive threshold by means of concatenation for the 9-qubit Bacon-Shor code. QEC cycles can be implemented fault-tolerantly given additional assumptions about the form of the noise \cite{crow2016improved, premakumar2020measurement}. However, devising \emph{practical} measurement-free QEC schemes compatible with the number of qubits and native gate operations available in current quantum computers that incorporate full fault tolerance remains a challenging task. 

In this manuscript we provide a QEC technique without measurements that is fully fault-tolerant towards circuit-level depolarizing noise on all circuit locations, inspired by Steane-type EC. We note that now logical qubit operations, i.e.~initialization, Clifford gates and QEC cycles, can be implemented without the need of measurements and real-time feedback.
In Sec.~\ref{sec:scheme} we present our novel measurement-free FT QEC scheme, discuss resource requirements and compare the scheme to conventional flag qubit assisted QEC \cite{Chao2018, Chamberland2018, Chao2020, reichardt2020fault}. Our scheme needs auxiliary logical qubit states, which can be prepared fault-tolerantly without measurements~\cite{heussen2023strategies, goto2023measurementfree}, as we show in Sec.~\ref{sec:stateprep}. In Sec.~\ref{sec:implementation} we provide protocols for practical implementations in state-of-the-art quantum hardware, via a) decompositions into two-qubit gates, b) native multi-qubit-controlled gates that were proposed for neutral-atom or ion trap architectures and c) multi-qubit Mølmer-Sørensen (MS) gates native to ion trap quantum processors. Contrary to general folklore, using all these multi-qubit operations does not invalidate the FT property of the scheme if used at suitable positions in the QEC circuits. Moreover, we propose a shuttling schedule to implement the proposed scheme in a state-of-the-art neutral-atom quantum processor. We provide conclusions and an outlook on future work in Sec.~\ref{sec:outlook}. \section{Measurement-free FT QEC cycle}\label{sec:scheme}

\begin{figure*}\includegraphics[width=0.99\linewidth]{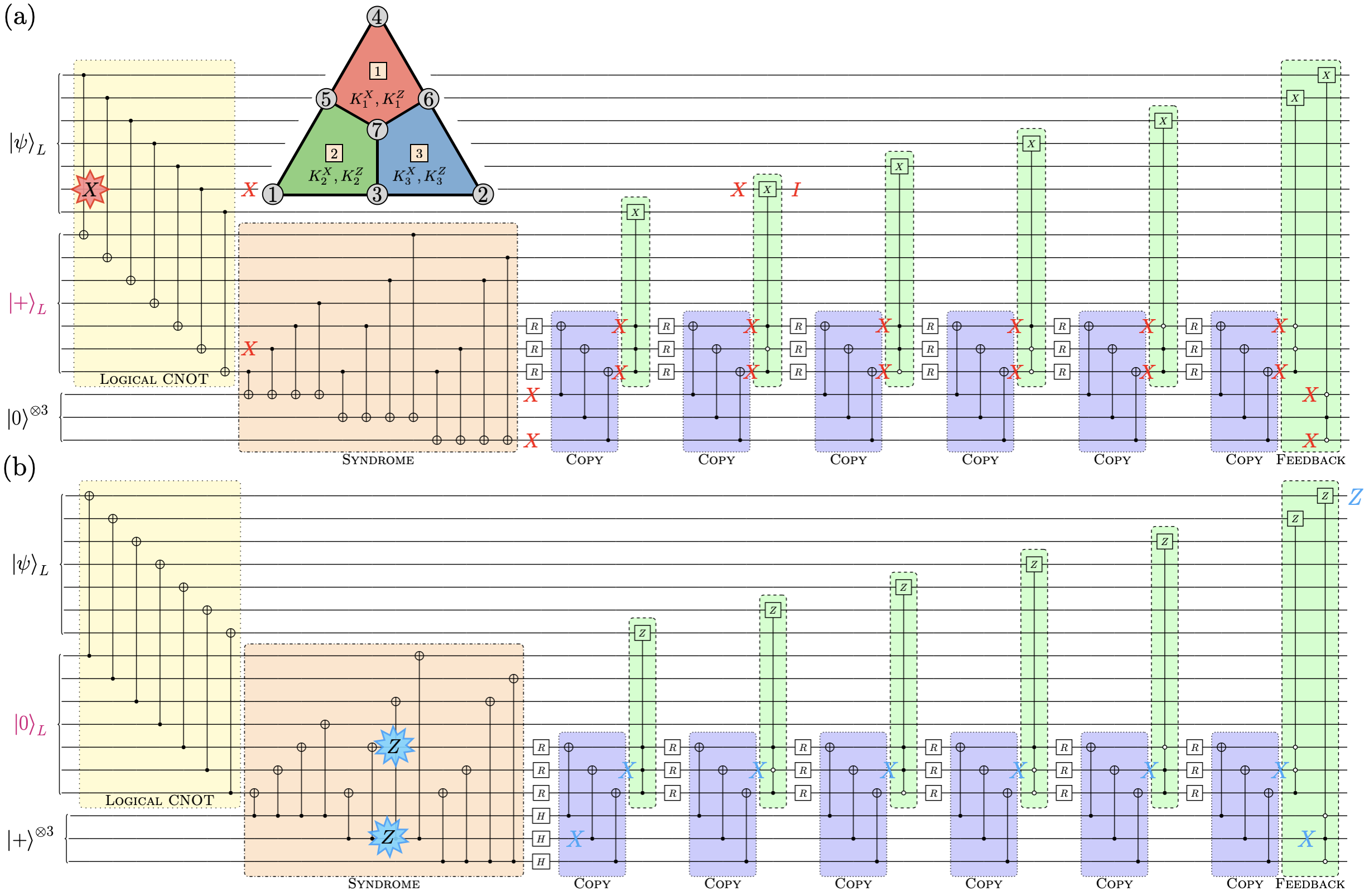}\caption{Detailed circuit on the physical qubit level of the measurement-free fault-tolerant quantum error correction cycle for the Steane code. The syndrome mapping (orange) does not require a distinctly FT routine because we perform intermediate reset operations $R$ to the state $\ket{0}$ that erase potentially dangerous faults. The control-qubits of the feedback gates (green) are conditioned on the physical qubits being in the $\ket{1}$ ($\ket{0}$) state for a black (white) circle. Any suitable procedure to prepare the logical auxiliary qubits fault-tolerantly can be used. We provide a measurement-free FT initialization circuit in Fig.~\ref{circ:ft_det_0}. For the last correction operation, copying from the third to the second block can be omitted. One can verify that no single error of any type on any circuit element will lead to an uncorrectable error. (a) $X$-correction block. A single bit-flip error on the logical data qubit, say $X_6$ (red star), is propagated to the logical auxiliary qubit. The sixth physical CNOT gate propagates the fault as illustrated by the two subsequent red $X$ markers. For a $Z$-fault, the propagation would happen reversely from the target to the control qubit of a CNOT gate. Propagation of $Y$-faults can be viewed as simultaneous occurrence of $X$- and $Z$-faults. Then, the syndrome is mapped to three physical qubits and copied to all multi-qubit-controlled feedback gates. The error is corrected by the second C$_3$NOT gate. For the first and all other C$_3$NOT gates the syndrome does not match the control structure. Thus, they have no additional effect on the logical data qubit. Note that, in contrast, for standard Steane-type EC, one would correct $X_6$ by measuring all qubits of the logical auxiliary qubit in the $Z$-basis after the logical CNOT gate and applying the feedback conditioned on the measurement result. (b) $Z$-correction block. As required for an FT circuit, single faults within the auxiliary system can never cause more than a weight-1 error on the logical data qubit. As an example, we show the single fault event $Z_{12}Z_{16}$ (blue stars), which propagates to a correctable weight-1 error on the logical data qubit through the last C$_3$Z gate. }
	\label{circ:cec}
\end{figure*}

Quantum error correcting codes are based on so-called stabilizer operators, whose eigenvalues must be measured in order to determine and correct for potential errors that might have happened to the logical qubit that is encoded in such a code \cite{nielsen2010quantum}. Two standard techniques to render QEC fault-tolerant are Shor- and Steane-type EC \cite{Shor1996, steane1997active}. The former verifies that a measured stabilizer expectation value is correct by fault-tolerantly encoding the syndrome into the parity of an auxiliary pre-verified FT GHZ-state. The readout procedure must be repeated until a majority vote determines the most likely value in order to protect against single measurement errors. With much fewer qubit and repetition overhead, the flag qubit paradigm was shown to efficiently realize FT QEC cycles \cite{Chamberland2018, Chao2018, Chao2020}. A small number of additional physical flag qubits act as heralds of errors that -- with non-FT QEC -- would lead to logical failure but can be corrected with flag-FT QEC. Measurement of syndrome and flag qubits is combined with classical processing and feedback conditioned on the in-sequence measurement information and, possibly, additional stabilizer measurements. This technique can be used in any QEC code. Steane-type EC on the other hand can be applied to the class of CSS quantum codes \cite{calderbank1996good}, which includes the well-known surface code \cite{bravyi1998quantum, dennis2002topological} and two-dimensional color codes \cite{bombin2006topological, bombin2006distillation, bombin2007topological, fowler2011two}. 

Steane-type EC sequentially corrects one type of Pauli errors (first $X$ then $Z$ or vice versa) by mapping faults from the data qubit register to a logical auxiliary qubit in the state $\ket{+}_L$ or $\ket{0}_L$ respectively. It then uses appropriate logical measurements of the logical auxiliary qubits to infer the most likely error on the logical data qubit, which can then be corrected by conditioning the classical recovery operation on the  measurement outcome of the logical auxiliary qubit.

We demonstrate that with sufficient qubit and gate overhead the need for measurements in Steane-type EC can be circumvented. In the following we first lay out our scheme for a general distance-3 CSS code and the $[[7,1,3]]$ Steane code \cite{steane1996error} -- the smallest representative of the family of two-dimensional color codes \cite{bombin2006topological, bombin2006distillation, bombin2007topological, fowler2011two} -- explicitly. We then discuss resources needed for implementation. Lastly, we demonstrate in which parameter regime we can expect an advantage of the scheme over conventional syndrome-measurement-based QEC.

\subsection{Scheme}

We illustrate one measurement-free FT QEC cycle of our protocol with logical building blocks in Fig.~\ref{circ:logical}. It requires three qubit registers: the first holds the logical data qubit formed of $n$ physical qubits, which we aim to correct, in an arbitrary logical state $\ket{\psi}_L$, potentially having suffered from some fault. The second register of equal size is used to initialize a logical auxiliary qubit, analogous to Steane-type EC, in the state $\ket{+}_L$ ($\ket{0}_L$) when correcting $X$($Z$)-errors. The third register contains $a$ unencoded physical qubits which are all initialized to $\ket{0}$ ($\ket{+}$) when correcting $X$($Z$)-errors. Our goal is to propagate faults, which are potentially present on the logical data qubit, through the circuit while at the same time preserving the logical qubit state during fault-free operation. In order to correct for $X$-errors on the state $\ket{\psi}_L$, a transversal, i.e.~bitwise, CNOT gate propagates them to the logical auxiliary qubit first as marked by the yellow block in Fig.~\ref{circ:logical}. For two logical qubits both encoded in the same CSS code, bitwise application of physical CNOT gates between their physical data qubits implements the logical CNOT gate. Transversal gates are naturally fault-tolerant since there are no couplings between two qubits of the same block. Thus, a single fault on one logical qubit or a physical CNOT gate can never propagate to an uncorrectable error on any logical qubit. Since $\ket{+}_L$ ($\ket{0}_L$) is the state on the target (control) of the logical CNOT when correcting $X$($Z$)-errors, there is no backaction on the control (target) state where we hold $\ket{\psi}_L$. This way, one does not learn about the logical state itself since no expectation values of logical operators are mapped to the logical auxiliary qubit but only individual faults. Then follows a coherent syndrome mapping $S$ from the second to the third register. The mapping can be done without special treatment to ensure fault tolerance, for example preventing uncorrectable errors on the logical auxiliary qubit, since the second register is anyways reset ($R$) afterwards. As a last step, the syndrome is coherently copied ($C$) back from the third to the second register and the feedback operation $F$ applies the correction on the data qubit in the first register that matches the syndrome. This last step needs to be repeated for all syndrome-correction pairs. Subsequently, the analogous procedure is applied to correct for $Z$-errors with the previous $X$- and $Z$-type states and operations interchanged. An additional step of Hadamard gates on the $a$ auxiliary qubits transforms the $\ket{+}$ states needed for the syndrome mapping to $Z$-eigenstates for the controlled feedback operation.

As an illustrative application example, we demonstrate our scheme using the [[7,1,3]] Steane code. Our scheme is equally applicable to the distance-3 surface code \cite{bravyi1998quantum, tomita2014low}, for which the treatment of single-qubit errors is analogous and we discuss the treatment of higher-weight errors in App.~\ref{sec:surface}. These two codes are the smallest instances of the leading approaches towards practical topological QEC. For the surface code, FT state preparation has recently been realized without measurements in Ref.~\cite{goto2023measurementfree}. The Steane code allows one to encode $k = 1$ logical qubit in the code space as defined as the joint +1 eigenspace of the six stabilizer generators
\begin{align}
    K_1^X &= X_4X_5X_6X_7 & K_1^Z &= Z_4Z_5Z_6Z_7 \notag \\
    K_2^X &= X_1X_3X_5X_7 & K_2^Z &= Z_1Z_3Z_5Z_7 \notag \\
    K_3^X &= X_2X_3X_6X_7 & K_3^Z &= Z_2Z_3Z_6Z_7 \label{eq:stabs}
\end{align}
on $n = 7$ physical qubits as shown as part of Fig.~\ref{circ:cec}a. The logical operators of the Steane code can be chosen as $X_L = X^{\otimes 7}$ and $Z_L = Z^{\otimes 7}$. The Steane code can correct $t = 1$ arbitrary Pauli error and thus has distance $d = 3$. For a logical measurement after the QEC cycle, all physical qubits can be measured transversally (and thus fault-tolerantly). 

The detailed circuits are depicted on the physical qubit level in Fig~\ref{circ:cec}. Here we remark that each data qubit correction is conditioned on its distinct three-bit syndrome, which is encoded into the control pattern of the C$_3$NOT (or C$_3$Z) gates that perform the corrections. Since the repetition for each data qubit correction starts by resetting the second register, the syndrome information can be copied anew from the third register using -- in this example -- $a = \frac{n-k}{2} = 3$ transversal CNOT gates without breaking fault tolerance. The feedback operations are quasi-transversal in the sense that a distinct syndrome is uniquely connected to a single physical data qubit. Due to the reset operations there is no connection between the individual syndromes. Consider, for example, the error $X_6$ in Fig.~\ref{circ:cec}a. It will be mapped to the three-qubit state $\ket{101}$ in the third register and, as a consequence, only the second C$_3$NOT gate with the 101-control structure will act non-trivially on the logical data qubit and correct the error. As long as only a single fault occurs on the second or third register, at most one (potentially erroneous) correction operator is applied to the first block. The input state $\ket{\psi}_L$ is assumed fault-free when a fault happens within the cycle because the scheme is FT towards a single fault only. This resulting single error on $\ket{\psi}_L$ will always be correctable by the QEC code. If $\ket{\psi}_L$ already carries a single error, it is guaranteed by the then fault-free QEC cycle that the correct syndrome is mapped to the third register and the appropriate correction is applied. Two faults are necessary to cause failure of the protocol but a single fault can never cause failure because of fault tolerance. Assuming a physical fault rate $p$, the probability of failure is of order $p^2$ while for a non-FT protocol the probability of failure is of order $p$. 

\begin{figure}\includegraphics[width=0.99\linewidth]{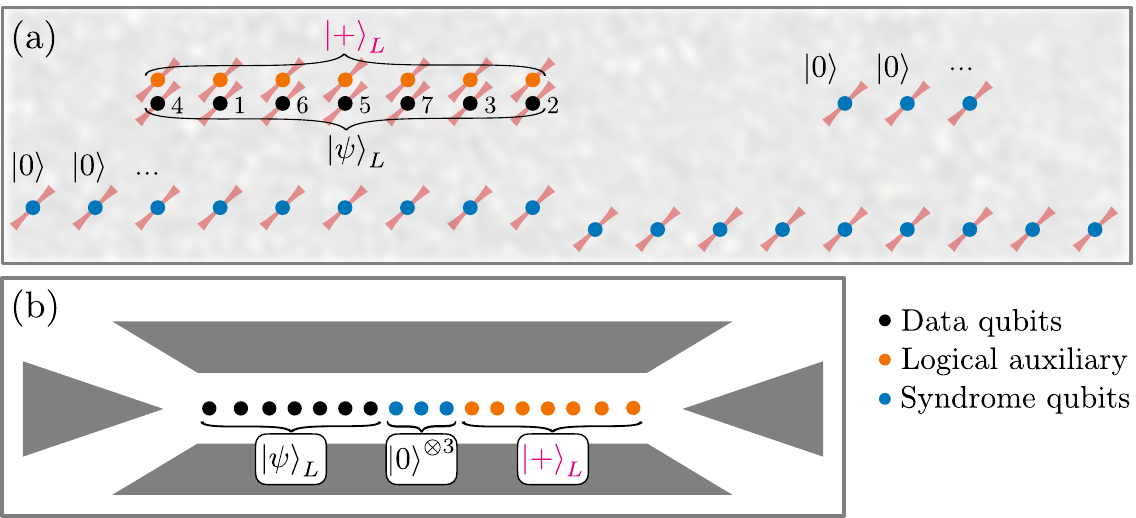}\caption{Sketched embedding of the scheme for the Steane code into (a) a neutral-atom tweezer array and (b) a static linear ion trap. For the tweezer array, we show a proposed initial configuration of the atoms. Entangling gates can by applied in parallel to atoms at close range. Between the application of entangling gates, atoms are shuttled to new locations, which we outline in more detail in Fig.~\ref{fig:Rydberg_schedule}.}
	\label{fig:schedule}
\end{figure}

Note that we only required that transversal CNOT operations between the first and second block (to propagate the errors) as well as between the second and third block (to coherently copy the syndrome) are available. The FT auxiliary qubit initialization, syndrome mapping and feedback can be implemented with any routine that is most suitable for the particular hardware under consideration. Possible implementations into a neutral-atom tweezer array and an ion trap are sketched in Fig.~\ref{fig:schedule}, which we elaborate in more detail in Sec.~\ref{sec:implementation}.

\begin{figure*}\includegraphics[width=0.99\linewidth]{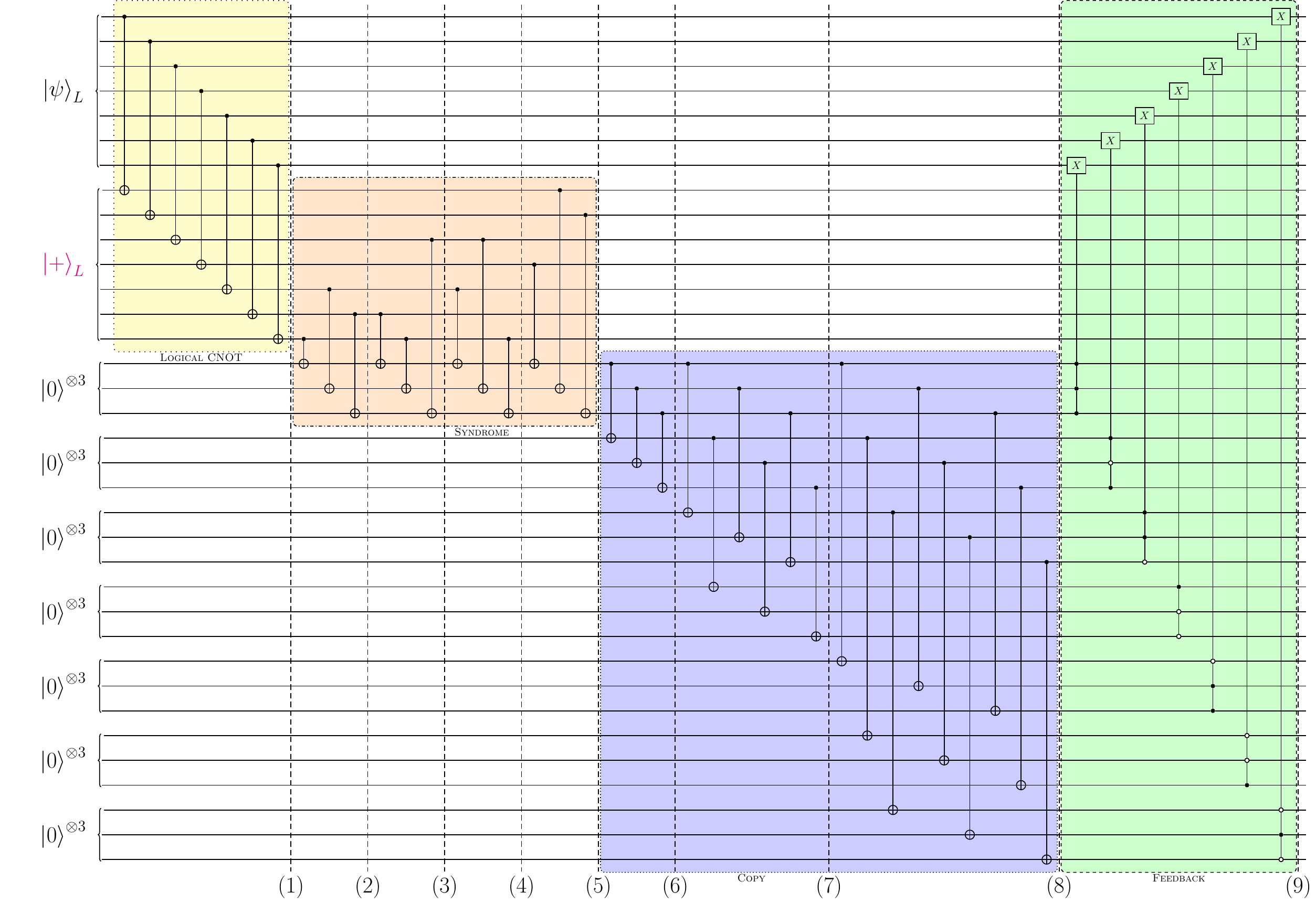}\caption{The $X$-correction block of the measurement-free FT QEC scheme can be scheduled in nine time steps (dashed vertical lines) when parallel gate operations and $N' = 2 \times 7 + 3 \times (2^3 - 1) = 35$ physical qubits are available. In the first step, the logical CNOT is applied. Then, from time step 2 to 5, the syndrome is mapped to a fresh set of auxiliary qubits. From time step 6 to 8, the syndrome is copied six times. In the last time step, the seven feedback operations are applied. It is possible to parallelize the $Z$-correction block analogously.}
	\label{circ:scheme_x_parallel}
\end{figure*}

\subsection{Resources}\label{sec:resources}

The scheme as presented in Fig.~\ref{circ:cec} requires $2n$ physical qubits for the two logical qubits and $a$ qubits to store the syndrome of one Pauli type. In total $N = 2n + a = \mathcal{O}(n)$ physical qubits are needed for an $[[n, k=1, d=3]]$ CSS code. If the $X$- and $Z$-syndromes have the same length, as for color codes or surface codes, we can take the number of syndrome qubits to be $a = \frac{n-k}{2}$. Note that some additional physical qubits might be required for the FT initialization of the logical auxiliary qubits depending on the specific code under consideration. The time overhead that is needed in order to perform the $2^a-1$ feedback operations can be transformed into a qubit overhead, which is useful e.g.~if the cycle time would be too long to implement the repeated copy steps or when no reset operation is available. Instead of repeatedly applying resets we can coherently copy the syndrome $2^a-2$ times to fresh auxiliary qubits. In this case the total number of qubits would increase to $N' = 2n + a\left(2^a-1\right) = \mathcal{O}(2^\frac{n}{2})$ because we still need two logical qubits using $n$ physical qubits each but then we also need $2^a-1$ blocks of $a$ qubits each to connect each feedback operation to a distinct syndrome block. 

Let us now count the number of CNOT gates that are needed to implement one QEC cycle, assuming for simplicity that the $X$-correction block and the $Z$-correction block are symmetric and thus require the same number of CNOT gates. Each logical CNOT gate amounts to $n$ physical CNOTs. Then, $a \times s$ CNOTs are needed for the syndrome mapping step where we assume, for simplicity, that all stabilizers have the same weight $s$, i.e.~the number of physical qubits that the stabilizers act non-trivially on. Coherently copying the syndrome $2^a-2$ times requires $a$ bitwise CNOT gates each. We need $m = 2^{a+1} - 3$ two-qubit gates to exactly decompose a single $a$-qubit-controlled feedback operation targeting a single physical data qubit \cite{barenco1995elementary}. For all $2^a-1$ feedback operations we need $m$ CNOT gates each. Therefore, in total, we require at most $2 \times \left(n + a(s + 2^a - 2) + m\left(2^a-1\right)\right)$ CNOT gates to implement the QEC cycle.

It is desirable to reduce the circuit depth of the QEC cycle as much as possible due to the limited coherence times in near-term devices. If CNOT gates can be executed in parallel, the transversal CNOT gate can be run in just one time step, the stabilizer readout needs $s$ time steps (again assuming all stabilizers have the same weight) and copying the syndrome from one block to $2^a-1$ blocks can be done in $a$ time steps. All feedback operations can in principle be executed in a single time step if the physical architecture permits.

For the Steane code, these requirements amount to a total of $256$ CNOT gates with $n = 7$ physical qubits, $a = 3$ stabilizers of each type that have weight $s = 4$. We elaborate a simplification of the multi-qubit-controlled gate decomposition in Sec.~\ref{sec:implementation} that will allow one to reduce $m$ from 13 to 8 and thus realize the Steane code QEC cycle with 186 CNOT gates. In total, the minimum circuit depth that can be achieved with the Steane code is $2 \times 9 = 18$ as shown in Fig.~\ref{circ:scheme_x_parallel}. Note that furthermore the $X$- and $Z$-type correction part of the QEC cycle (see Fig.~\ref{circ:cec}a and \ref{circ:cec}b) could also be largely carried out in parallel if one disposes of two simultaneously operated logical auxiliary logical qubits and registers of additional bare physical qubits.

Let us remark that the $a$ physical auxiliary qubits are not strictly necessary to map the syndrome from the logical auxiliary qubit. Instead one may use an appropriate decoding circuit to obtain the syndrome on a subset of the physical qubits forming the logical auxiliary qubit and perform the quantum feedback as we discuss in more detail in App.~\ref{sec:decoding}.

\subsection{Measurement-free advantage} \label{sec:advantage}

In the following, we analyze under which conditions the measurement-free (MF) EC protocol can be expected to yield lower logical failure rates than conventional EC involving syndrome measurements (SM). We provide an analytical estimation for advantageous use of the MFEC scheme and compare it to numerical statevector simulations.

We assume that all operations, i.e.~gates, qubit initializations and measurements, in the circuits of the protocol, compiled into CNOT gates, are prone to depolarizing noise of strength $p$ (see App.~\ref{sec:noise} for details on the noise model). Also, we consider an idling error rate $p_{\text{idle,m}}$ for idling during measurements for the measurement-based protocol and an idling error rate $p_{\text{idle,op}}$ for idling during operations for the MFEC protocol. These are the two dominant sources of idling noise for both protocols, assuming that the time to perform a qubit measurement is much longer than the time to perform a gate operation or qubit initialization/reset. The finite duration $t$ of physical operations causes an idling time on those qubits that are not targeted by these operations. The idling error rate for a qubit with coherence time $T_2$ that is prone to pure Markovian dephasing during an idling time $t$ is
\begin{align}
    p_\text{idle} = \frac{1}{2}\left( 1 - \exp\left(-\frac{t}{T_2}\right)\right). \label{eq:idlrate}
\end{align}
The rate $p_\text{idle}$ is linearly proportional to the idling time $t$ if $t/T_2 \ll 1$. 

Denoting the logical failure rates of the two protocols $p_L^\text{MF}$ and $p_L^\text{SM}$ respectively, MFEC is advantageous when the ratio of the failure rates $p_L^\text{MF}/p_L^\text{SM} \leq 1$. We estimate in App.~\ref{sec:appmeasfreeadv} that the MFEC protocol is advantageous when the ratio of measurement to operation time fulfills the inequality
\begin{align}
    \frac{t_\text{meas}}{t_\text{ops}} &\geq \sqrt{\frac{\Tilde{c}^2}{\Tilde{c}'^2} + \frac{c \Tilde{c}}{\Tilde{c}'^2} \frac{p}{p_{\text{idle,op}}} + \frac{c^2}{\Tilde{c}'^2} \frac{p^2}{p_{\text{idle,op}}^2}}. \label{eq:mfadv_main}
\end{align}
Here, $c$ ($\Tilde{c}$) is the number of operations (idling locations) in the MFEC circuit and $\Tilde{c}'$ is the number of idling locations during measurements in the conventional EC protocol.

\begin{figure}\centering
    \includegraphics[width=0.99\linewidth]{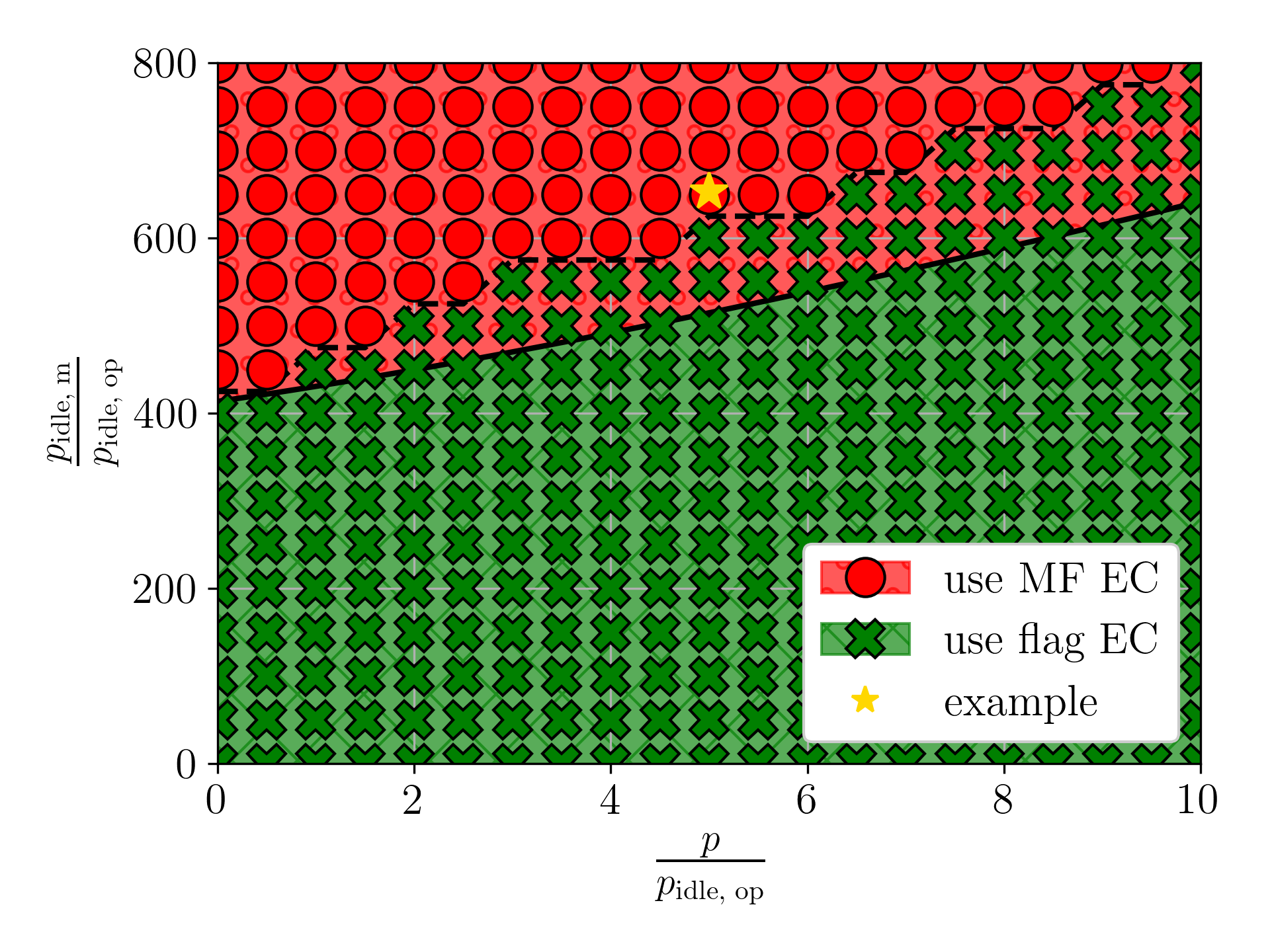}
	\caption{Utility diagram for measurement-free (MF) EC and flag EC. Regions of advantage based on numerical simulation (markers) and the estimation in Eq.~\eqref{eq:mfadv_main} (filled area) are shown in terms of the parameter ratios $p/p_{\text{idle,op}}$ and $p_{\text{idle,m}}/p_{\text{idle,op}}$, which approximates $t_\text{meas}/t_\text{ops}$, with $\Tilde{c}^2/\Tilde{c}'^2 = 171041,\,c \Tilde{c}/\Tilde{c}'^2 = 13471$ and $c^2/\Tilde{c}'^2 = 1061$ in accordance with Eqs.~\eqref{eq:c_ops}-\eqref{eq:cprime_idlemeas}. The lines mark the estimated boundaries between regions of advantage. In the limit of vanishing operation errors, $p = 0$, MFEC yields an advantage over flag EC if measurements are at least 400 times slower than operations. The star marker represents the parameters given in Eqs.~\eqref{eq:advantage_p}-\eqref{eq:advantage_t}. At $p/p_{\text{idle,op}} = 1$ and $p_{\text{idle,m}}/p_{\text{idle,op}} = 1$, we use absolute values of $p = p_{\text{idle,m}} = p_{\text{idle,op}} = 10^{-4}$ in the numerical simulations.}
	\label{fig:utility}
\end{figure}

As an example of a measurement-based protocol, we choose a state-of-the-art flag-qubit-based EC protocol \cite{reichardt2020fault}, described in App.~\ref{sec:flag}. We pick a set of noise parameters which satisfies Eq.~\eqref{eq:mfadv_main} with a large margin, given the constants for the flag EC circuits in Eqs.~\eqref{eq:c_ops}-\eqref{eq:cprime_idlemeas}, as
\begin{align}
    p &= 5 \times 10^{-4} \label{eq:advantage_p}\\
    p_{\text{idle,op}} &= 10^{-4} \stackrel{T_2 = 10^{-2}\,\text{s}}{\Longleftrightarrow} t_\text{ops} = 2 \times 10^{-6}\,\text{s}\\ 
    p_{\text{idle,m}} &= 6.5 \times 10^{-2} \Longleftrightarrow t_\text{meas} = 1.4 \times 10^{-3}\,\text{s}. \label{eq:advantage_t}
\end{align}
For these parameters we find via Monte Carlo simulation that MFEC achieves a logical failure rate of $p_L^\text{MF} = (3.44 \pm 0.25)\%$ when compiled into CNOT gates while flag EC fails for $p_L^\text{FL} = (3.96 \pm 0.26)\%$ of the runs. The numerical data in Fig.~\ref{fig:utility} is obtained by sampling logical failure rates of both schemes in Monte Carlo simulation until the uncertainty intervals of the two estimators allow one to distinguish which of the two schemes is advantageous.
While gate error rates of $p = 5 \times 10^{-4}$ are experimentally demanding, we stress that our scheme offers the possibility to perform FT QEC in physical systems that currently cannot support the measurement duration necessary for conventional QEC schemes.

\section{Deterministic FT logical state preparation}\label{sec:stateprep}

Our MFEC scheme needs logical auxiliary qubits whose encoding must be FT in order to render the full scheme FT. In this section we describe how to fault-tolerantly initialize the logical auxiliary qubit (and also the logical data qubit) without measurements. This way, our FTEC scheme can be performed in a fully measurement-free setting. Let us remark nevertheless that the measurement-based encoding protocol from Ref.~\cite{goto2016minimizing} has been realized recently in ion-trap platforms \cite{Ryan-Anderson2021, postler2022demonstration, ryan2022implementing}.

\begin{figure}[!b]\includegraphics[width=0.99\linewidth]{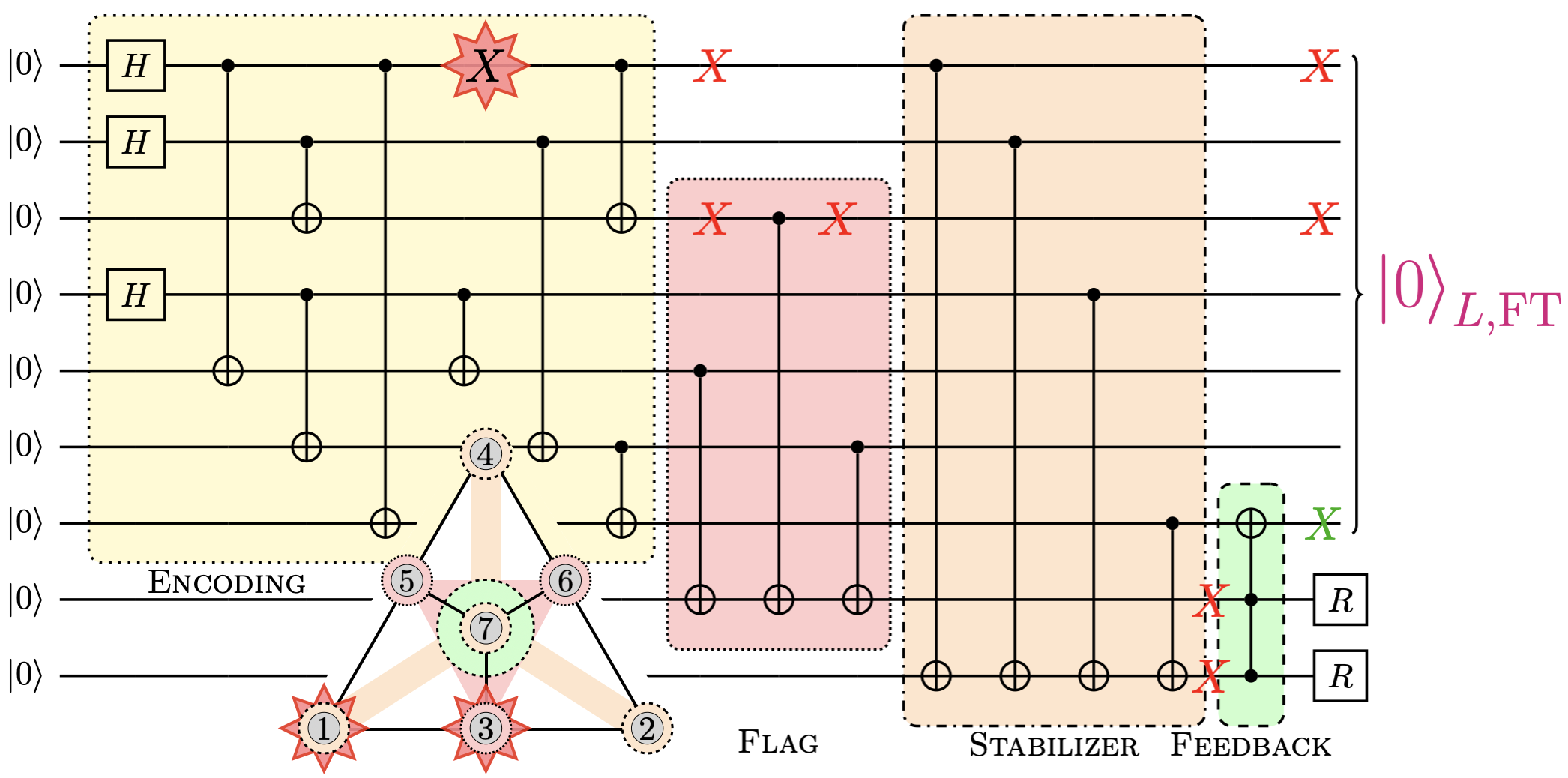}
	\caption{Circuit to fault-tolerantly initialize the logical zero state of the Steane code without measurements. The first eight CNOT gates prepare the state non-fault-tolerantly. The subsequent three CNOT gates map the logical operator $Z_3Z_5Z_6$ to a flag qubit that heralds successful preparation (red dotted circles on code graph). The last four CNOT gates map the complementary stabilizer $Z_1Z_2Z_4Z_7$ (orange dashed circles on code graph) to a second auxiliary qubit. Only if both measurements yield the $-1$-eigenvalue, the correction $X_7$ is applied via the Toffoli gate (large green circle on code graph). In the end, both auxiliary qubits are reset ($R$). The dangerous fault $X_1$ after the fourth CNOT gate (red star) propagates to both auxiliary qubits and triggers the Toffoli feedback (green box). The resulting operator $X_1X_3X_7$ is stabilizer-equivalent to the correctable error $X_5$ via application of $K_2^X$.}
	\label{circ:ft_det_0}
\end{figure}

In Ref.~\cite{heussen2023strategies} some of us suggest an extension of the prescription for logical qubit initialization in Ref.~\cite{goto2016minimizing} by making use of the flag qubit information instead of discarding the state. The circuit for measurement-free FT initialization to $\ket{0}_L$ is shown in Fig.~\ref{circ:ft_det_0}. By mapping the two eigenvalues of both the logical operator $Z_3Z_5Z_6$ (red block, ``flag'') and the complementary stabilizer $Z_1Z_2Z_4Z_7$ (orange block, ``stabilizer'') to two auxiliary qubits, all dangerous weight-2 errors at the end of the circuit can be transformed into correctable errors. 

In fact, there are only two dangerous errors, namely $X_1X_3$ and $X_4X_5$ ($X_6X_7$ is stabilizer-equivalent to $X_4X_5$ via application of $K_1^X$), which flip the first auxiliary qubit from $\ket{0}$ to $\ket{1}$ since their support has odd overlap with $Z_3Z_5Z_6$. Of course, $X_1X_3$ and $X_4X_5$ also have odd overlap with the qubits that take part in the subsequent stabilizer mapping step so the second auxiliary qubit is also flipped from $\ket{0}$ to $\ket{1}$. The dangerous errors will lead to both auxiliary qubits being in the $\ket{1}$ state. Correctable weight-1 errors only flip one of the two from $\ket{0}$ to $\ket{1}$. No single fault during the mapping of either $Z$-operator can result in the auxiliary qubits being in the state $\ket{11}$. The operator $X_7$ is applied coherently via the Toffoli gate if both the flag qubit and the extra stabilizer qubit are in the $\ket{1}$ state (green block, ``feedback''). This way, any dangerous weight-2 error will be transformed into a correctable weight-1 error by multiplication with $X_7$. Treatment of the $X_1X_3$ error is sketched as an example in Fig.~\ref{circ:ft_det_0}. 

The auxiliary qubit state before applying the Toffoli gate can only be different from $\ket{00}$ if a fault has happened at some circuit location before. An additional fault during the Toffoli gate would render the total fault configuration to be of order $p^2$. If instead the circuit up to the Toffoli gate has been fault-free and the $t=1$ fault now occurs within the Toffoli gate with probability $p$, it can at most propagate to a correctable error since the Toffoli is only connected to a single data qubit. 

\begin{figure}\centering
    \includegraphics[width=0.99\linewidth]{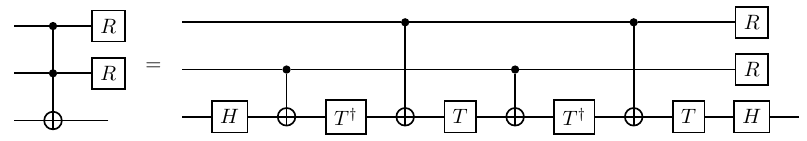}\caption{Decomposition of the Toffoli gate followed by reset of the control qubits into Hadamard gates $H$, the standard $T,\,T^\dagger$ and CNOT gates as well as reset $R$.}
	\label{circ:tof3_decomp}
\end{figure}

The Toffoli gate can be decomposed into a sequence of single- and two-qubit gates.\footnote{We note that the reduced Toffoli gate decompositions given in Refs.~\cite{barenco1995elementary, maslov2016advantages} are not feasible for use in our scheme. There, the Toffoli operation is performed up to a relative phase between certain computational basis states, which would lead to erroneous phase flips when applied to a superposition state such as $\ket{0}_L$.} Since the auxiliary qubit state is discarded at the end of the circuit anyways, we can modify the well-known decomposition into six CNOT gates from Ref.~\cite{nielsen2010quantum}. Figure \ref{circ:tof3_decomp} shows the decomposition of the Toffoli gate followed by reset into only four CNOT gates.

In summary, since the state preparation scheme can be performed without qubit measurements, it qualifies to prepare logical auxiliary qubits and thereby completes our measurement-free FT QEC scheme. 

\section{Practical implementation}\label{sec:implementation}

\begin{figure*}\centering
    \includegraphics[width=0.99\linewidth]{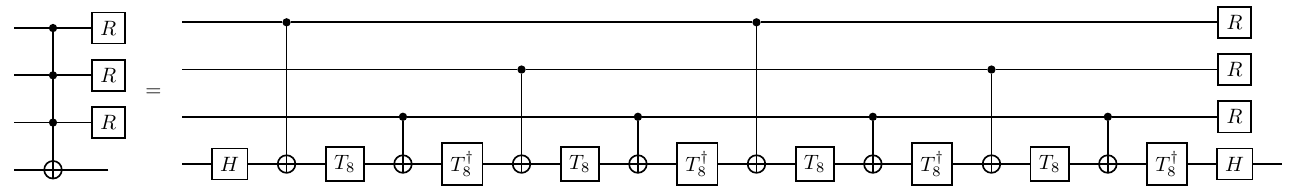}\caption{Decomposition of a triple-controlled-NOT gate followed by reset $R$ of the control qubits into a sequence of Hadamard gates ($H$), CNOT gates and $T_8 = \exp(-\ii\frac{\pi}{16}Z)$ gates followed by reset $R$ of the control qubits.}
	\label{circ:tof4_decomp}
\end{figure*}

\begin{figure}\centering
    \includegraphics[width=0.99\linewidth]{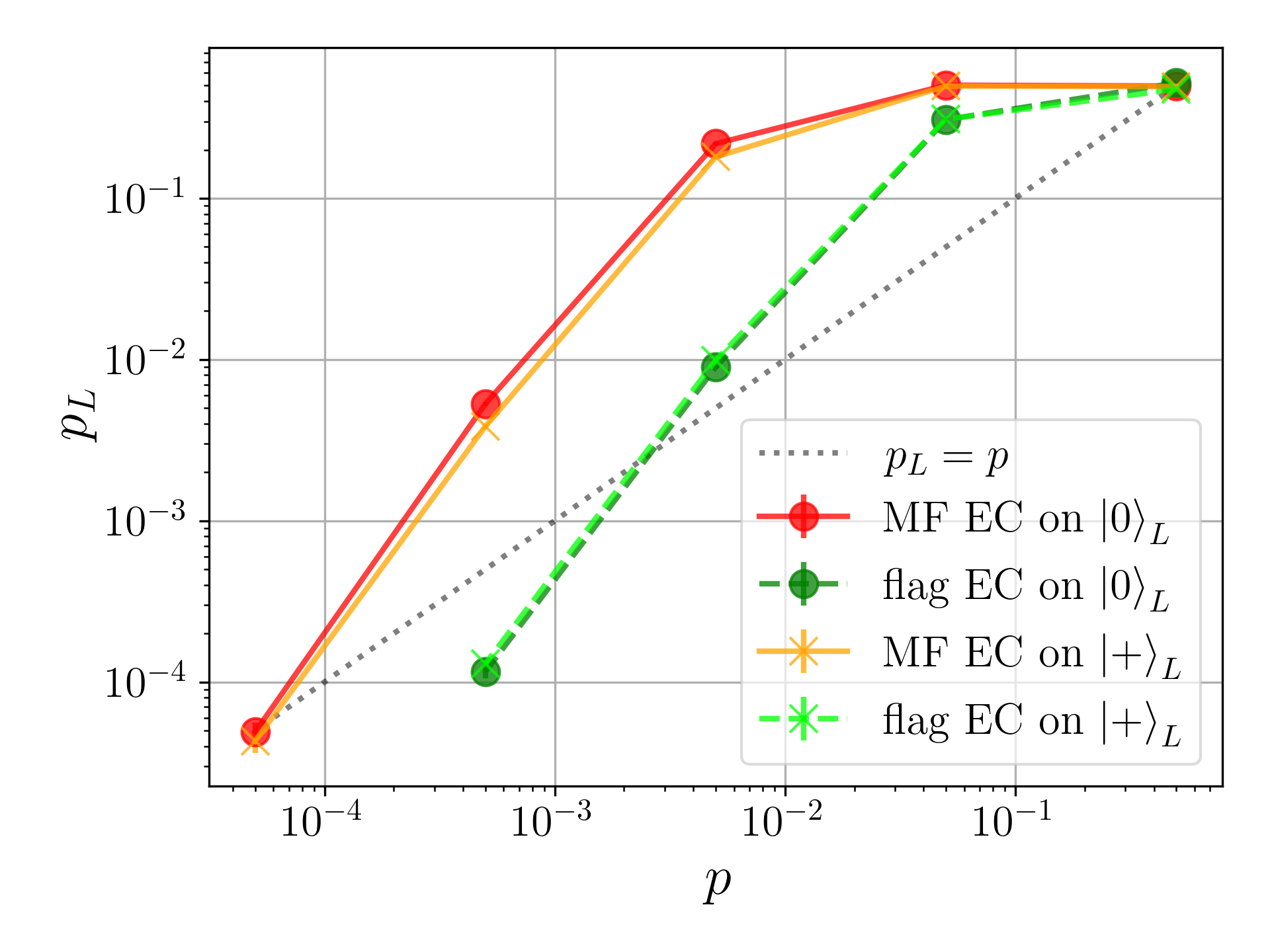}
	\caption{Logical failure rates for the measurement-free (MF) EC scheme compared to conventional flag EC for two logical input states $\ket{0}_L$ and $\ket{+}_L$ with a single-parameter depolarizing noise model of strength $p$. The MFEC scheme is compiled into CNOT gates. Threshold values for outperforming the physical error rate $p_L = p$ lie at approximately $p = 3 \times 10^{-3}$ for the flag scheme and $p = 6 \times 10^{-5}$ for the measurement-free scheme.}
	\label{fig:singleP}
\end{figure}

The suggested measurement-free fault-tolerant quantum error correction protocol can be implemented using various sets of basis gates that are native to different quantum computing platforms.
In the following, we provide compilations of our 17-qubit scheme (Fig.~\ref{circ:cec}) into CNOT gates, native multi-qubit-controlled gates as well as multi-qubit Mølmer-Sørensen gates. The latter are widely-used entangling gates in trapped-ion systems \cite{molmer1999multiparticle}.

If native multi-qubit gates are practically available and if their physical error rates are lower than the expected overall error of their decompositions, they are preferential to implement the feedback operation (green boxes in Fig.~\ref{circ:cec}) over a gate decomposition into single- and two-qubit gates in order to minimize circuit depth. For the syndrome mapping step (orange boxes in Fig.~\ref{circ:cec}), multi-qubit gates can still be used to decrease the gate count. It is crucial that the transversal CNOT gates used for fault propagation (yellow boxes in Fig.~\ref{circ:cec}) cannot be replaced by multi-qubit gates. These would destroy the FT property of the scheme under a general error channel (see Eq.~\eqref{eq:depol}) since single faults could result in higher-weight errors that are uncorrectable in the Steane code. 

To be truly fault-tolerant, every qubit operation used in an actual implementation acting on a respective qubit state $\rho$ must be assumed to be prone to noise. Here we employ the standard depolarizing noise model (see App.~\ref{sec:noise} for details) with different noise strengths than in Sec.~\ref{sec:advantage}: For the decomposition into two-qubit gates, we show simulations for a single-parameter depolarizing noise model with noise strength $p$ on operations. On idling locations, we choose a corresponding noise strength $p/100$, which is the order of magnitude reached for two-qubit entangling gates in state-of-the-art ion trap systems \cite{pino2021demonstration}. This model has been assumed before for simulation of QEC blocks \cite{Chamberland2018, chamberland2019fault}. For the decompositions into multi-qubit gates, we use a multi-parameter noise model with different noise strengths $\vec{p} = (p_1, p_2, ...)$ for the respective multi-qubit depolarizing channel (see Eq.~\eqref{eq:depol}) on operations. The parameters are based on values that either can be achieved experimentally in quantum processors already \cite{pino2021demonstration, Ryan-Anderson2021} or are based on theory proposals \cite{bermudez2017assessing, rasmussen2020single, espinoza2021high, pelegri_2022_highfidelity}. Since dephasing is the dominant source of noise on idling qubits in atomic architectures, we make use of the dephasing channel (see Eq.~\eqref{eq:deph}) with noise strengths assumed to be $\vec{p}/100$ for
simplicity\footnote{The order of magnitude is estimated from gate and coherence times of alkali-atom-platforms. With $p_2 = 5 \times 10^{-3}$, $t_2 \approx \SI{250}{ns}$ \cite{evered2023high} and $T_2^* \approx \SI{4}{ms}$ ($T_2 \approx \SI{1}{s}$ when spin echo techniques are applied) \cite{bluvstein2022aquantum, graham2022multiqubit} we find a ratio of $p_2/p_\text{idle,2} \approx 160\,(4 \times 10^4)$. For ion traps, multi-qubit gate parameters are given in Ref.~\cite{bermudez2017assessing}. For example for the five-qubit MS gate, the noise strength $p_5 = 5 \times 10^{-2}$ and duration of $t_5 = \SI{60}{\micro s}$ would correspond to a factor of $p_5/p_\text{idle,5} \approx 166$ with a coherence time $T_2 = \SI{100}{ms}$. In other ion trap platforms \cite{pino2021demonstration, moses2023race} the ratio can be somewhat higher, for example for the two-qubit gate with $p_2 = 2 \times 10^{-3}$ that takes time $t_2 = \SI{25}{\micro s}$ and corresponds to a factor of $p_2/p_\text{idle,2} \approx 320$ with a coherence time $T_2 = \SI{2}{s}$.}
on the respective idling locations for the multi-parameter noise model. To assess the break-even point where $p_L = p_2$, we scale all physical error rates uniformly with a scaling parameter $\lambda$ like 
\begin{align}
    (p_1, p_2, ...) \rightarrow \lambda \cdot (p_1, p_2, ...).
\end{align}
Numerical statevector simulations of these noisy circuits are performed with a modified version of the python package PECOS \cite{ryan2018quantum, pecos}.

\subsection{Decomposition in two-qubit gates}\label{sec:cnotimpl}

A large extent of the scheme is already expressed in terms of CNOT gates. The only components left to decompose are the multi-qubit-controlled feedback gates C$_3$NOT and C$_3$Z. The authors of Ref.~\cite{barenco1995elementary} state that 13 two-qubit gates are necessary to exactly decompose the full gate. However, since the state of the control qubits is reset after the gate anyways, decomposition is possible with fewer CNOT gates. Figure \ref{circ:tof4_decomp} shows the circuit equivalence of C$_3$NOT and a sequence of eight CNOT gates and standard single-qubit rotations when the state of the three control qubits after the gate is irrelevant. Alternating rotations of angle $\pm \pi/8$ cancel each other exactly if the control qubit state is different from $\ket{111}$. Only if all CNOT gates are activated, the intermediate $X$-flips make the eight $Z$-rotations align in order to realize a full $X(\pi)$-rotation in combination with the outermost Hadamard gates on the target qubit. 

In Fig.~\ref{fig:singleP} we compare the logical failure rate of the MFEC scheme with CNOT gates to the conventional flag EC from Sec.~\ref{sec:advantage}. Quadratic scaling behavior $p_L \sim p^2$ as $p \rightarrow 0$ for both logical input states $\ket{0}_L$ and $\ket{+}_L$ is clearly visible in Fig.~\ref{fig:singleP}. This is expected for the two FT schemes since there exist no single fault events that occur with probability $p$ and contribute to the logical failure rate $p_L$. Our MFEC scheme achieves logical failure rates $p_L$ approximately one order of magnitude larger than the flag scheme for a given physical error rate $p$.

To narrow the gap between the two schemes and achieve lower logical failure rates with the MFEC scheme, we now look into possible improvements using multi-qubit gates.

\subsection{Use of multi-qubit gates}\label{sec:multiimpl}

Depending on the physical architecture under consideration, specific multi-qubit gates might be available for practical operation of the scheme.
In neutral-atom platforms, the Rydberg blockade can be utilized to perform native multi-qubit gates.
CCZ gates that require only global laser pulses have already been realized in experiments \cite{levine2019parallel, evered2023high}.
Moreover, there are multiple theoretical proposals to implement, e.g., $\mathrm{C}_n\mathrm{NOT}$ gates \cite{evered2023high, isenhower_2011_multibit, pelegri_2022_highfidelity, khazali_2020_fast}, $\mathrm{C}\mathrm{NOT}_n$ gates \cite{mueller_2009_mesoscopic, khazali_2020_fast} or $\mathrm{C}_n\mathrm{NOT}_m$ gates \cite{young_2021_asymmetric}, up to single-qubit rotations.
In ion traps the \emph{iToffoli} gate with varying number of control qubits can be implemented directly \cite{monz2009realization, espinoza2021high}.
Additionally, one can realize multi-ion MS gates in these systems \cite{molmer1999multiparticle, schindler2013quantum}. Also in superconducting architectures there are proposals and demonstrated implementations of native \emph{iToffoli} \cite{kim2022high, baker2022single} and CCPHASE gates \cite{glaser2023controlled}.

In the following we provide decompositions and numerical simulations of the MFEC scheme using different sets of multi-qubit gates where possible. Logical failure rates of these compilations are compared via numerical simulation to the scheme that uses only CNOT gates.

\textbf{Native multi-qubit-controlled feedback in neutral-atom platforms.}
We now describe the usage of native $\mathrm{C}_3\mathrm{NOT}$ gates for feedback operations (green boxes in Fig.~\ref{circ:cec}). With the large fidelities and fast gate times of such gates (see Table~\ref{tab:multiparams} and Refs.~\cite{bermudez2017assessing, rasmussen2020single, espinoza2021high, pelegri_2022_highfidelity}) an improvement of logical failure rate can be expected over a decomposition into two-qubit gates, also due to the reduction of idling locations. From inspection of Fig.~\ref{circ:tof4_decomp} we notice that due to the reset operations only the target qubit can cause erroneous output of the multi-qubit-controlled feedback gate. In this decomposed version, eight two-qubit gate locations and ten single-qubit gate locations can cause an error in first order in $p$. We thus estimate that using a native multi-qubit-controlled feedback gate to be advantageous over the decomposition as long as $p_4 < 8p_2 + 10p_1$ at least, which is fulfilled for the parameters in Table~\ref{tab:multiparams}.

\begin{figure}\centering
    \includegraphics[width=0.99\linewidth]{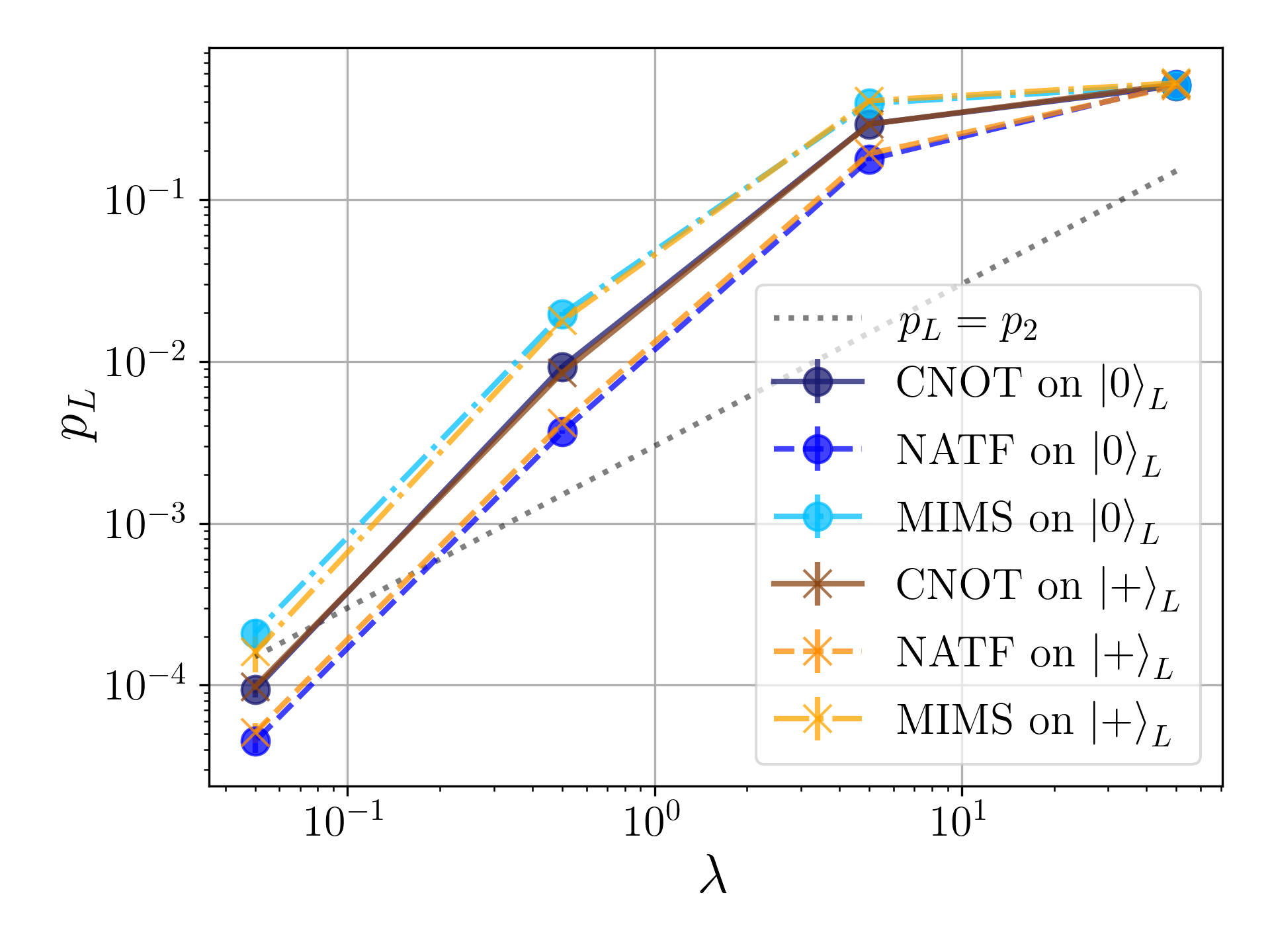}
	\caption{Logical failure rates for the measurement-free scheme acting on both logical input states $\ket{0}_L$ and $\ket{+}_L$ with multi-parameter noise and a scaling parameter $\lambda$ that uniformly varies all physical error rates. Compilation of the scheme into CNOT gates is compared to implementations using native multi-qubit-controlled feedback operations (NATF) and multi-ion Mølmer-Sørensen gates (MIMS) where applicable. Respective physical error rates are listed in Table \ref{tab:multiparams}. Depending on the implementation, logical failure rates can vary to up to one order of magnitude.}
	\label{fig:multiP}
\end{figure}

\textbf{Coherent syndrome mapping and quantum feedback in ion traps.} The 17-qubit scheme for the Steane code from Fig.~\ref{circ:cec} can be embedded into a trapped ion quantum processor, as sketched in Fig.~\ref{fig:schedule}b, hosting a static one-dimensional ion crystal as in Ref.~\cite{pogorelov2021compact}. In these systems, the native entangling gate can be implemented by a laser-driven $X$-type MS gate described by the unitary
\begin{align}
    \ms_q(\theta) &= \exp \left( -\ii \frac{\theta}{4} \left( \sum_{k=1}^q X_k \right)^2 \right) \label{eq:ms}
\end{align}
targeting $q$ ions simultaneously \cite{molmer1999multiparticle, sorensen2000entanglement}. Such multi-qubit MS gates can be used for the syndrome mapping (orange boxes in Fig.~\ref{circ:cec}) and feedback steps \cite{muller2011simulating, barreiro2011open}. Previously it has been found by exhaustive count of gate combinations in Ref.~\cite{martinez2016compiling} that the Toffoli gate is equivalent to a sequence of local rotations and $\ms_3$ gates (reproduced in Fig.~\ref{circ:ms3_decomp} of App.~\ref{sec:ms4}), which we do not improve further for use in our QEC scheme. An application of $\ms_5$ gates for mapping the expectation value of a weight-4 stabilizer to a single auxiliary qubit has been given in Refs.~\cite{muller2011simulating, bermudez2017assessing} (see App.~\ref{sec:ms4} for an example in Fig.~\ref{circ:stab_ms5}). Six $\ms_5$ gates are needed to map the syndrome to the physical auxiliary qubits. Recall that high-weight Pauli faults that are generated by the depolarizing noise channel of the $\ms_5$ gate do not break fault-tolerance since they only act on the logical auxiliary qubit, which is reset after syndrome mapping.

\begin{table}\begin{center}
\begin{tabular}{c || c | c | c | c | c | c | c || c}
Impl. & $p_1$ [\%] & $p_2$ & $p_3$ & $p_4$ & $p_5$ & $p_\text{i}$ & $p_\text{m}$ & $p_\text{idle}$\\ \hline
CNOT & $0.007$ & $0.3$ & - & - & - & $0.0002$ & $0.3$ & $p/100$ \\
NATF & $0.007$ & $0.3$ & $1$ & $1$ & $1$ & $0.0002$ & $0.3$ & $p/100$\\
MIMS & $0.007$ & $0.3$ & $1$ & $1$ & $1$ & $0.0002$ & $0.3$ & $p/100$\\
\end{tabular} 
\end{center}
\caption{List of physical error rates (all in \%) used for simulation of three different gate implementations at scaling parameter $\lambda = 1$. The error rates for idling positions during an operation are assumed to be $1\%$ of the respective operation's error rate. Multi-qubit gate error rates ($p_3,p_4,p_5$) are based on theory proposals \cite{bermudez2017assessing, rasmussen2020single, espinoza2021high, pelegri_2022_highfidelity} while single qubit operation ($p_1,p_i,p_m$) and two-qubit gate error rates ($p_2$) are state-of-the-art values \cite{Ryan-Anderson2021}.}
\label{tab:multiparams}
\end{table}

Using a variational circuit ansatz in pennylane \cite{bergholm2018pennylane}, we found a decomposition of the C$_3$NOT gate followed by reset of the control qubits into four $\ms_4$ gates that we depict in Fig.~\ref{circ:ms4_decomp}. The remaining CNOT gates are compiled into $\ms_2$ gates and local rotations using Eq.~\eqref{eq:msdecomp}. 

Figure \ref{fig:multiP} shows logical failure rates for all three previously described implementations for MFEC for both logical Steane code input states $\ket{0}_L$ and $\ket{+}_L$. Notably, all implementations are capable of achieving logical failure rates $p_L < p_2$ lower than the two-qubit error rate with improvements of the scaling factor $\lambda$ of approximately one order of magnitude. For low physical error rates, where the six lines in Fig.~\ref{fig:multiP} are (almost) parallel, native multi-qubit-controlled feedback operations yield approximately a factor five of improvement over the CNOT compilation of the scheme. Multi-ion MS gates perform approximately a factor of two worse than the CNOT version in this regime\footnote{The syndrome mapping step in this variant is expected to be more noisy since $6p_5 > 12p_2$ as compared to the CNOT version with the parameters from Table \ref{tab:multiparams}. Also note that the decomposition into MS gates requires additional single-qubit rotations compared to the CNOT version, which we did not optimize for these simulations.}.

\subsection{Implementation with neutral atoms}
\label{sec:Rydberg}

As mentioned above, measurements in neutral-atom platforms are slow as compared to gates and it is challenging to perform measurements without atom loss and with real-time feedback.
While multi-qubit gates can be performed within roughly $100$ -- $\SI{500}{ns}$ \cite{levine2019parallel, evered2023high, pagano_2022_error, jandura_2022_timeoptimal}, recently demonstrated mid-circuit measurements in free space take $3.5$ -- $\SI{25}{ms}$ \cite{lis2023midcircuit, norcia2023midcircuit, huie2023repetitive, graham2023mid}.
Such values correspond to ratios $t_\text{meas}/t_\text{ops}$ between $7 \times 10^{3}$ and $2.5 \times 10^{5}$.

With a coherence time of $T_2^* = \SI{4}{ms}$, we estimate the logical failure rate of MFEC using native multi-qubit-controlled feedback gates via numerical simulation to be $p_L^\text{MF} = (5.9 \pm 0.3)\%$. With measurement times of approximately $t_{\text{meas}} \in \{\SI{500}{\micro s}, \SI{1}{ms}\}$, flag EC achieves respective logical failure rates of $p_L^\text{FL} \in \{(4.9 \pm 0.2)\%, (11.8 \pm 0.4)\%\}$. Assuming an anticipated improvement of future operation error rates $p_1,\,p_2,\,p_3,\,p_4$ and $p_\text{i}$ by a factor of 2, the logical failure rate of MFEC drops to $p_L^\text{MF} = (1.7 \pm 0.1)\%$, i.e.~by about a factor of 4, in agreement with the expectations for an FT protocol.

Using the substantially longer coherence time $T_2 = \SI{1}{s}$, which can be achieved by involving spin-echo techniques, the failure rate of MFEC is reduced only slightly to $p_L^\text{MF} = (5.6 \pm 0.1)\%$. This can be understood because the performance of the MFEC protocol is in this parameter regime not limited by its overall duration, but operational error rates. 
In contrast, the performance of the flag EC protocol improves more strongly, with the scheme benefiting more from an extended coherence time, resulting in a logical failure rate of $p_L^\text{FL}= (0.66 \pm 0.04)\%$ for a measurement time $t_\text{meas} = \SI{1}{ms}$. Improving the operation error rates by a factor of 2 yields a predicted MFEC logical failure rate of $p_L^\text{MF} = (1.7 \pm 0.1)\%$.
The simulated\footnote{For the simulations, we choose realistic operation error rates $p_1 = 3 \times 10^{-4},\,p_2 = 5 \times 10^{-3},\,p_3 = 2 \times 10^{-2},\,p_4 = 3 \times 10^{-2},\,p_\text{i} = p_\text{m} = 2 \times 10^{-3}$ and operation times $t_1 = t_\text{i} = \SI{1}{\micro s},\,t_2 = \SI{250}{ns},\,t_3 = t_4 = \SI{500}{ns}$. With coherence time $T_2^* = \SI{4}{ms}\,(T_2 = \SI{1}{s})$, this leads to idling error rates $p_{\text{idle}, 1} = p_{\text{idle,i}} = 1.2 \times 10^{-4},\,p_{\text{idle}, 2} = 3.1 \times 10^{-5}$ and $p_{\text{idle}, 3} = p_{\text{idle}, 4} = 6.2 \times 10^{-5}\,(p_{\text{idle}, 1} = p_{\text{idle,i}} = 5 \times 10^{-7},\,p_{\text{idle}, 2} = 1.25 \times 10^{-7}$ and $p_{\text{idle}, 3} = p_{\text{idle}, 4} = 2.5 \times 10^{-7}$) using Eq.~\eqref{eq:idlrate} \cite{bluvstein2022aquantum, evered2023high}. For measurement times $t_\text{meas} \in \{\SI{500}{\micro s}, \SI{1}{ms}\}$ we employ idling error rates $p_{\text{idle,m}} \in \{5.9 \times 10^{-2}, 11.8 \times 10^{-2}\}$ ($p_{\text{idle,m}} \in \{2.5 \times 10^{-4}, 5 \times 10^{-4}\}$) that follow from the values of $T_2^*$ ($T_2$) stated above.}
failure rates are summarized in Table \ref{tab:na_comp}.

\begin{table}\begin{center}
\begin{tabular}{c || c | c }
Protocol (varied parameters) & $T_2^* = \SI{4}{ms}$ & $T_2 = \SI{1}{s}$ \\ \hline 
MFEC (\textit{realistic}) & $(5.9 \pm 0.3)\%$ & $(5.6 \pm 0.1)\%$ \\
MFEC (\textit{improved}) & $(1.7 \pm 0.1)\%$ & $(1.7 \pm 0.1)\%$ \\
Flag EC ($t_{\text{meas}} = \SI{1}{ms}$) & $(11.8 \pm 0.4)\%$ & $(0.66 \pm 0.04)\%$ \\
Flag EC ($t_{\text{meas}} = \SI{500}{\micro s}$) & $(4.9 \pm 0.2)\%$ & $(0.65 \pm 0.04)\%$
\end{tabular} 
\end{center}
\caption{Comparison of simulated logical failure rates for MFEC and flag EC, assuming that either the coherence times $T_2^*$ or $T_2$ can be achieved. We list MFEC failure rates for \textit{realistic} operation times and error rates, as well as anticipated future error rates that are \textit{improved} by a factor of 2. For flag EC, we consider measurement times of $\SI{500}{\micro s}$ and $\SI{1}{ms}$ as representative values. While flag EC yields the lowest failure rate if coherence time $T_2$ can be achieved, MFEC outperforms flag EC with improved operations and if coherence times are limited to $T_2^*$.}
\label{tab:na_comp}
\end{table}

\begin{figure*}[t]
    \centering
    \includegraphics[width=0.99\linewidth]{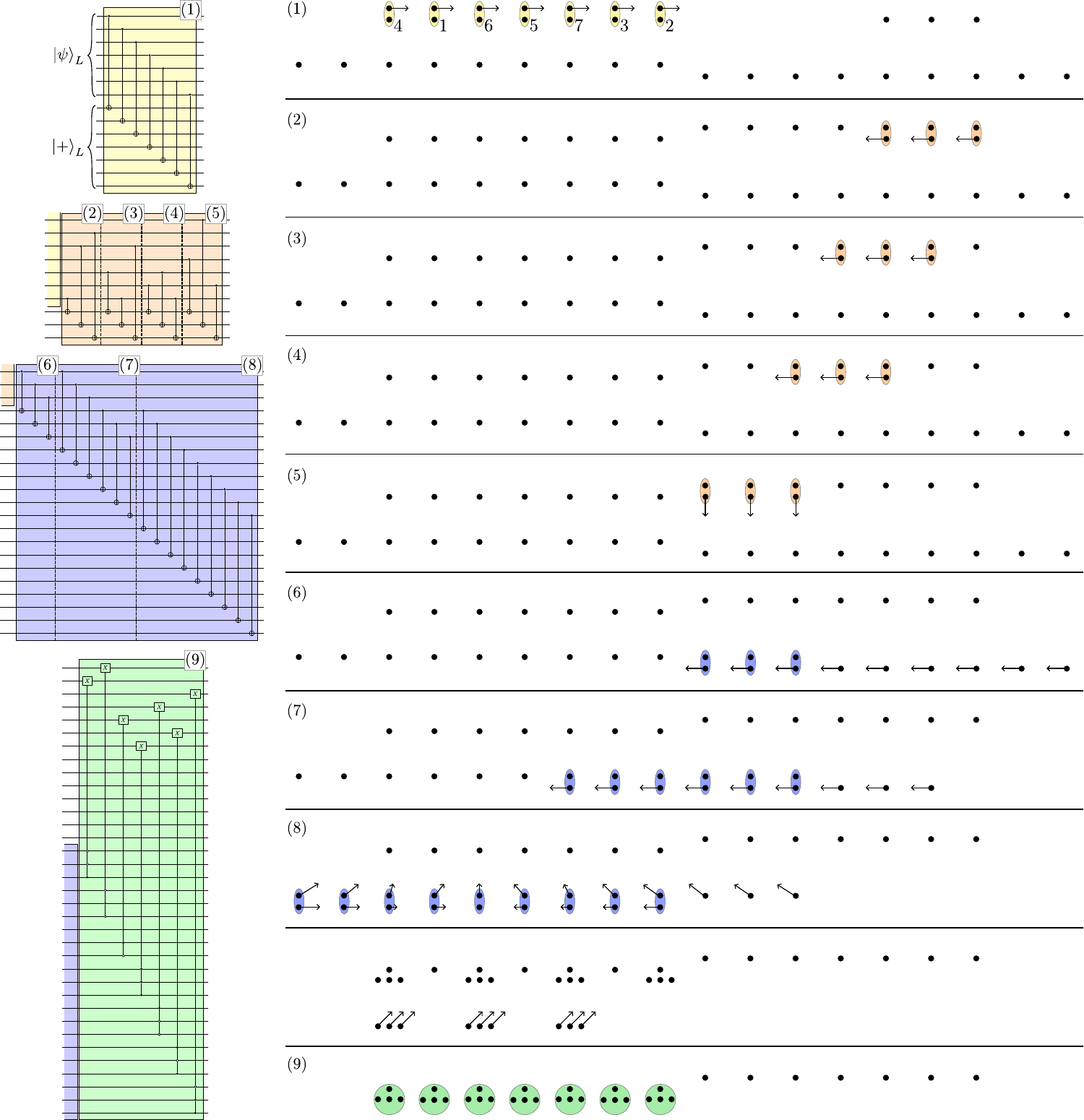}\caption{Proposal for the implementation of the measurement-free FT QEC scheme with mobile neutral atoms in optical tweezers. The steps (1)--(9) correspond to the time steps as shown on the left-hand side of this figure and in Fig.~\ref{circ:scheme_x_parallel}. Ellipses that encircle sets of atoms correspond to the parallel application of entangling gates. Arrows indicate shuttling moves after the gate executions. Single-qubit gates are not shown.}
    \label{fig:Rydberg_schedule}
\end{figure*}

Our proposed scheme might thus open a competitive pathway for this platform to achieve beneficial FT QEC before fast in-sequence measurements and in-sequence logic will become widely available.
Although leakage out of the computational subspace is a dominant error source in neutral-atom platforms, such errors can be handled either by converting them into Pauli $Z$-errors \cite{cong2022hardware} or erasures \cite{wu2022erasure}.
To execute the quantum circuits presented in this work, a certain connectivity between qubits is required.
In a static array of atoms with nearest-neighbour or even next-to-nearest-neighbour interactions this would require many SWAP gates.
In neutral-atom quantum processors, however, individual atoms can be dynamically rearranged during a computation, which yields a very good effective qubit connectivity \cite{barredo_2016_an_atom, endres_2016_atom, bluvstein2022aquantum} but can increase effective gate times. 
Moreover, the multi-qubit gates required for our scheme are natively available in this platform \cite{evered2023high}.

In Fig.~\ref{fig:Rydberg_schedule} we propose a layout of atoms in a tweezer array, together with shuttling moves, to realize the measurement-free FT QEC scheme in an experiment.
We assume a near-term neutral-atom platform with a global Rydberg laser illuminating all atoms in the tweezer array.
This allows to perform two-qubit gates or multi-qubit gates on sets of atoms that are located within the blockade radius of each other.
If those sets of atoms are placed sufficiently far away from each other, such gates can be performed in parallel \cite{bluvstein2022aquantum, evered2023high}.
Furthermore, we assume that atoms can be locally addressed to perform individual single-qubit gates.
If also single-qubit gates can only be performed globally, additional shuttling moves of subsets of atoms into dedicated single-qubit gate operation zones are necessary.
The left-hand side of Fig.~\ref{fig:Rydberg_schedule} shows a circuit that realizes the measurement-free FT correction cycle for $X$-errors.
On the right-hand side we first depict the initial layout of 35 atoms in the tweezer array.
As described in Sec.~\ref{sec:scheme}, the higher number of qubits avoids that qubits have to be reset during the computation and allows for a minimal circuit depth.
If in-sequence qubit reset is available, it is also possible to work with 17 atoms in a tweezer array, realizing the circuit shown in Fig.~\ref{circ:cec}.
We sketch the application of parallel two- and multi-qubit gates as well as shuttling moves, which are performed between the application of entangling gates.
The application of single-qubit gates is not shown.
We choose the atom layout in a way such that the total number of shuttling operations remains small and many parallel qubit moves are possible.
The scheme requires a static 2D tweezer array, generated e.g.~by a spatial light modulator (SLM), and a movable tweezer array realized with a single 2D acousto-optic deflector (AOD).
Parallel moves of rows and columns are possible for atoms placed in the AOD array while atoms placed in the SLM tweezers remain fixed.
Between shuttling operations, atoms can be relocated from static SLM tweezers into the dynamic AOD tweezers and vice versa~\cite{henriet_2020_quantum, kaufman2021quantum}.
Our proposed shuttling protocol thus requires 9 parallel moves of atoms, which is comparable in complexity to already demonstrated experiments~\cite{bluvstein2022aquantum}.

\section{Conclusions \& Outlook}\label{sec:outlook}

In this work, we have presented a novel scheme for fault-tolerant quantum error correction without the need to measure individual physical qubits to read out the syndrome.

As we showed by numerical simulation, the measurement-free FT QEC scheme achieves logical failure rates approximately one order of magnitude higher than the corresponding flag error correction protocol with single-parameter circuit-level depolarizing noise. Additionally, compilations of our scheme into different native gate sets can lead to variations and reductions in logical failure rates of up to one order of magnitude for the physical error rates we considered. This offers room for optimization to bridge the gap between the measurement-free and the conventional FT QEC schemes. We expect that a platform, which can realize an advantageous compilation, for instance using native multi-qubit-controlled gates with sufficient gate fidelities, can in this way at least partly compensate the extra infidelity introduced by the additional overhead in the measurement-free scheme compared to conventional syndrome-measurement EC. For a set of realistic parameters in a neutral atom setup, we showed via numerical simulation that the measurement-free FT QEC cycle can outperform flag-FT EC in the regime where system performance is limited by coherence time.
Moreover, in neutral-atom platforms the outlined scheme may prove particularly useful due to the challenges posed by fast, low-loss and fully parallelized measurements and real-time feedback. Furthermore, neutral atoms natively feature the possibility to realize multi-qubit gates required for our scheme. Many of the required key components have been demonstrated recently in experiments, including mid-circuit shuttling of atoms and parallel application of two-qubit and multi-qubit gates \cite{bluvstein2022aquantum, evered2023high}.
Hardware-specific noise characteristics such as biased noise or bias-preserving gates \cite{cong2022hardware}, could even further compensate for the overhead of the measurement-free scheme.
In this sense our simulations with depolarizing noise might be overly pessimistic.

However, also an embedding of the scheme into a solid-state platform is not futile since we do not require full all-to-all qubit connectivity. Optimizing the compilation of a scheme to hardware constraints like qubit connectivity in a systematic way could also further improve the scheme. It is an open question how additional physical qubits could be used for an embedding with reduced connectivity without breaking fault tolerance, for instance by using the techniques of Refs. \cite{lao2020fault, chamberland2020very}.

Extending our scheme to larger distance codes poses additional requirements for the construction of suitable fault-tolerant circuits on the auxiliary system. This is required to ensure that multiple faults do not cause a logical failure, which could be subject of future work, e.g.~adapting concepts proposed in Refs.~\cite{Chamberland2018,Chao2020}. Using concatenation for scale-up could provide an alternative route worth exploring.

Additionally, developing new measurement-free versions of FT logical gates or FT gadgets such as code switching or lattice surgery would further enlarge the toolbox of measurement-free FT quantum computing protocols and thereby assist in enabling error corrected universal quantum computation in future hardware platforms. 

\section*{Code availability}
All codes used for data analysis are available from the corresponding author upon reasonable request.

\section*{Author contributions}
S.H. devised the scheme, its implementations, performed numerical simulations and analyzed the data. D.L. developed the neutral-atom implementation, embedding and schedules. All authors contributed to theory modelling. S.H. and D.L. wrote the manuscript with contributions from M.M. who supervised the project.

\section*{Acknowledgements}
We thank Johannes Zeiher for fruitful discussions and feedback on the manuscript. We gratefully acknowledge support by the EU Quantum Technology Flagship grant under Grant Agreement No.820495 (AQTION), the BMBF project MUNIQC-ATOMS, the U.S. Army Research Office through Grant No. W911NF-21-1-0007, the European Union’s Horizon Europe research and innovation program under Grant Agreement No. 101046968 (BRISQ), the ERC Starting Grant QNets through Grant No. 804247 and by the Office of the Director of National Intelligence (ODNI), Intelligence Advanced Research Projects Activity (IARPA), via the U.S. Army Research Office through Grant No. W911NF-16-1-0070. Furthermore, the project leading to this publication has received funding from the European Union’s Horizon Europe research and innovation programme under grant agreement No 101114305 (“MILLENION-SGA1” EU Project). This research is also part of the Munich Quantum Valley (K-8), which is supported by the Bavarian state government with funds from the Hightech Agenda Bayern Plus. The views and conclusions contained herein are those of the authors and should not be interpreted as necessarily representing the official policies or endorsements, either expressed or implied, of the ODNI, IARPA, or the U.S. Government. The U.S. Government is authorized to reproduce and distribute reprints for governmental purposes notwithstanding any copyright annotation thereon. Any opinions, findings, and conclusions or recommendations expressed in this material are those of the author(s) and do not necessarily reflect the view of the U.S. Army Research office. The authors gratefully acknowledge funding by the Deutsche Forschungsgemeinschaft (DFG, German Research Foundation) under Germany’s Excellence Strategy ‘Cluster of Excellence Matter and Light for Quantum Computing (ML4Q) EXC 2004/1’ 390534769 and the computing time provided to them at the NHR Center NHR4CES at RWTH Aachen University (Project No. p0020074). This is funded by the Federal Ministry of Education and Research and the state governments participating on the basis of the resolutions of the GWK for national high performance computing at universities.
 
\appendix

\begin{figure*}\includegraphics[width=0.99\linewidth]{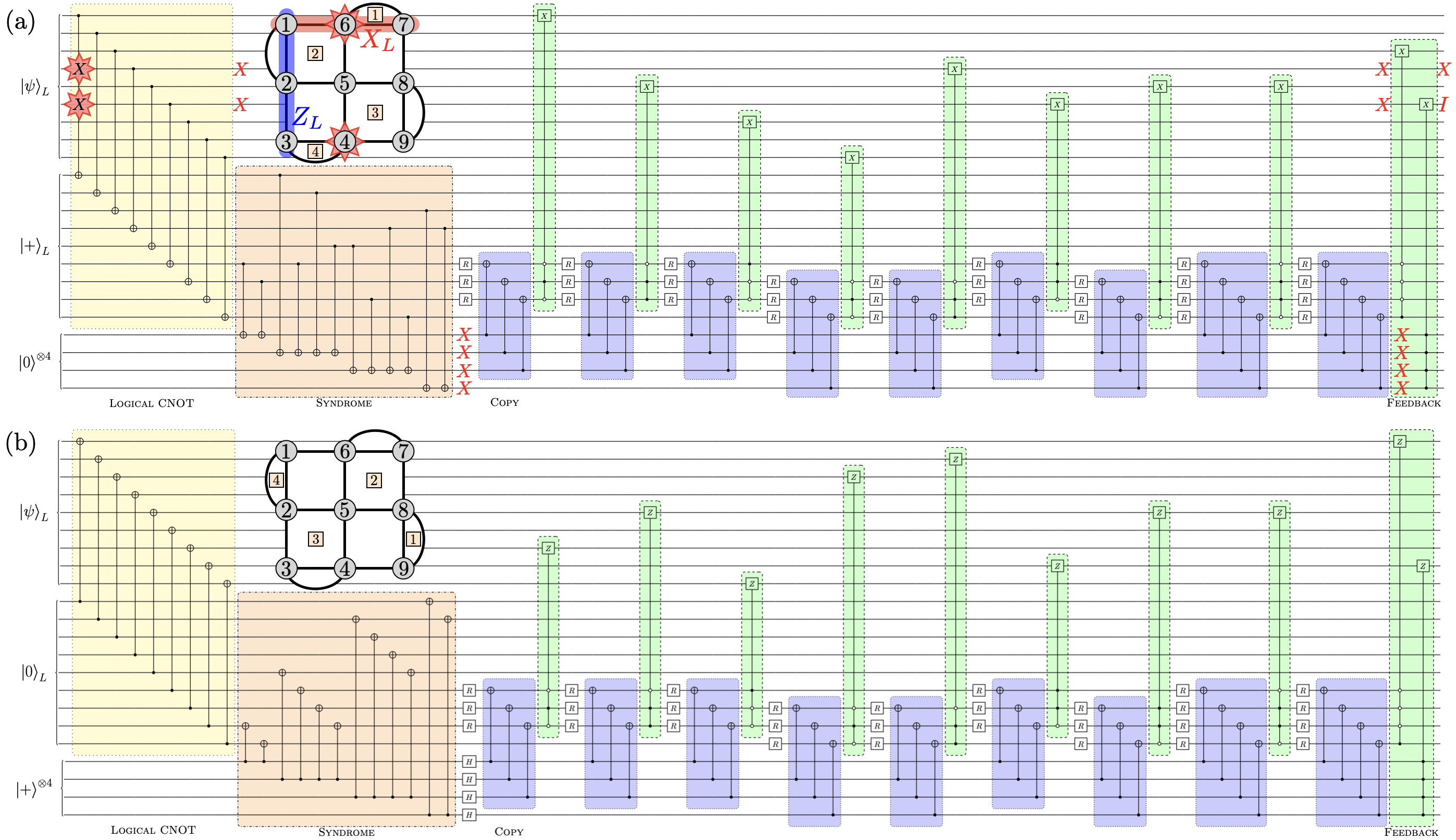}\caption{Measurement-free FT QEC cycle for the distance-3 surface code based on the look up table in Table \ref{tab:surface_lut_red}. (a) $X$-correction block with the correctable weight-2 error $X_4X_6$ (red stars), which is transformed to $X_4$. (b) $Z$-correction block. Note that 7 (3) out of 10 feedback operations in each block are conditional 3-qubit-controlled (4-qubit-controlled) multi-qubit gates.}
	\label{fig:surface}
\end{figure*}

\section{Correcting weight-2 errors in the $d=3$ surface code}\label{sec:surface}

The distance-3 surface code has the property that in addition to all weight-1 errors some weight-2 errors can be corrected, e.g.~with a look-up table decoder \cite{tomita2014low}. The code is defined by the stabilizers
\begin{align}
    K_1^X &= X_8X_9,~~~~~~~~~~~ K_1^Z = Z_6Z_7, \notag \\
    K_2^X &= X_5X_6X_7X_8,~~~ K_2^Z = Z_1Z_2Z_5Z_6, \notag \\
    K_3^X &= X_2X_3X_4X_5,~~~ K_3^Z = Z_4Z_5Z_8Z_9, \notag \\
    K_4^X &= X_1X_2,~~~~~~~~~~~ K_4^Z = Z_3Z_4
\end{align}
and its logical operators can be chosen as $X_L = X_1X_6X_7$ and $Z_L = Z_1Z_2Z_3$ as shown in the insets of Fig.~\ref{fig:surface}. The look up table that is used in our measurement-free EC protocol to correct errors in the surface code is given in Table \ref{tab:surface_lut_red}. For example, the error $X_4X_6$ is correctable because its $Z$-syndrome $\{-1,-1,-1,-1\}$ is not taken up by any other weight-1 $X$-error. Due to the asymmetric arrangement of plaquettes, the error $Z_4Z_6$ is not correctable. Its $X$-syndrome $\{+1,-1,-1,+1\}$ is already in use to correct the weight-1 error $Z_5$. This is why applying a multi-controlled-multi-target gate for the correction of $X_4X_6$ would destroy fault tolerance. In our scheme, however, a multi-controlled feedback gate will only couple to a single data qubit in order to preserve fault tolerance. Weight-2 errors could still be corrected by copying the respective syndrome twice and applying two distinct weight-1 feedback operations conditioned on this same syndrome instead of applying one weight-2 feedback operation. For instance, the $Z$-syndrome $\{-1,-1,-1,-1\}$ must be copied twice to correct $X_4$ and $X_6$ distinctly. This, however, would increase the circuit depth and gate count of the coherent feedback circuit block considerably. Instead of applying weight-2 corrections, it is sufficient to transform weight-2 errors into correctable weight-1 errors -- these will be corrected for in the subsequent EC round. This is sufficient to have a fully FT EC protocol. This conversion of weight-2 into weight-1 errors can be achieved for $X$- and $Z$-errors by the recovery operations given in Table \ref{tab:surface_lut_red}, which are translated into the feedback structure of the circuits in Fig.~\ref{fig:surface}. Note that no weight-2 recoveries need to be applied. For the error $X_4X_6$ for instance, the circuit applies the recovery $X_6$ and leaves the correctable error $X_4$ on the logical data qubit. Note that, at the same time, all nine possible weight-1 errors will be corrected by the protocol, as required. Overall, this provides a compact fully FT and MF protocol, with the respective EC half-cycles for $X$- and $Z$-type correction implemented by the circuits in Fig.~\ref{fig:surface}a and b, respectively. Measurement-free and FT initialization of the auxiliary logical qubit is possible using, e.g., the encoding protocol recently demonstrated in Ref.~\cite{goto2023measurementfree}.

\begin{table}\begin{center}
\begin{tabular}{c | c | c | c}
Syndrome \\ ($K_1^Z, K_2^Z, K_3^Z, K_4^Z$) & Error $E$ & Recovery $R$ & Outcome $RE$ \\ \hline 
$++++$ & $I$ & $I$ & $I$ \\
$+++-$ & $X_3$ & $X_3$ & $I$ \\
$++-+$ & $X_8$ (or $X_9$) & $X_9$ & $K_1^X$ (or $I$) \\ 
$++--$ & $X_4$ & $X_4$ & $I$ \\
$+-++$ & $X_1$ (or $X_2$) & $X_1$ & $I$ (or $K_4^X$)\\ 
$+-+-$ & $X_1X_3$ & $X_1$ & $X_3$ \\
$+--+$ & $X_5$ & $X_5X_5X_5$ & $I$ \\
$+---$ & $X_3X_5$ & $X_5$ & $X_3$ \\
$-+++$ & $X_7$ & $X_7$ & $I$ \\
$-++-$ & $X_3X_7$ & $X_7$ & $X_3$ \\
$-+-+$ & $X_7X_9$ & $X_9$ & $X_7$ \\ 
$-+--$ & $X_4X_7$ & $X_4$ & $X_7$ \\
$--++$ & $X_6$ & $X_6$ & $I$ \\
$--+-$ & $X_3X_6$ & $X_6$ & $X_3$ \\
$---+$ & $X_5X_7$ & $X_5$ & $X_7$ \\
$----$ & $X_4X_6$ & $X_6$ & $X_4$ \\ \hline\hline
\rule{0pt}{3ex} Syndrome \\ ($K_1^X, K_2^X, K_3^X, K_4^X$) & Error $E$ & Recovery $R$ & Outcome $RE$ \\ \hline 
$++++$ & $I$ & $I$ & $I$ \\
$+++-$ & $Z_1$& $Z_1$ & $I$ \\
$++-+$ & $Z_4$ (or $Z_3$) & $Z_3$ & $K_4^Z$ (or $I$) \\ 
$++--$ & $Z_2$ & $Z_2$ & $I$ \\
$+-++$ & $Z_7$ (or $Z_6$)& $Z_7$ & $I$ (or $K_1^Z$)\\
$+-+-$ & $Z_1Z_7$ & $Z_7$ & $Z_1$ \\
$+--+$ & $Z_5$& $Z_5Z_5Z_5$ & $I$ \\
$+---$ & $Z_1Z_5$& $Z_5$ & $Z_1$ \\
$-+++$ & $Z_9$& $Z_9$ & $I$ \\
$-++-$ & $Z_1Z_9$& $Z_9$ & $Z_1$ \\
$-+-+$ & $Z_3Z_9$& $Z_3$ & $Z_9$ \\ 
$-+--$ & $Z_2Z_9$& $Z_2$ & $Z_9$ \\
$--++$ & $Z_8$& $Z_8$ & $I$ \\
$--+-$ & $Z_1Z_8$& $Z_8$ & $Z_1$ \\
$---+$ & $Z_5Z_9$& $Z_5$ & $Z_9$ \\
$----$ & $Z_2Z_8$& $Z_8$ & $Z_2$
\end{tabular} 
\end{center}
\caption{The 4-bit $Z$-syndrome (upper half) of the surface code shown in the inset of Fig.~\ref{fig:surface}a allows one to correct not only all 9 single-qubit $X$-errors but also some additional weight-2 $X$-errors. Note that, for instance, applying $X_8$ corrects both errors $X_8$ and $X_9$ since $K_1^X = X_8X_9$ is a stabilizer. The third column contains the recovery operations that are applied by the circuit in Fig.~\ref{fig:surface} and the last column shows that the result $RE$ is always an error of at most weight-1. The $X$-syndromes and corresponding $Z$-type operations are given in the lower half of the table and the $X$-stabilizers are shown in the inset of Fig.~\ref{fig:surface}b.}
\label{tab:surface_lut_red}
\end{table}

\section{Decoding variant}\label{sec:decoding}

\begin{figure}\includegraphics[width=0.99\linewidth]{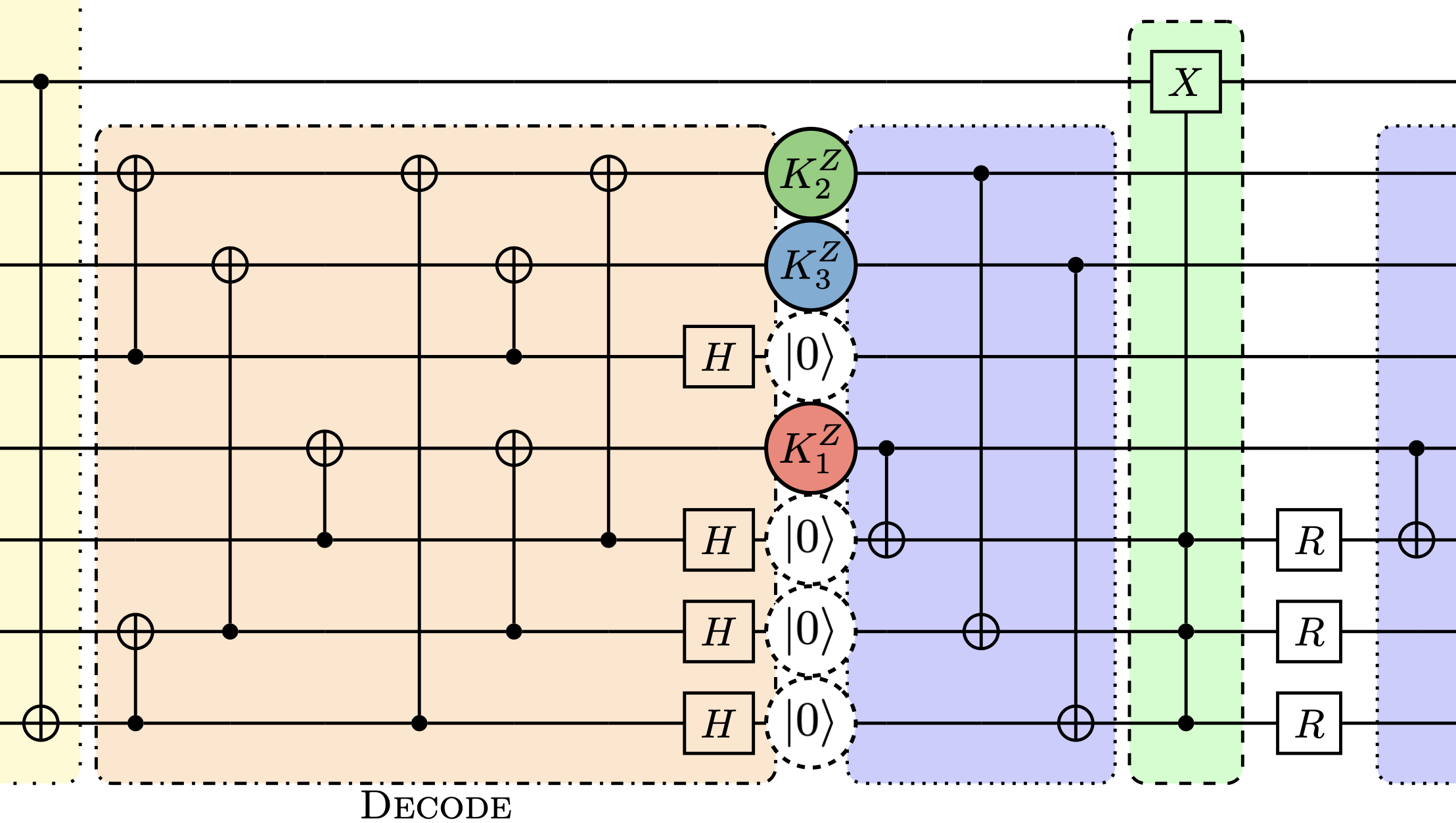}\caption{Performing an inverse encoding circuit for the $\ket{+}_L$ state leaves the $Z$-syndrome $(K_1^Z, K_2^Z, K_3^Z)$ on qubits 4, 1 and 2 of the orange block as indicated respectively by the RGB-colored circles. The syndrome mapping step (orange box in Fig.~\ref{circ:cec}a) is replaced with the decoding block and the copy and feedback steps are adjusted accordingly. In the absence of noise, each of the four other physical qubits is in the $\ket{0}$ state after decoding (dashed white circles).}
	\label{circ:scheme_x_dec}
\end{figure}

The syndrome mapping step from the $n$ physical qubits that hold the logical auxiliary qubit state to the $a$ physical qubits that hold the syndrome information can be simplified because the logical auxiliary qubit is discarded after the syndrome mapping anyways. Instead of supplying fresh auxiliary qubits, one may replace the syndrome mapping step (orange box in Fig.~\ref{circ:cec}a) with a decoding circuit. This can reduce the required number of qubits to perform the MFEC cycle to $N = 2n$. The decoding circuits can be constructed by inverting encoding circuits of, for instance, surface code or color code states with similar numbers of CNOT gates as the syndrome mapping \cite{lang2012minimal, amaro2020scalable}. An example for the $X$-correction block of the Steane code, where the logical auxiliary qubit is initialized to the $\ket{+}_L$ state, is shown in Fig.~\ref{circ:scheme_x_dec}. Here, the physical qubits 1, 2 and 4 carry the expectation value of the $Z$-type stabilizer generators $K_2^Z$, $K_3^Z$ and $K_1^Z$ respectively. The expectation value is $+1(-1)$ if the physical qubit is in the $\ket{0}(\ket{1})$ state.

Note that, as for the syndrome mapping described in the main text, the decoding block itself does not need any fault tolerance overhead for the full MFEC scheme to be fault-tolerant. If there are not enough physical qubits after the decoding step to coherently copy the syndrome information, additional auxiliary qubits must be supplied.

The minimal circuit depth is unchanged with this modification of the orange block since the decoding in Fig.~\ref{circ:scheme_x_dec} needs four time steps, just as the syndrome mapping. However, one could use, for instance, a sequence of only three MS$_4$ gates, if practically available, to perform the decoding step \cite{nigg2014quantum}.

\section{Noise model}\label{sec:noise}

We consider the depolarizing channel of strength $p$ 
\begin{align}
	\curlE_p(\rho) = (1-p)\rho + \frac{p}{4^q-1} \sum_{\sigma \in \Lambda} \left( \bigotimes_{j=1}^q \sigma_j\right) \rho \left( \bigotimes_{j=1}^q \sigma_j \right) \label{eq:depol}
\end{align}
where $\Lambda = \{I,\,X,\,Y,\,Z\}^{\otimes q} \backslash \{I^{\otimes q}\}$ and $q$ is the number of qubits the noise channel acts on. It is the most general noise channel in the sense that the $4^q$ Pauli operators (and the identity operation) form the basis of the $q$-qubit Pauli group. Thus, any fault that can happen in a physical gate can be expressed in this Pauli basis. As a consequence, if the circuit is FT under the channel in Eq.~\eqref{eq:depol}, then it is FT towards any noise channel on the $q$ qubits.

For the single-parameter noise model in Secs.~\ref{sec:advantage} and \ref{sec:cnotimpl} we use the conventional depolarizing noise model where
\begin{enumerate}
    \item a single-qubit gate is followed by a Pauli fault drawn uniformly and independently from $\{X,Y,Z\}$ with probability $p/3$,
    \item a two-qubit gate is followed by a two-Pauli fault drawn uniformly and independently from $\{I,X,Y,Z\}^{\otimes 2} \backslash I \otimes I$ with probability $p/15$,
    \item qubit initialization is flipped ($\ket{0} \rightarrow \ket{1}$) with probability $2p/3$,
    \item qubit measurements yield a flipped result ($\pm 1 \rightarrow \mp 1$) with probability $2p/3$ and
    \item idling locations are followed by a Pauli fault drawn uniformly and independently from $\{X,Y,Z\}$ with probability $p/100$.
\end{enumerate}

Additionally, for the multi-parameter noise model in Secs.~\ref{sec:multiimpl} and \ref{sec:Rydberg}, $q$-qubit gates are followed by Pauli faults drawn randomly and uniformly from $\{I, X, Y, Z\}^{\otimes q} \backslash I^{\otimes q}$ with respective probabilities $p_q$. For modelling of dephasing noise on idling locations during multi-qubit gates, initializations and measurements in Secs.~\ref{sec:multiimpl} and \ref{sec:Rydberg}, we use the single-qubit channel
\begin{align}
    \curlE_{p_{\text{idle},q}}(\rho) &= (1-p_{\text{idle},q}) \rho + p_{\text{idle},q}Z\rho Z \label{eq:deph}
\end{align}
with respective idling error rates $p_{\text{idle},q}$ during $q$-qubit gates and a rate $p_{\text{idle},i}\,(p_{\text{idle},m})$ during physical qubit initialization (measurement).

\section{Details of MFEC performance}\label{sec:appmeasfreeadv}

In this section, we provide the analytical estimation of a parameter region for advantageous use of the MFEC scheme, given by Eq.~\eqref{eq:mfadv_main}. First, we look at the two limits were only one type of noise, either on physical operations or idling locations, is present in the system. Then we interpolate between these limits and estimate parameter regions of advantageous use of either the MFEC or the SMEC protocol.

Let us consider the limit of vanishing physical operation error rates $p \rightarrow 0$ first so that only idling noise is present in the system. Then, MFEC is advantageous over SMEC when its QEC cycle time $\tau_\text{MF} < \tau_\text{SM}$ is smaller than the syndrome measurement EC cycle time. For their logical failure rates this means that then $p_L^\text{MF}/p_L^\text{SM} < 1$. For an FT protocol, the ratio of logical failure rates is proportional to the squared cycle time ratio
\begin{align}
    \frac{p_L^\text{MF}}{p_L^\text{SM}} &= \left( \frac{\tau_\text{MF}}{\tau_\text{SM}}\right)^2
\end{align}
if idling were the only source of failure and we operate in a regime where $\tau \ll T_2$. Single faults of probability $p_\text{idle} \sim \tau$ cannot lead to failure due to the FT circuit design of the protocol.

Assuming that a single operation in the MF protocol takes time $t_\text{ops}$, we can estimate the total cycle time as 
\begin{align}
    \tau_\text{MF} &\simeq t_\text{ops} \cdot \text{\#operations}
\end{align}
in case all operations are executed sequentially. When the SMEC protocol cycle time is dominated by the time $t_\text{meas}$ to perform a qubit measurement, i.e.~we assume that measurements take much more time than gate operations, we can estimate
\begin{align}
    \tau_\text{SM} &\simeq t_\text{meas} \cdot \text{\#measurements}.
\end{align}

As another limiting case, we assume no idling noise at all, i.e.~the coherence time $T_2 \rightarrow \infty$, so that the operation error rate $p \neq 0$ is the only non-vanishing physical error rate. In this case, the logical failure rates scale like
\begin{align}
    p_L^\text{MF} &\sim (c p)^2 \label{eq:pLmf_exp1} \\
    p_L^\text{SM} &\sim (c' p)^2 \label{eq:pLfl_exp1}
\end{align}
for $p \ll 1$ since no single fault of probability $p$ can cause a logical failure. The two constants $c^2$ and $c'^2$ are determined by the number of \emph{bad} locations, i.e.~fault locations that can lead to failure, for each respective protocol. Their ratio determines the (dis-)advantage of MFEC over SMEC since 
\begin{align}
    \frac{p_L^\text{MF}}{p_L^\text{SM}} &= \left(\frac{c}{c'}\right)^2. \label{eq:pLratio}
\end{align}

\begin{figure*}
    \includegraphics[width=0.99\linewidth]{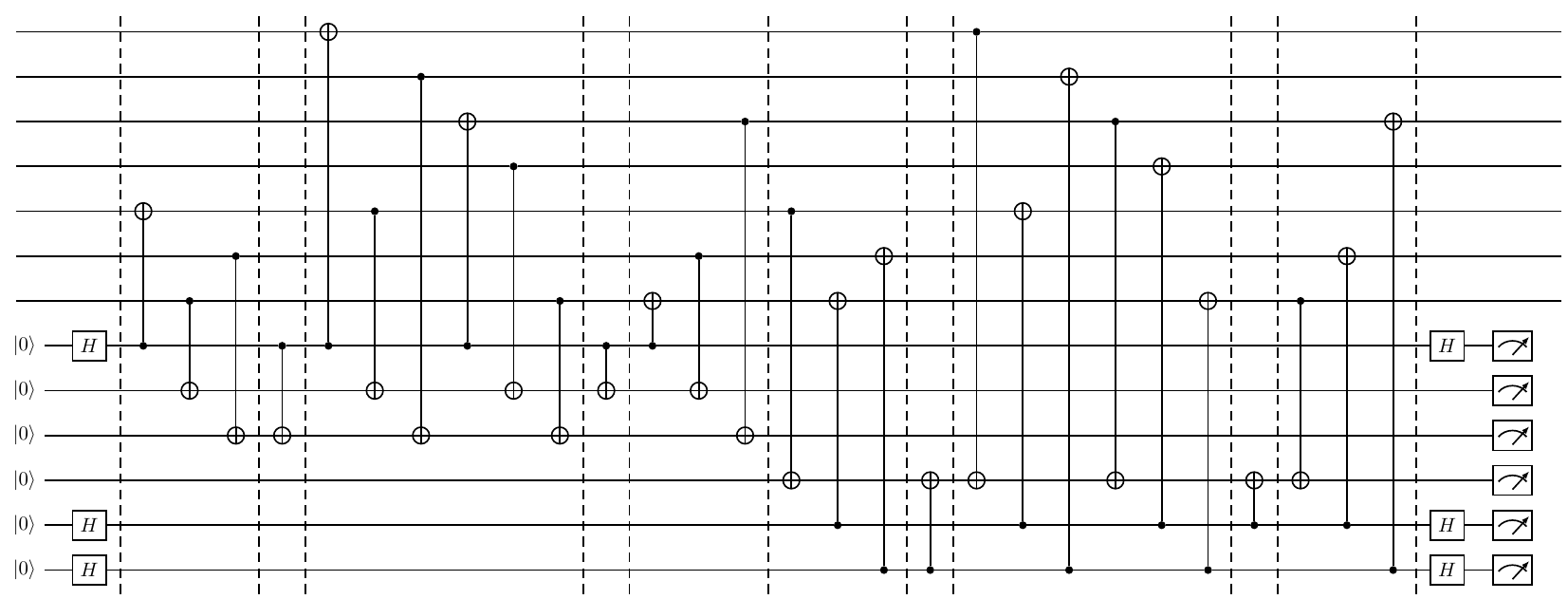}\caption{Parallel measurement of three stabilizers in two sequential steps realizes the flag fault-tolerant circuit for error detection \cite{reichardt2020fault}. Due to the interleaved scheduling of CNOT gates, all six auxiliary qubits act as measurement and flag qubits at the same time.}
    \label{circ:flaginter}
\end{figure*}

Now, in a realistic scenario where both idling and operations are prone to noise, the small-$p$ behavior of the logical failure rate will be
\begin{align}
    p_L^\text{MF} &\sim c^2p^2 + \Tilde{c}^2p_{\text{idle,op}}^2 + bp p_{\text{idle,op}}\label{eq:pLmf_exp11}\\
    p_L^\text{SM} &\sim c'^2p^2 + \Tilde{c}'^2p_{\text{idle},m}^2 + b' p p_{\text{idle},m}\label{eq:pLfl_exp11}
\end{align}
for $p,\,p_\text{idle} \ll 1$ since at least two faults in total are needed to cause logical failure for both protocols: either two faults on operations with probability $p$ each or two faults on idling locations with probability $p_{\text{idle}}$ each or one operation fault with probability $p$ and another idling fault with probability $p_{\text{idle}}$ can cause logical failure. The constants $\Tilde{c},\Tilde{c}',b$ and $b'$ are determined analogously to $c$ and $c'$ by the number of bad locations of these respective fault combinations. Here we included the respective dominant source of idling noise for both protocols; idling during operations with rate $p_{\text{idle,op}}$ for the MF protocol and idling during measurements with rate $p_{\text{idle},m}$ for the SM protocol. Again, we assume for the latter that measurements are much slower than operations.

We can upper-bound Eqs.~\eqref{eq:pLmf_exp11} and \eqref{eq:pLfl_exp11} by assuming that the constants $c,\Tilde{c},b,c',\Tilde{c}'$ and $b'$ represent the \emph{total} number of respective circuit locations for two operation faults, two idling faults or both one operation and one idling fault for either protocol. Then $b$ and $b'$ can be expressed as the products $c\Tilde{c}$ and $c'\Tilde{c}'$ respectively. In principle, one could also determine these constants by exhaustively counting the numbers of bad locations, i.e.~placing all possible combinations of two fault operators on operation and idling locations, or estimate the fraction of bad locations to total locations via Monte Carlo simulation. The ratio of logical failure rates from Eq.~\eqref{eq:pLratio} is then extended to read
\begin{align}
    \frac{p_L^\text{MF}}{p_L^\text{SM}} &\sim \frac{(c p)^2 + (\Tilde{c} p_{\text{idle,op}})^2 + c \Tilde{c} p p_{\text{idle,op}}}{(c' p)^2 + (\Tilde{c}' p_{\text{idle},m})^2 + c' \Tilde{c}' p p_{\text{idle},m}} \label{eq:pLmf_exp2}
\end{align}
where all contributions come with their own constants $c, c', \Tilde{c}, \Tilde{c}'$. 

We can expand Eq.~\eqref{eq:pLmf_exp2} around $p = 0$ and $p_{\text{idle,op}} = 0$ so that we include all second order terms:
\begin{align}
    \frac{p_L^\text{MF}}{p_L^\text{SM}} &= \left(\frac{\Tilde{c}^2}{\Tilde{c}'^2} \frac{p_{\text{idle,op}}^2}{p_{\text{idle},m}^2} + \mathcal{O}(p_{\text{idle,op}}^3) \right) \notag \\
    &+ p \left( \frac{c\Tilde{c}p_{\text{idle,op}}}{\Tilde{c}'^2 p_{\text{idle},m}^2} + \mathcal{O}(p_{\text{idle,op}}^2) \right) \notag \\ 
    &+ p^2 \left( \frac{c^2}{\Tilde{c}'^2 p_{\text{idle},m}^2} + \mathcal{O}(p_{\text{idle,op}}) \right) + \mathcal{O}(p^3).
\end{align}

\begin{figure*}
    \includegraphics[width=0.99\linewidth]{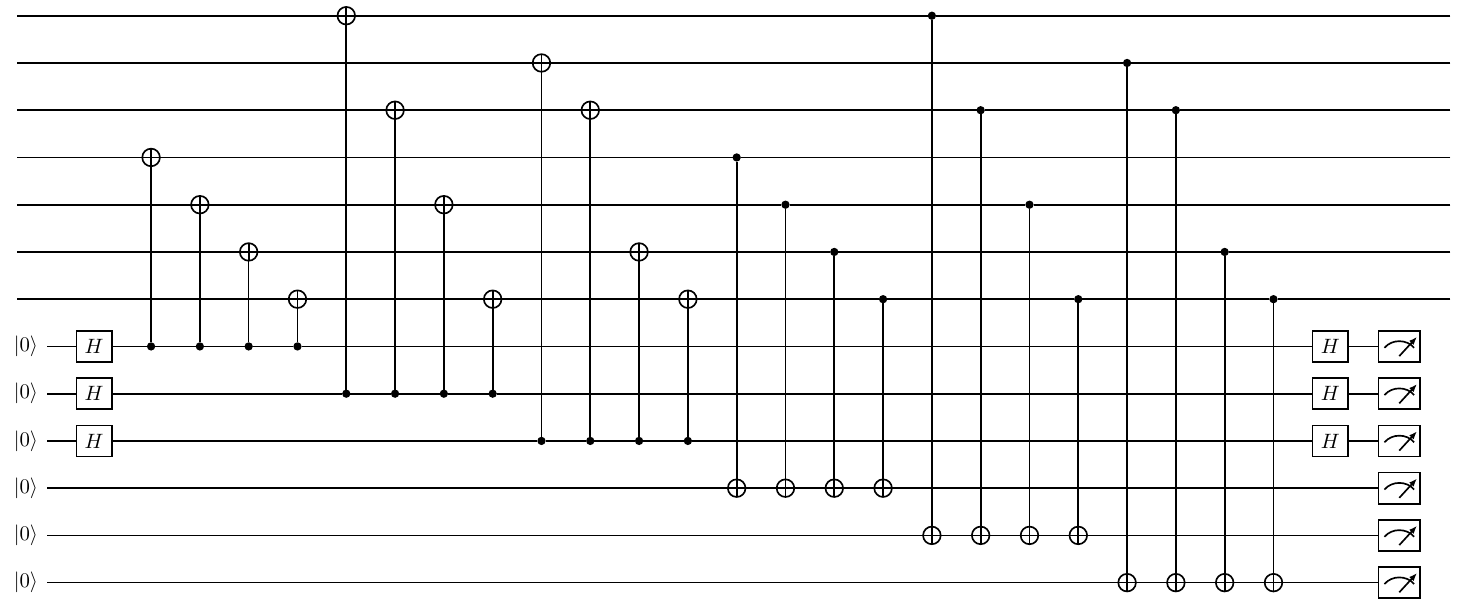}\caption{Non-flagged circuit to readout all six stabilizers sequentially. Combined with the flag information from the circuit in Fig.~\ref{circ:flaginter}, this circuit is used for FT syndrome readout.}
    \label{circ:flagnfs}
\end{figure*}

We roughly estimate the MF advantage $p_L^\text{MF}/p_L^\text{SM} \leq 1$, as given by Eq.~\eqref{eq:mfadv_main}, whenever the ratio of idling times (or error rates, provided they are sufficiently small due to $t/T_2 \ll 1$) 
\begin{align}
    \frac{t_\text{meas}}{t_\text{ops}} \approx \frac{p_{\text{idle},m}}{p_{\text{idle,op}}} &\geq \sqrt{\frac{\Tilde{c}^2}{\Tilde{c}'^2} + \frac{c \Tilde{c}}{\Tilde{c}'^2} \frac{p}{p_{\text{idle,op}}} + \frac{c^2}{\Tilde{c}'^2} \frac{p^2}{p_{\text{idle,op}}^2}} \label{eq:mfadv}
\end{align}
is larger then the bound set by our estimation of the error rate ratios, which translates to 
\begin{align}
    \frac{t_\text{meas}}{t_\text{ops}} &\gtrsim \sqrt{\frac{\Tilde{c}^2}{\Tilde{c}'^2} + 2T_2 \frac{c \Tilde{c}}{\Tilde{c}'^2} \frac{p}{t_\text{ops}} + 4T_2^2\frac{c^2}{\Tilde{c}'^2} \frac{p^2}{t_\text{ops}^2}}.
\end{align}
using Eq.~\eqref{eq:idlrate}. Note again that we have assumed the ratios $\Tilde{c}^2/\Tilde{c}'^2$, $c \Tilde{c}/\Tilde{c}'^2$ and $c^2/\Tilde{c}'^2$ for the numbers of \emph{total} locations to be approximately equal to the ratios using the numbers of \emph{bad} locations. 

The boundary between regions of advantageous use of either MFEC or SMEC is estimated approximately by Eq.~\eqref{eq:mfadv}, which is illustrated in Fig.~\ref{fig:utility} for comparison to a state-of-the-art flag EC protocol \cite{reichardt2020fault}. The flag scheme consists of sequentially running two blocks of three parallel stabilizer measurements. In case a non-trivial measurement occurs, an additional round of non-flagged syndrome readout is performed and the correction is inferred from the flag error set and the Steane code's look up table (see App.~\ref{sec:flag} for details). All gate operations are executed sequentially, i.e.~with only one gate per time step. Physical qubit initializations and measurements are executed in parallel in the simulation.
For the MF scheme the numbers of locations are
\begin{align}
    c &= \text{\#operations (MF)} = 456 \label{eq:c_ops} \\
    \Tilde{c} &= \text{\#idling during operations (MF)} = 5790
\end{align}
and for the flag scheme we use
\begin{align}
    c' &= \text{\#operations (FL)} = 88 \\
    \Tilde{c}' &= \text{\#idling during measurement (FL)} = 14. \label{eq:cprime_idlemeas}
\end{align}
The parameter regions shown in the utility diagram of Fig.~\ref{fig:utility} correspond to these values.

\section{Circuits for flag EC}\label{sec:flag}

The flag EC scheme used for comparison in Secs.~\ref{sec:advantage}, \ref{sec:cnotimpl} and \ref{sec:Rydberg} was suggested in Ref.~\cite{reichardt2020fault} and has been recently implemented with trapped ions \cite{Ryan-Anderson2021, ryan2022implementing}. It consists of application of the two circuits shown in Figs.~\ref{circ:flaginter} and \ref{circ:flagnfs}.

As a first step, all stabilizers are measured in an interleaved way with the help of six auxiliary qubits. When all six qubits are measured as $+1$, we know that no uncorrectable error is present on the data qubits. When any of the qubits is measured as $-1$ however, we cannot tell whether faults have propagated from within the measurement circuit to the data qubits or there have been faults on the input state already. The additional round of syndrome readout, performed as a second step, is needed to clarify the syndrome of the faulty state. If the syndrome is different from the previously measured one, we interpret the measurement outcome of the first block as a flag event. This means that we apply the appropriate two-qubit correction according to the flag error set if the syndrome is consistent with the possible two-qubit errors. Otherwise, or if the two measured syndromes agree, we apply the single-qubit correction according to the Steane code's look up table. This way, logical failures can only happen with probability $\mathcal{O}(p^2)$.

\section{Multi-ion MS gate circuits} \label{sec:ms4}

In the following, we list the components needed to compile the MFEC scheme in Fig.~\ref{circ:cec} into multi-ion MS gates, as discussed in Sec.~\ref{sec:multiimpl}.

\textbf{CNOT gate.} For the decompositions of CNOT gates into MS$_2$ gates and local rotations $R_\sigma = \exp \left(-\ii \frac{\theta}{2} \sigma \right)$ with $\sigma \in \{X,Y,Z\}$ we use the identity 
\begin{align}
    &\text{C}_i\text{NOT}_j \label{eq:msdecomp} \\ 
    &= R_Y^{(i)}(-\pi/2) R_X^{(i)}(\pi/2) R_X^{(j)}(\pi/2) \ms_2^{(ij)}(-\pi/2) R_Y^{(i)}(\pi/2) \notag
\end{align}
for a gate acting on qubits $i$ and/or $j$ \cite{maslov2017basic}.

\begin{figure*}
    \centering
    \includegraphics[width=0.99\linewidth]{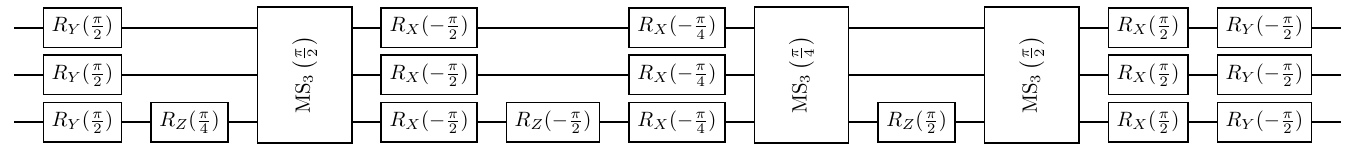}\caption{The Toffoli gate used for the quantum feedback in Fig.~\ref{circ:ft_det_0} can be compiled to multi-ion MS gates given by Eq.~\eqref{eq:ms} and local rotations $R_\sigma = \exp \left(-\ii \frac{\theta}{2} \sigma \right)$ with $\sigma \in \{X,Y,Z\}$ \cite{martinez2016compiling}. The first two wires are the control qubits and the third wire is the target qubit. The decomposition is exact. The last $X$- and $Y$-rotations on the control qubits could be omitted in our case due to the subsequent reset.}
    \label{circ:ms3_decomp}
\end{figure*}

\textbf{Toffoli gate.} The Toffoli gate decomposition into three MS$_3$ gates and local rotations from Ref.~\cite{martinez2016compiling} is reproduced in Fig.~\ref{circ:ms3_decomp}.

\begin{figure*}
    \centering
    \includegraphics[width=0.49\linewidth]{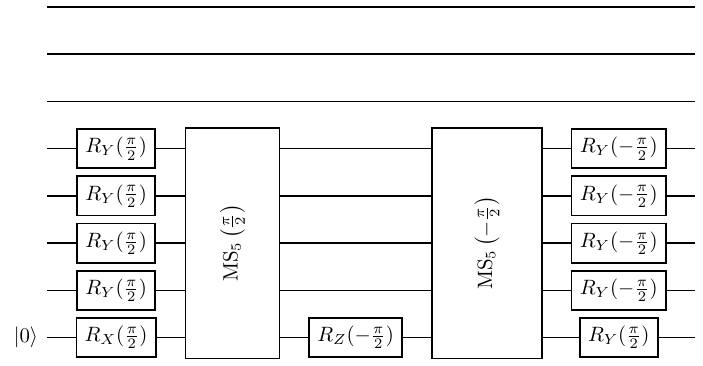}\caption{Circuit to read out the stabilizer $K_1^Z = Z_4Z_5Z_6Z_7$ using two 5-qubit MS gates with $\theta = \pm\frac{\pi}{2}$ and single qubit rotations \cite{muller2011simulating, bermudez2017assessing}. The stabilizer eigenvalue is mapped to the last qubit by the gate sequence. To read out the corresponding $X$-type stabilizer, the $Y$-rotations on the data qubits must be omitted.}
    \label{circ:stab_ms5}
\end{figure*}

\textbf{Syndrome mapping.} The circuit that uses two MS$_5$ gates for mapping of a stabilizer expectation value to a single auxiliary qubit from Refs.~\cite{muller2011simulating, bermudez2017assessing} is reproduced in Fig.~\ref{circ:stab_ms5} for the stabilizer $K_1^Z$. The circuits for the other $Z$-type stabilizers are analogous. For the mapping of $X$-type stabilizers, no $Y$-rotations on the data qubits must be applied.

\begin{figure*}[!htbp]
    \centering
    \includegraphics[width=0.99\linewidth]{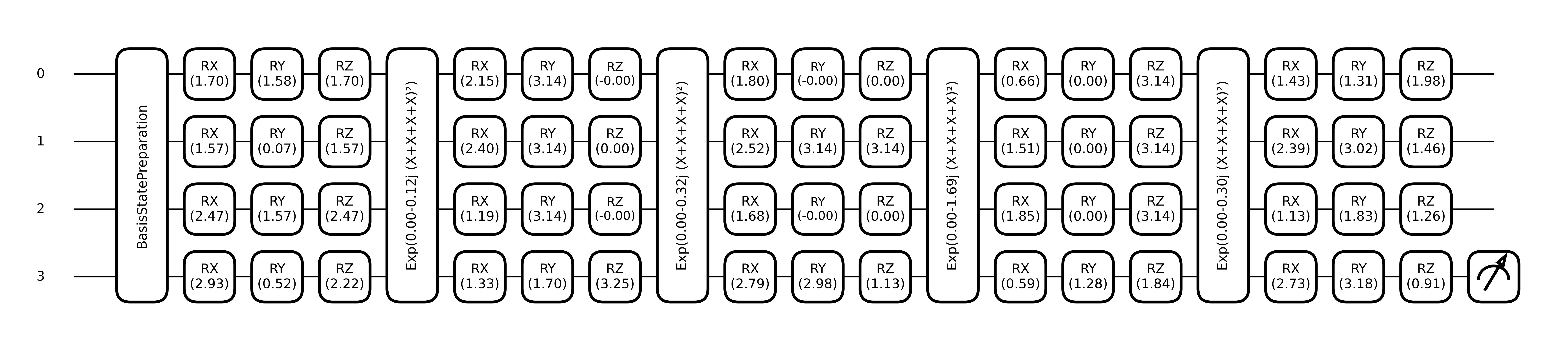}\caption{Compiled circuit with rounded rotation angles that implements the C$_3$NOT gate followed by resetting the control qubits. The first three wires are the control qubits and the fourth wire is the target qubit.}
    \label{circ:ms4_decomp}
\end{figure*}

\begin{figure*}[!htbp]
    \centering
    \includegraphics[width=0.45\linewidth]{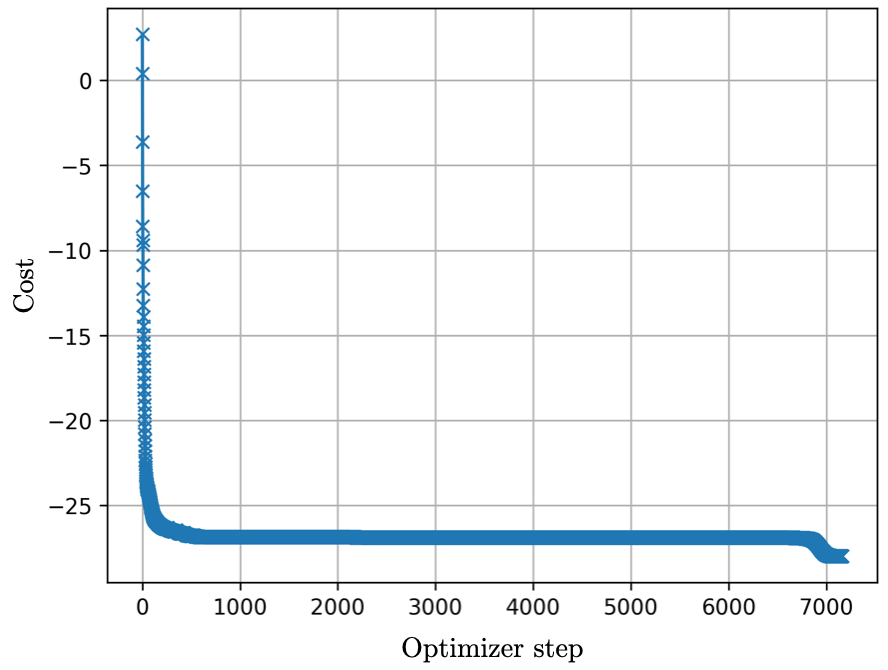}\caption{The cost function converged to the target value of $-28$ for the parametrized circuit in Fig.~\ref{circ:ms4_decomp}.}
    \label{fig:ms4_cost}
\end{figure*}

\textbf{C$_3$NOT gate.} The C$_3$NOT gate followed by reset of the control qubits can be decomposed into four $\ms_4$ gates as shown in Fig.~\ref{circ:ms4_decomp}. We found the decomposition using a parametrized circuit ansatz in pennylane \cite{bergholm2018pennylane}. The parameters $\vec{x}$ are the 64 rotation angles that parametrize the four $\ms_4$ gates and layers of arbitrary $X$-, $Y$- and $Z$-rotations on each qubit before and after the $\ms_4$ gates. As a cost function, we use the weighted average of the Pauli-$X$ and -$Z$ expectation value of the target qubit for application of the parametrized circuit unitary $U(\vec{x})$ to all 16 computational basis states and additional 16 states with the control qubits in the computational basis and the target qubit in the polar basis. The expectation values of the states $U(\vec{x})\ket{1110},\,U(\vec{x})\ket{1111},\,U(\vec{x})\ket{111+}$ and $U(\vec{x})\ket{111-}$ are multiplied by $-7/2$ while all other states have weight $-1/2$ so that the minimal cost function value is $-28$. We keep optimizing the circuit parameters using the \texttt{AdagradOptimizer} in pennylane until the minimal value is found with an absolute tolerance of $10^{-4}$ (see Fig.~\ref{fig:ms4_cost}). The converged angles for all 64 gates in the parametrized circuit are given in Table \ref{tab:ms4}.

\begin{table*}\begin{center}
\begin{tabular}{c | c | c || c | c | c}
time step & rotation & angle $\theta$ & time step & rotation & angle $\theta$ \\ \hline
$ 1 $ & $R_X^{(1)}$ & $ 1.7016306974989441 $ & $ 40 $ & $R_X^{(1)}$ & $ 0.6597193711414078 $\\ 
$ 2 $ & $R_Y^{(1)}$ & $ 1.5795679517126637 $ & $ 41 $ & $R_Y^{(1)}$ & $ 0.000127747992253436 $\\ 
$ 3 $ & $R_Z^{(1)}$ & $ 1.7016364361060135 $ & $ 42 $ & $R_Z^{(1)}$ & $ 3.141326508787639 $\\ 
$ 4 $ & $R_X^{(2)}$ & $ 1.5706691398072636 $ & $ 43 $ & $R_X^{(2)}$ & $ 1.5086374816792918 $\\ 
$ 5 $ & $R_Y^{(2)}$ & $ 0.07488703273604898 $ & $ 44 $ & $R_Y^{(2)}$ & $ 0.0003195950440647524 $\\ 
$ 6 $ & $R_Z^{(2)}$ & $ 1.5708162845196991 $ & $ 45 $ & $R_Z^{(2)}$ & $ 3.141618529495557 $\\ 
$ 7 $ & $R_X^{(3)}$ & $ 2.4733498110482617 $ & $ 46 $ & $R_X^{(3)}$ & $ 1.851040827063517 $\\ 
$ 8 $ & $R_Y^{(3)}$ & $ 1.570978087668805 $ & $ 47 $ & $R_Y^{(3)}$ & $ 2.5330048710337304\cdot 10^{-5}$\\ 
$ 9 $ & $R_Z^{(3)}$ & $ 2.473338798284629 $ & $ 48 $ & $R_Z^{(3)}$ & $ 3.141620662584241 $\\ 
$ 10 $ & $R_X^{(4)}$ & $ 2.925198421469473 $ & $ 49 $ & $R_X^{(4)}$ & $ 0.5851416542930774 $\\ 
$ 11 $ & $R_Y^{(4)}$ & $ 0.5184554247895697 $ & $ 50 $ & $R_Y^{(4)}$ & $ 1.2817745160285958 $\\ 
$ 12 $ & $R_Z^{(4)}$ & $ 2.2155872069814624 $ & $ 51 $ & $R_Z^{(4)}$ & $ 1.8434910130310886 $\\ 
$ 13 $ & MS & $ 0.47618449796926465 $  & $ 52 $ & MS & $ 1.1852630143343514 $ \\ \hline
$ 14 $ & $R_X^{(1)}$ & $ 2.1467351234080563 $ & $ 53 $ & $R_X^{(1)}$ & $ 1.4334708713034108 $\\ 
$ 15 $ & $R_Y^{(1)}$ & $ 3.141888995985973 $ & $ 54 $ & $R_Y^{(1)}$ & $ 1.3149000591854494 $\\ 
$ 16 $ & $R_Z^{(1)}$ & $ -0.0007547728309335194 $ & $ 55 $ & $R_Z^{(1)}$ & $ 1.9799754514097745 $\\ 
$ 17 $ & $R_X^{(2)}$ & $ 2.400499912253685 $ & $ 56 $ & $R_X^{(2)}$ & $ 2.385180572177377 $\\ 
$ 18 $ & $R_Y^{(2)}$ & $ 3.1415882232243812 $ & $ 57 $ & $R_Y^{(2)}$ & $ 3.0169184347659774 $\\ 
$ 19 $ & $R_Z^{(2)}$ & $ 6.7559904951046\cdot 10^{-5}$ & $ 58 $ & $R_Z^{(2)}$ & $ 1.4620953315141385 $\\ 
$ 20 $ & $R_X^{(3)}$ & $ 1.185074722683019 $ & $ 59 $ & $R_X^{(3)}$ & $ 1.1258670832503659 $\\ 
$ 21 $ & $R_Y^{(3)}$ & $ 3.141691576333686 $ & $ 60 $ & $R_Y^{(3)}$ & $ 1.8289082653712643 $\\ 
$ 22 $ & $R_Z^{(3)}$ & $ -3.562993822558777\cdot 10^{-5}$ & $ 61 $ & $R_Z^{(3)}$ & $ 1.257061399856186 $\\ 
$ 23 $ & $R_X^{(4)}$ & $ 1.3267674839439592 $ & $ 62 $ & $R_X^{(4)}$ & $ 2.726979125875662 $\\ 
$ 24 $ & $R_Y^{(4)}$ & $ 1.697674066992338 $ & $ 63 $ & $R_Y^{(4)}$ & $ 3.182404050384048 $\\ 
$ 25 $ & $R_Z^{(4)}$ & $ 3.2542541431385135 $ & $ 64 $ & $R_Z^{(4)}$ & $ 0.9113006716946862 $\\ 
$ 26 $ & MS & $ 1.2961753988001792 $ & & & \\\hline
$ 27 $ & $R_X^{(1)}$ & $ 1.80332241033805 $& & & \\
$ 28 $ & $R_Y^{(1)}$ & $ -9.14307205449397\cdot 10^{-6}$& & & \\
$ 29 $ & $R_Z^{(1)}$ & $ 0.00013217255162876435 $& & & \\
$ 30 $ & $R_X^{(2)}$ & $ 2.5206093101178855 $& & & \\
$ 31 $ & $R_Y^{(2)}$ & $ 3.141617687717394 $& & & \\
$ 32 $ & $R_Z^{(2)}$ & $ 3.141423898780103 $& & & \\
$ 33 $ & $R_X^{(3)}$ & $ 1.6829531013144419 $& & & \\
$ 34 $ & $R_Y^{(3)}$ & $ -1.2744701675664226\cdot 10^{-5}$& & & \\
$ 35 $ & $R_Z^{(3)}$ & $ 1.1892895790349708\cdot 10^{-5}$& & & \\
$ 36 $ & $R_X^{(4)}$ & $ 2.7949214025522897 $& & & \\
$ 37 $ & $R_Y^{(4)}$ & $ 2.9814542086299864 $& & & \\
$ 38 $ & $R_Z^{(4)}$ & $ 1.134224555725302 $& & & \\
$ 39 $ & MS & $ 6.779626443860337 $ & & &
\end{tabular} 
\end{center}
\caption{Angle parameters for the C$_3$NOT gate decomposition in Fig.~\ref{circ:ms4_decomp}.}
\label{tab:ms4}
\end{table*}
 
\clearpage
\bibliographystyle{bibstyle}
\bibliography{references}

\begin{thebibliography}{109}%
\makeatletter
\providecommand \@ifxundefined [1]{%
 \@ifx{#1\undefined}
}%
\providecommand \@ifnum [1]{%
 \ifnum #1\expandafter \@firstoftwo
 \else \expandafter \@secondoftwo
 \fi
}%
\providecommand \@ifx [1]{%
 \ifx #1\expandafter \@firstoftwo
 \else \expandafter \@secondoftwo
 \fi
}%
\providecommand \natexlab [1]{#1}%
\providecommand \enquote  [1]{``#1''}%
\providecommand \bibnamefont  [1]{#1}%
\providecommand \bibfnamefont [1]{#1}%
\providecommand \citenamefont [1]{#1}%
\providecommand \href@noop [0]{\@secondoftwo}%
\providecommand \href [0]{\begingroup \@sanitize@url \@href}%
\providecommand \@href[1]{\@@startlink{#1}\@@href}%
\providecommand \@@href[1]{\endgroup#1\@@endlink}%
\providecommand \@sanitize@url [0]{\catcode `\\12\catcode `\$12\catcode
  `\&12\catcode `\#12\catcode `\^12\catcode `\_12\catcode `\%12\relax}%
\providecommand \@@startlink[1]{}%
\providecommand \@@endlink[0]{}%
\providecommand \url  [0]{\begingroup\@sanitize@url \@url }%
\providecommand \@url [1]{\endgroup\@href {#1}{\urlprefix }}%
\providecommand \urlprefix  [0]{URL }%
\providecommand \Eprint [0]{\href }%
\providecommand \doibase [0]{https://doi.org/}%
\providecommand \selectlanguage [0]{\@gobble}%
\providecommand \bibinfo  [0]{\@secondoftwo}%
\providecommand \bibfield  [0]{\@secondoftwo}%
\providecommand \translation [1]{[#1]}%
\providecommand \BibitemOpen [0]{}%
\providecommand \bibitemStop [0]{}%
\providecommand \bibitemNoStop [0]{.\EOS\space}%
\providecommand \EOS [0]{\spacefactor3000\relax}%
\providecommand \BibitemShut  [1]{\csname bibitem#1\endcsname}%
\let\auto@bib@innerbib\@empty
\bibitem [{\citenamefont {Campbell}\ \emph {et~al.}(2017)\citenamefont
  {Campbell}, \citenamefont {Terhal},\ and\ \citenamefont
  {Vuillot}}]{Campbell2017}%
  \BibitemOpen
  \bibfield  {author} {\bibinfo {author} {E.~T. Campbell}, \bibinfo {author}
  {B.~M. Terhal},\ and\ \bibinfo {author} {C.~Vuillot},\ }\emph {{Roads towards
  fault-tolerant universal quantum computation}},\ \href
  {https://doi.org/10.1038/nature23460} {\bibfield  {journal} {\bibinfo
  {journal} {Nature}\ }\textbf {\bibinfo {volume} {549}},\ \bibinfo {pages}
  {172} (\bibinfo {year} {2017})}\BibitemShut {NoStop}%
\bibitem [{\citenamefont {Knill}\ \emph {et~al.}(1996)\citenamefont {Knill},
  \citenamefont {Laflamme},\ and\ \citenamefont {Zurek}}]{knill1996threshold}%
  \BibitemOpen
  \bibfield  {author} {\bibinfo {author} {E.~Knill}, \bibinfo {author}
  {R.~Laflamme},\ and\ \bibinfo {author} {W.~Zurek},\ }\href@noop {} {\emph
  {Threshold Accuracy for Quantum Computation}} (\bibinfo {year} {1996}),\
  \Eprint {https://arxiv.org/abs/quant-ph/9610011} {arXiv:quant-ph/9610011
  [quant-ph]} \BibitemShut {NoStop}%
\bibitem [{\citenamefont {Aharonov}\ and\ \citenamefont
  {Ben-Or}(2008)}]{Aharonov2008}%
  \BibitemOpen
  \bibfield  {author} {\bibinfo {author} {D.~Aharonov}\ and\ \bibinfo {author}
  {M.~Ben-Or},\ }\emph {Fault-tolerant quantum computation with constant error
  rate},\ \href {https://doi.org/10.1137/S0097539799359385} {\bibfield
  {journal} {\bibinfo  {journal} {SIAM Journal on Computing}\ }\textbf
  {\bibinfo {volume} {38}},\ \bibinfo {pages} {1207} (\bibinfo {year}
  {2008})}\BibitemShut {NoStop}%
\bibitem [{\citenamefont {Preskill}(1998)}]{preskill1998reliable}%
  \BibitemOpen
  \bibfield  {author} {\bibinfo {author} {J.~Preskill},\ }\emph {{Reliable
  quantum computers}},\ \href {https://doi.org/10.1098/rspa.1998.0167}
  {\bibfield  {journal} {\bibinfo  {journal} {Proceedings of the Royal Society
  of London. Series A: Mathematical, Physical and Engineering Sciences}\
  }\textbf {\bibinfo {volume} {454}},\ \bibinfo {pages} {385} (\bibinfo {year}
  {1998})}\BibitemShut {NoStop}%
\bibitem [{\citenamefont {Linke}\ \emph {et~al.}(2017)\citenamefont {Linke},
  \citenamefont {Gutierrez}, \citenamefont {Landsman}, \citenamefont {Figgatt},
  \citenamefont {Debnath}, \citenamefont {Brown},\ and\ \citenamefont
  {Monroe}}]{linke2017fault}%
  \BibitemOpen
  \bibfield  {author} {\bibinfo {author} {N.~M. Linke}, et~al.,\ }\emph
  {{Fault-tolerant quantum error detection}},\ \href
  {https://doi.org/10.1126/sciadv.1701074} {\bibfield  {journal} {\bibinfo
  {journal} {Science Advances}\ }\textbf {\bibinfo {volume} {3}},\ \bibinfo
  {pages} {e1701074} (\bibinfo {year} {2017})}\BibitemShut {NoStop}%
\bibitem [{\citenamefont {Takita}\ \emph {et~al.}(2017)\citenamefont {Takita},
  \citenamefont {Cross}, \citenamefont
  {C{\ifmmode\acute{o}\else\'{o}\fi}rcoles}, \citenamefont {Chow},\ and\
  \citenamefont {Gambetta}}]{Takita2017}%
  \BibitemOpen
  \bibfield  {author} {\bibinfo {author} {M.~Takita}, \bibinfo {author} {A.~W.
  Cross}, \bibinfo {author} {A.~D. C{\ifmmode\acute{o}\else\'{o}\fi}rcoles},
  \bibinfo {author} {J.~M. Chow},\ and\ \bibinfo {author} {J.~M. Gambetta},\
  }\emph {{Experimental demonstration of fault-tolerant state preparation with
  superconducting qubits}},\ \href
  {https://doi.org/10.1103/PhysRevLett.119.180501} {\bibfield  {journal}
  {\bibinfo  {journal} {Physical Review Letters}\ }\textbf {\bibinfo {volume}
  {119}},\ \bibinfo {pages} {180501} (\bibinfo {year} {2017})}\BibitemShut
  {NoStop}%
\bibitem [{\citenamefont {Andersen}\ \emph {et~al.}(2020)\citenamefont
  {Andersen}, \citenamefont {Remm}, \citenamefont {Lazar}, \citenamefont
  {Krinner}, \citenamefont {Lacroix}, \citenamefont {Norris}, \citenamefont
  {Gabureac}, \citenamefont {Eichler},\ and\ \citenamefont
  {Wallraff}}]{andersen2020repeated}%
  \BibitemOpen
  \bibfield  {author} {\bibinfo {author} {C.~K. Andersen}, et~al.,\ }\emph
  {{Repeated quantum error detection in a surface code}},\ \href
  {https://doi.org/10.1038/s41567-020-0920-y} {\bibfield  {journal} {\bibinfo
  {journal} {Nature Physics}\ }\textbf {\bibinfo {volume} {16}},\ \bibinfo
  {pages} {875} (\bibinfo {year} {2020})}\BibitemShut {NoStop}%
\bibitem [{\citenamefont {Egan}\ \emph {et~al.}(2021)\citenamefont {Egan},
  \citenamefont {Debroy}, \citenamefont {Noel}, \citenamefont {Risinger},
  \citenamefont {Zhu}, \citenamefont {Biswas}, \citenamefont {Newman},
  \citenamefont {Li}, \citenamefont {Brown}, \citenamefont {Cetina},\ and\
  \citenamefont {Monroe}}]{Egan2021}%
  \BibitemOpen
  \bibfield  {author} {\bibinfo {author} {L.~Egan}, et~al.,\ }\emph
  {{Fault-tolerant control of an error-corrected qubit}},\ \href
  {https://doi.org/10.1038/s41586-021-03928-y} {\bibfield  {journal} {\bibinfo
  {journal} {Nature}\ }\textbf {\bibinfo {volume} {598}},\ \bibinfo {pages}
  {281} (\bibinfo {year} {2021})}\BibitemShut {NoStop}%
\bibitem [{\citenamefont {Erhard}\ \emph {et~al.}(2021)\citenamefont {Erhard},
  \citenamefont {Poulsen~Nautrup}, \citenamefont {Meth}, \citenamefont
  {Postler}, \citenamefont {Stricker}, \citenamefont {Stadler}, \citenamefont
  {Negnevitsky}, \citenamefont {Ringbauer}, \citenamefont {Schindler},
  \citenamefont {Briegel}, \citenamefont {Blatt}, \citenamefont {Friis},\ and\
  \citenamefont {Monz}}]{Erhard2021}%
  \BibitemOpen
  \bibfield  {author} {\bibinfo {author} {A.~Erhard}, et~al.,\ }\emph
  {{Entangling logical qubits with lattice surgery}},\ \href
  {https://doi.org/10.1038/s41586-020-03079-6} {\bibfield  {journal} {\bibinfo
  {journal} {Nature}\ }\textbf {\bibinfo {volume} {589}},\ \bibinfo {pages}
  {220} (\bibinfo {year} {2021})}\BibitemShut {NoStop}%
\bibitem [{\citenamefont {Abobeih}\ \emph {et~al.}(2022)\citenamefont
  {Abobeih}, \citenamefont {Wang}, \citenamefont {Randall}, \citenamefont
  {Loenen}, \citenamefont {Bradley}, \citenamefont {Markham}, \citenamefont
  {Twitchen}, \citenamefont {Terhal},\ and\ \citenamefont
  {Taminiau}}]{abobeih2022fault}%
  \BibitemOpen
  \bibfield  {author} {\bibinfo {author} {M.~H. Abobeih}, et~al.,\ }\emph
  {Fault-tolerant operation of a logical qubit in a diamond quantum
  processor},\ \href {https://doi.org/10.1038/s41586-022-04819-6} {\bibfield
  {journal} {\bibinfo  {journal} {Nature}\ }\textbf {\bibinfo {volume} {606}},\
  \bibinfo {pages} {884} (\bibinfo {year} {2022})}\BibitemShut {NoStop}%
\bibitem [{\citenamefont {Postler}\ \emph {et~al.}(2022)\citenamefont
  {Postler}, \citenamefont {Heu{\ss}en}, \citenamefont {Pogorelov},
  \citenamefont {Rispler}, \citenamefont {Feldker}, \citenamefont {Meth},
  \citenamefont {Marciniak}, \citenamefont {Stricker}, \citenamefont
  {Ringbauer}, \citenamefont {Blatt}, \citenamefont {Schindler}, \citenamefont
  {Müller},\ and\ \citenamefont {Monz}}]{postler2022demonstration}%
  \BibitemOpen
  \bibfield  {author} {\bibinfo {author} {L.~Postler}, et~al.,\ }\emph
  {Demonstration of fault-tolerant universal quantum gate operations},\ \href
  {https://doi.org/10.1038/s41586-022-04721-1} {\bibfield  {journal} {\bibinfo
  {journal} {Nature}\ }\textbf {\bibinfo {volume} {605}},\ \bibinfo {pages}
  {675} (\bibinfo {year} {2022})}\BibitemShut {NoStop}%
\bibitem [{\citenamefont {Hilder}\ \emph {et~al.}(2022)\citenamefont {Hilder},
  \citenamefont {Pijn}, \citenamefont {Onishchenko}, \citenamefont {Stahl},
  \citenamefont {Orth}, \citenamefont {Lekitsch}, \citenamefont
  {Rodriguez-Blanco}, \citenamefont {M{\"u}ller}, \citenamefont
  {Schmidt-Kaler},\ and\ \citenamefont {Poschinger}}]{hilder2022fault}%
  \BibitemOpen
  \bibfield  {author} {\bibinfo {author} {J.~Hilder}, et~al.,\ }\emph
  {Fault-tolerant parity readout on a shuttling-based trapped-ion quantum
  computer},\ \href
  {https://doi.org/https://doi.org/10.1103/PhysRevX.12.011032} {\bibfield
  {journal} {\bibinfo  {journal} {Physical Review X}\ }\textbf {\bibinfo
  {volume} {12}},\ \bibinfo {pages} {011032} (\bibinfo {year}
  {2022})}\BibitemShut {NoStop}%
\bibitem [{\citenamefont {Ryan-Anderson}\ \emph {et~al.}(2022)\citenamefont
  {Ryan-Anderson}, \citenamefont {Brown}, \citenamefont {Allman}, \citenamefont
  {Arkin}, \citenamefont {Asa-Attuah}, \citenamefont {Baldwin}, \citenamefont
  {Berg}, \citenamefont {Bohnet}, \citenamefont {Braxton}, \citenamefont
  {Burdick} \emph {et~al.}}]{ryan2022implementing}%
  \BibitemOpen
  \bibfield  {author} {\bibinfo {author} {C.~Ryan-Anderson}, et~al.,\
  }\href@noop {} {\emph {Implementing fault-tolerant entangling gates on the
  five-qubit code and the color code}} (\bibinfo {year} {2022}),\ \Eprint
  {https://arxiv.org/abs/2208.01863} {arXiv:2208.01863 [quant-ph]} \BibitemShut
  {NoStop}%
\bibitem [{\citenamefont {Marques}\ \emph {et~al.}(2022)\citenamefont
  {Marques}, \citenamefont {Varbanov}, \citenamefont {Moreira}, \citenamefont
  {Ali}, \citenamefont {Muthusubramanian}, \citenamefont {Zachariadis},
  \citenamefont {Battistel}, \citenamefont {Beekman}, \citenamefont {Haider},
  \citenamefont {Vlothuizen}, \citenamefont {Bruno}, \citenamefont {Terhal},\
  and\ \citenamefont {DiCarlo}}]{marques2022logical}%
  \BibitemOpen
  \bibfield  {author} {\bibinfo {author} {J.~F. Marques}, et~al.,\ }\emph
  {{Logical-qubit operations in an error-detecting surface code}},\ \href
  {https://doi.org/10.1038/s41567-021-01423-9} {\bibfield  {journal} {\bibinfo
  {journal} {Nature Physics}\ }\textbf {\bibinfo {volume} {18}},\ \bibinfo
  {pages} {80} (\bibinfo {year} {2022})}\BibitemShut {NoStop}%
\bibitem [{\citenamefont {Ryan-Anderson}\ \emph {et~al.}(2021)\citenamefont
  {Ryan-Anderson}, \citenamefont {Bohnet}, \citenamefont {Lee}, \citenamefont
  {Gresh}, \citenamefont {Hankin}, \citenamefont {Gaebler}, \citenamefont
  {Francois}, \citenamefont {Chernoguzov}, \citenamefont {Lucchetti},
  \citenamefont {Brown}, \citenamefont {Gatterman}, \citenamefont {Halit},
  \citenamefont {Gilmore}, \citenamefont {Gerber}, \citenamefont {Neyenhuis},
  \citenamefont {Hayes},\ and\ \citenamefont {Stutz}}]{Ryan-Anderson2021}%
  \BibitemOpen
  \bibfield  {author} {\bibinfo {author} {C.~Ryan-Anderson}, et~al.,\ }\emph
  {{Realization of real-time fault-tolerant quantum error correction}},\ \href
  {https://doi.org/10.1103/PhysRevX.11.041058} {\bibfield  {journal} {\bibinfo
  {journal} {Physical Review X}\ }\textbf {\bibinfo {volume} {11}},\ \bibinfo
  {pages} {041058} (\bibinfo {year} {2021})}\BibitemShut {NoStop}%
\bibitem [{\citenamefont {{Google Quantum AI}}(2021)}]{google2021exponential}%
  \BibitemOpen
  \bibfield  {author} {\bibinfo {author} {{Google Quantum AI}},\ }\emph
  {Exponential suppression of bit or phase errors with cyclic error
  correction},\ \href
  {https://doi.org/https://doi.org/10.1038/s41586-021-03588-y} {\bibfield
  {journal} {\bibinfo  {journal} {Nature}\ }\textbf {\bibinfo {volume} {595}},\
  \bibinfo {pages} {383} (\bibinfo {year} {2021})}\BibitemShut {NoStop}%
\bibitem [{\citenamefont {Krinner}\ \emph {et~al.}(2022)\citenamefont
  {Krinner}, \citenamefont {Lacroix}, \citenamefont {Remm}, \citenamefont
  {Di~Paolo}, \citenamefont {Genois}, \citenamefont {Leroux}, \citenamefont
  {Hellings}, \citenamefont {Lazar}, \citenamefont {Swiadek}, \citenamefont
  {Herrmann} \emph {et~al.}}]{krinner2022realizing}%
  \BibitemOpen
  \bibfield  {author} {\bibinfo {author} {S.~Krinner}, et~al.,\ }\emph
  {Realizing repeated quantum error correction in a distance-three surface
  code},\ \href {https://doi.org/https://doi.org/10.1038/s41586-022-04566-8}
  {\bibfield  {journal} {\bibinfo  {journal} {Nature}\ }\textbf {\bibinfo
  {volume} {605}},\ \bibinfo {pages} {669} (\bibinfo {year}
  {2022})}\BibitemShut {NoStop}%
\bibitem [{\citenamefont {Zhao}\ \emph {et~al.}(2022)\citenamefont {Zhao},
  \citenamefont {Ye}, \citenamefont {Huang}, \citenamefont {Zhang},
  \citenamefont {Wu}, \citenamefont {Guan}, \citenamefont {Zhu}, \citenamefont
  {Wei}, \citenamefont {He}, \citenamefont {Cao} \emph
  {et~al.}}]{zhao2022realization}%
  \BibitemOpen
  \bibfield  {author} {\bibinfo {author} {Y.~Zhao}, et~al.,\ }\emph
  {Realization of an error-correcting surface code with superconducting
  qubits},\ \href
  {https://doi.org/https://doi.org/10.1103/PhysRevLett.129.030501} {\bibfield
  {journal} {\bibinfo  {journal} {Physical Review Letters}\ }\textbf {\bibinfo
  {volume} {129}},\ \bibinfo {pages} {030501} (\bibinfo {year}
  {2022})}\BibitemShut {NoStop}%
\bibitem [{\citenamefont {{Google Quantum AI}}(2023)}]{google2023suppressing}%
  \BibitemOpen
  \bibfield  {author} {\bibinfo {author} {{Google Quantum AI}},\ }\emph
  {Suppressing quantum errors by scaling a surface code logical qubit},\ \href
  {https://doi.org/https://doi.org/10.1038/s41586-022-05434-1} {\bibfield
  {journal} {\bibinfo  {journal} {Nature}\ }\textbf {\bibinfo {volume} {614}},\
  \bibinfo {pages} {676} (\bibinfo {year} {2023})}\BibitemShut {NoStop}%
\bibitem [{\citenamefont {Bluvstein}\ \emph {et~al.}(2022)\citenamefont
  {Bluvstein}, \citenamefont {Levine}, \citenamefont {Semeghini}, \citenamefont
  {Wang}, \citenamefont {Ebadi}, \citenamefont {Kalinowski}, \citenamefont
  {Keesling}, \citenamefont {Maskara}, \citenamefont {Pichler}, \citenamefont
  {Greiner}, \citenamefont {Vuleti{\'{c}}},\ and\ \citenamefont
  {Lukin}}]{bluvstein2022aquantum}%
  \BibitemOpen
  \bibfield  {author} {\bibinfo {author} {D.~Bluvstein}, et~al.,\ }\emph {A
  quantum processor based on coherent transport of entangled atom arrays},\
  \href {https://doi.org/10.1038/s41586-022-04592-6} {\bibfield  {journal}
  {\bibinfo  {journal} {Nature}\ }\textbf {\bibinfo {volume} {604}},\ \bibinfo
  {pages} {451} (\bibinfo {year} {2022})}\BibitemShut {NoStop}%
\bibitem [{\citenamefont {Cong}\ \emph {et~al.}(2022)\citenamefont {Cong},
  \citenamefont {Levine}, \citenamefont {Keesling}, \citenamefont {Bluvstein},
  \citenamefont {Wang},\ and\ \citenamefont {Lukin}}]{cong2022hardware}%
  \BibitemOpen
  \bibfield  {author} {\bibinfo {author} {I.~Cong}, et~al.,\ }\emph
  {Hardware-efficient, fault-tolerant quantum computation with Rydberg atoms},\
  \href {https://doi.org/10.1103/PhysRevX.12.021049} {\bibfield  {journal}
  {\bibinfo  {journal} {Physical Review X}\ }\textbf {\bibinfo {volume} {12}},\
  \bibinfo {pages} {021049} (\bibinfo {year} {2022})}\BibitemShut {NoStop}%
\bibitem [{\citenamefont {Wu}\ \emph {et~al.}(2022)\citenamefont {Wu},
  \citenamefont {Kolkowitz}, \citenamefont {Puri},\ and\ \citenamefont
  {Thompson}}]{wu2022erasure}%
  \BibitemOpen
  \bibfield  {author} {\bibinfo {author} {Y.~Wu}, \bibinfo {author}
  {S.~Kolkowitz}, \bibinfo {author} {S.~Puri},\ and\ \bibinfo {author} {J.~D.
  Thompson},\ }\emph {Erasure conversion for fault-tolerant quantum computing
  in alkaline earth Rydberg atom arrays},\ \href
  {https://doi.org/10.1038/s41467-022-32094-6} {\bibfield  {journal} {\bibinfo
  {journal} {Nature Communications}\ }\textbf {\bibinfo {volume} {13}},\
  \bibinfo {pages} {4657} (\bibinfo {year} {2022})}\BibitemShut {NoStop}%
\bibitem [{\citenamefont {Scholl}\ \emph {et~al.}(2023)\citenamefont {Scholl},
  \citenamefont {Shaw}, \citenamefont {Tsai}, \citenamefont {Finkelstein},
  \citenamefont {Choi},\ and\ \citenamefont {Endres}}]{scholl2023erasure}%
  \BibitemOpen
  \bibfield  {author} {\bibinfo {author} {P.~Scholl}, et~al.,\ }\href@noop {}
  {\emph {Erasure conversion in a high-fidelity Rydberg quantum simulator}}
  (\bibinfo {year} {2023}),\ \Eprint {https://arxiv.org/abs/2305.03406}
  {arXiv:2305.03406 [quant-ph]} \BibitemShut {NoStop}%
\bibitem [{\citenamefont {Ma}\ \emph {et~al.}(2023)\citenamefont {Ma},
  \citenamefont {Liu}, \citenamefont {Peng}, \citenamefont {Zhang},
  \citenamefont {Jandura}, \citenamefont {Claes}, \citenamefont {Burgers},
  \citenamefont {Pupillo}, \citenamefont {Puri},\ and\ \citenamefont
  {Thompson}}]{ma2023highfidelity}%
  \BibitemOpen
  \bibfield  {author} {\bibinfo {author} {S.~Ma}, et~al.,\ }\href@noop {}
  {\emph {High-fidelity gates with mid-circuit erasure conversion in a
  metastable neutral atom qubit}} (\bibinfo {year} {2023}),\ \Eprint
  {https://arxiv.org/abs/2305.05493} {arXiv:2305.05493 [quant-ph]} \BibitemShut
  {NoStop}%
\bibitem [{\citenamefont {Pogorelov}\ \emph {et~al.}(2021)\citenamefont
  {Pogorelov}, \citenamefont {Feldker}, \citenamefont {Marciniak},
  \citenamefont {Postler}, \citenamefont {Jacob}, \citenamefont
  {Krieglsteiner}, \citenamefont {Podlesnic}, \citenamefont {Meth},
  \citenamefont {Negnevitsky}, \citenamefont {Stadler} \emph
  {et~al.}}]{pogorelov2021compact}%
  \BibitemOpen
  \bibfield  {author} {\bibinfo {author} {I.~Pogorelov}, et~al.,\ }\emph
  {Compact ion-trap quantum computing demonstrator},\ \href
  {https://doi.org/https://doi.org/10.1103/PRXQuantum.2.020343} {\bibfield
  {journal} {\bibinfo  {journal} {PRX Quantum}\ }\textbf {\bibinfo {volume}
  {2}},\ \bibinfo {pages} {020343} (\bibinfo {year} {2021})}\BibitemShut
  {NoStop}%
\bibitem [{\citenamefont {Saffman}(2016)}]{saffman2016quantum}%
  \BibitemOpen
  \bibfield  {author} {\bibinfo {author} {M.~Saffman},\ }\emph {Quantum
  computing with atomic qubits and Rydberg interactions: progress and
  challenges},\ \href {https://doi.org/10.1088/0953-4075/49/20/202001}
  {\bibfield  {journal} {\bibinfo  {journal} {Journal of Physics B: Atomic,
  Molecular and Optical Physics}\ }\textbf {\bibinfo {volume} {49}},\ \bibinfo
  {pages} {202001} (\bibinfo {year} {2016})}\BibitemShut {NoStop}%
\bibitem [{\citenamefont {Blinov}\ \emph {et~al.}(2002)\citenamefont {Blinov},
  \citenamefont {Deslauriers}, \citenamefont {Lee}, \citenamefont {Madsen},
  \citenamefont {Miller},\ and\ \citenamefont
  {Monroe}}]{blinov2002sympathetic}%
  \BibitemOpen
  \bibfield  {author} {\bibinfo {author} {B.~Blinov}, et~al.,\ }\emph
  {Sympathetic cooling of trapped ${\mathrm{Cd}}^{+}$ isotopes},\ \href
  {https://doi.org/10.1103/PhysRevA.65.040304} {\bibfield  {journal} {\bibinfo
  {journal} {Physical Review A}\ }\textbf {\bibinfo {volume} {65}},\ \bibinfo
  {pages} {040304} (\bibinfo {year} {2002})}\BibitemShut {NoStop}%
\bibitem [{\citenamefont {Barrett}\ \emph {et~al.}(2003)\citenamefont
  {Barrett}, \citenamefont {DeMarco}, \citenamefont {Schaetz}, \citenamefont
  {Meyer}, \citenamefont {Leibfried}, \citenamefont {Britton}, \citenamefont
  {Chiaverini}, \citenamefont {Itano}, \citenamefont
  {Jelenkovi\ifmmode~\acute{c}\else \'{c}\fi{}}, \citenamefont {Jost},
  \citenamefont {Langer}, \citenamefont {Rosenband},\ and\ \citenamefont
  {Wineland}}]{barrett2003sympathetic}%
  \BibitemOpen
  \bibfield  {author} {\bibinfo {author} {M.~D. Barrett}, et~al.,\ }\emph
  {Sympathetic cooling of ${}^{9}{\mathrm{Be}}^{+}$ and
  ${}^{24}{\mathrm{Mg}}^{+}$ for quantum logic},\ \href
  {https://doi.org/10.1103/PhysRevA.68.042302} {\bibfield  {journal} {\bibinfo
  {journal} {Phys. Rev. A}\ }\textbf {\bibinfo {volume} {68}},\ \bibinfo
  {pages} {042302} (\bibinfo {year} {2003})}\BibitemShut {NoStop}%
\bibitem [{\citenamefont {Kielpinski}\ \emph {et~al.}(2002)\citenamefont
  {Kielpinski}, \citenamefont {Monroe},\ and\ \citenamefont
  {Wineland}}]{kielpinski2002architecture}%
  \BibitemOpen
  \bibfield  {author} {\bibinfo {author} {D.~Kielpinski}, \bibinfo {author}
  {C.~Monroe},\ and\ \bibinfo {author} {D.~J. Wineland},\ }\emph {Architecture
  for a large-scale ion-trap quantum computer},\ \href
  {https://doi.org/https://doi.org/10.1038/nature00784} {\bibfield  {journal}
  {\bibinfo  {journal} {Nature}\ }\textbf {\bibinfo {volume} {417}},\ \bibinfo
  {pages} {709} (\bibinfo {year} {2002})}\BibitemShut {NoStop}%
\bibitem [{\citenamefont {Chiaverini}\ \emph {et~al.}(2005)\citenamefont
  {Chiaverini}, \citenamefont {Blakestad}, \citenamefont {Britton},
  \citenamefont {Jost}, \citenamefont {Langer}, \citenamefont {Leibfried},
  \citenamefont {Ozeri},\ and\ \citenamefont
  {Wineland}}]{chiaverini2005surface}%
  \BibitemOpen
  \bibfield  {author} {\bibinfo {author} {J.~Chiaverini}, et~al.,\ }\href@noop
  {} {\emph {Surface-Electrode Architecture for Ion-Trap Quantum Information
  Processing}} (\bibinfo {year} {2005}),\ \Eprint
  {https://arxiv.org/abs/quant-ph/0501147} {arXiv:quant-ph/0501147 [quant-ph]}
  \BibitemShut {NoStop}%
\bibitem [{\citenamefont {Moses}\ \emph {et~al.}(2023)\citenamefont {Moses},
  \citenamefont {Baldwin}, \citenamefont {Allman}, \citenamefont {Ancona},
  \citenamefont {Ascarrunz}, \citenamefont {Barnes}, \citenamefont
  {Bartolotta}, \citenamefont {Bjork}, \citenamefont {Blanchard}, \citenamefont
  {Bohn} \emph {et~al.}}]{moses2023race}%
  \BibitemOpen
  \bibfield  {author} {\bibinfo {author} {S.~Moses}, et~al.,\ }\href@noop {}
  {\emph {A race track trapped-ion quantum processor}} (\bibinfo {year}
  {2023}),\ \Eprint {https://arxiv.org/abs/2305.03828} {arXiv:2305.03828
  [quant-ph]} \BibitemShut {NoStop}%
\bibitem [{\citenamefont {Graham}\ \emph {et~al.}(2023)\citenamefont {Graham},
  \citenamefont {Phuttitarn}, \citenamefont {Chinnarasu}, \citenamefont {Song},
  \citenamefont {Poole}, \citenamefont {Jooya}, \citenamefont {Scott},
  \citenamefont {Scott}, \citenamefont {Eichler},\ and\ \citenamefont
  {Saffman}}]{graham2023mid}%
  \BibitemOpen
  \bibfield  {author} {\bibinfo {author} {T.~Graham}, et~al.,\ }\href@noop {}
  {\emph {Mid-circuit measurements on a neutral atom quantum processor}}
  (\bibinfo {year} {2023}),\ \Eprint {https://arxiv.org/abs/2303.10051}
  {arXiv:2303.10051 [quant-ph]} \BibitemShut {NoStop}%
\bibitem [{\citenamefont {Norcia}\ \emph {et~al.}(2023)\citenamefont {Norcia},
  \citenamefont {Cairncross}, \citenamefont {Barnes}, \citenamefont
  {Battaglino}, \citenamefont {Brown}, \citenamefont {Brown}, \citenamefont
  {Cassella}, \citenamefont {Chen}, \citenamefont {Coxe}, \citenamefont {Crow},
  \citenamefont {Epstein}, \citenamefont {Griger}, \citenamefont {Jones},
  \citenamefont {Kim}, \citenamefont {Kindem}, \citenamefont {King},
  \citenamefont {Kondov}, \citenamefont {Kotru}, \citenamefont {Lauigan},
  \citenamefont {Li}, \citenamefont {Lu}, \citenamefont {Megidish},
  \citenamefont {Marjanovic}, \citenamefont {McDonald}, \citenamefont
  {Mittiga}, \citenamefont {Muniz}, \citenamefont {Narayanaswami},
  \citenamefont {Nishiguchi}, \citenamefont {Notermans}, \citenamefont {Paule},
  \citenamefont {Pawlak}, \citenamefont {Peng}, \citenamefont {Ryou},
  \citenamefont {Smull}, \citenamefont {Stack}, \citenamefont {Stone},
  \citenamefont {Sucich}, \citenamefont {Urbanek}, \citenamefont {van~de
  Veerdonk}, \citenamefont {Vendeiro}, \citenamefont {Wilkason}, \citenamefont
  {Wu}, \citenamefont {Xie},\ and\ \citenamefont
  {Bloom}}]{norcia2023midcircuit}%
  \BibitemOpen
  \bibfield  {author} {\bibinfo {author} {M.~A. Norcia}, et~al.,\ }\href@noop
  {} {\emph {Mid-circuit qubit measurement and rearrangement in a $^{171}$Yb
  atomic array}} (\bibinfo {year} {2023}),\ \Eprint
  {https://arxiv.org/abs/2305.19119} {arXiv:2305.19119 [quant-ph]} \BibitemShut
  {NoStop}%
\bibitem [{\citenamefont {Lis}\ \emph {et~al.}(2023)\citenamefont {Lis},
  \citenamefont {Senoo}, \citenamefont {McGrew}, \citenamefont {Rönchen},
  \citenamefont {Jenkins},\ and\ \citenamefont {Kaufman}}]{lis2023midcircuit}%
  \BibitemOpen
  \bibfield  {author} {\bibinfo {author} {J.~W. Lis}, et~al.,\ }\href@noop {}
  {\emph {Mid-circuit operations using the omg-architecture in neutral atom
  arrays}} (\bibinfo {year} {2023}),\ \Eprint
  {https://arxiv.org/abs/2305.19266} {arXiv:2305.19266 [quant-ph]} \BibitemShut
  {NoStop}%
\bibitem [{\citenamefont {Huie}\ \emph {et~al.}(2023)\citenamefont {Huie},
  \citenamefont {Li}, \citenamefont {Chen}, \citenamefont {Hu}, \citenamefont
  {Jia}, \citenamefont {Sun},\ and\ \citenamefont
  {Covey}}]{huie2023repetitive}%
  \BibitemOpen
  \bibfield  {author} {\bibinfo {author} {W.~Huie}, et~al.,\ }\href@noop {}
  {\emph {Repetitive readout and real-time control of nuclear spin qubits in
  $^{171}$Yb atoms}} (\bibinfo {year} {2023}),\ \Eprint
  {https://arxiv.org/abs/2305.02926} {arXiv:2305.02926 [quant-ph]} \BibitemShut
  {NoStop}%
\bibitem [{\citenamefont {Singh}\ \emph {et~al.}(2023)\citenamefont {Singh},
  \citenamefont {Bradley}, \citenamefont {Anand}, \citenamefont {Ramesh},
  \citenamefont {White},\ and\ \citenamefont {Bernien}}]{singh2023midcircuit}%
  \BibitemOpen
  \bibfield  {author} {\bibinfo {author} {K.~Singh}, et~al.,\ }\emph
  {Mid-circuit correction of correlated phase errors using an array of
  spectator qubits},\ \href {https://doi.org/10.1126/science.ade5337}
  {\bibfield  {journal} {\bibinfo  {journal} {Science}\ }\textbf {\bibinfo
  {volume} {380}},\ \bibinfo {pages} {1265} (\bibinfo {year}
  {2023})}\BibitemShut {NoStop}%
\bibitem [{\citenamefont {Bochmann}\ \emph {et~al.}(2010)\citenamefont
  {Bochmann}, \citenamefont {M\"ucke}, \citenamefont {Guhl}, \citenamefont
  {Ritter}, \citenamefont {Rempe},\ and\ \citenamefont
  {Moehring}}]{bochmann2010lossless}%
  \BibitemOpen
  \bibfield  {author} {\bibinfo {author} {J.~Bochmann}, et~al.,\ }\emph
  {Lossless state detection of single neutral atoms},\ \href
  {https://doi.org/10.1103/PhysRevLett.104.203601} {\bibfield  {journal}
  {\bibinfo  {journal} {Physical Review Letters}\ }\textbf {\bibinfo {volume}
  {104}},\ \bibinfo {pages} {203601} (\bibinfo {year} {2010})}\BibitemShut
  {NoStop}%
\bibitem [{\citenamefont {Deist}\ \emph {et~al.}(2022)\citenamefont {Deist},
  \citenamefont {Lu}, \citenamefont {Ho}, \citenamefont {Pasha}, \citenamefont
  {Zeiher}, \citenamefont {Yan},\ and\ \citenamefont
  {Stamper-Kurn}}]{deist2022fast}%
  \BibitemOpen
  \bibfield  {author} {\bibinfo {author} {E.~Deist}, et~al.,\ }\emph
  {Mid-circuit cavity measurement in a neutral atom array},\ \href
  {https://doi.org/10.1103/PhysRevLett.129.203602} {\bibfield  {journal}
  {\bibinfo  {journal} {Physical Review Letters}\ }\textbf {\bibinfo {volume}
  {129}},\ \bibinfo {pages} {203602} (\bibinfo {year} {2022})}\BibitemShut
  {NoStop}%
\bibitem [{\citenamefont {Singh}\ \emph {et~al.}(2022)\citenamefont {Singh},
  \citenamefont {Anand}, \citenamefont {Pocklington}, \citenamefont {Kemp},\
  and\ \citenamefont {Bernien}}]{singh2022dual}%
  \BibitemOpen
  \bibfield  {author} {\bibinfo {author} {K.~Singh}, \bibinfo {author}
  {S.~Anand}, \bibinfo {author} {A.~Pocklington}, \bibinfo {author} {J.~T.
  Kemp},\ and\ \bibinfo {author} {H.~Bernien},\ }\emph {Dual-element,
  two-dimensional atom array with continuous-mode operation},\ \href
  {https://doi.org/10.1103/PhysRevX.12.011040} {\bibfield  {journal} {\bibinfo
  {journal} {Physical Review X}\ }\textbf {\bibinfo {volume} {12}},\ \bibinfo
  {pages} {011040} (\bibinfo {year} {2022})}\BibitemShut {NoStop}%
\bibitem [{\citenamefont {Schindler}\ \emph {et~al.}(2011)\citenamefont
  {Schindler}, \citenamefont {Barreiro}, \citenamefont {Monz}, \citenamefont
  {Nebendahl}, \citenamefont {Nigg}, \citenamefont {Chwalla}, \citenamefont
  {Hennrich},\ and\ \citenamefont {Blatt}}]{schindler2011experimental}%
  \BibitemOpen
  \bibfield  {author} {\bibinfo {author} {P.~Schindler}, et~al.,\ }\emph
  {Experimental repetitive quantum error correction},\ \href
  {https://doi.org/https://doi.org/10.1126/science.1203329} {\bibfield
  {journal} {\bibinfo  {journal} {Science}\ }\textbf {\bibinfo {volume}
  {332}},\ \bibinfo {pages} {1059} (\bibinfo {year} {2011})}\BibitemShut
  {NoStop}%
\bibitem [{\citenamefont {Barreiro}\ \emph {et~al.}(2011)\citenamefont
  {Barreiro}, \citenamefont {M{\"u}ller}, \citenamefont {Schindler},
  \citenamefont {Nigg}, \citenamefont {Monz}, \citenamefont {Chwalla},
  \citenamefont {Hennrich}, \citenamefont {Roos}, \citenamefont {Zoller},\ and\
  \citenamefont {Blatt}}]{barreiro2011open}%
  \BibitemOpen
  \bibfield  {author} {\bibinfo {author} {J.~T. Barreiro}, et~al.,\ }\emph {An
  open-system quantum simulator with trapped ions},\ \href
  {https://doi.org/https://doi.org/10.1038/nature09801} {\bibfield  {journal}
  {\bibinfo  {journal} {Nature}\ }\textbf {\bibinfo {volume} {470}},\ \bibinfo
  {pages} {486} (\bibinfo {year} {2011})}\BibitemShut {NoStop}%
\bibitem [{\citenamefont {Schindler}\ \emph {et~al.}(2013)\citenamefont
  {Schindler}, \citenamefont {M{\"u}ller}, \citenamefont {Nigg}, \citenamefont
  {Barreiro}, \citenamefont {Martinez}, \citenamefont {Hennrich}, \citenamefont
  {Monz}, \citenamefont {Diehl}, \citenamefont {Zoller},\ and\ \citenamefont
  {Blatt}}]{schindler2013quantum}%
  \BibitemOpen
  \bibfield  {author} {\bibinfo {author} {P.~Schindler}, et~al.,\ }\emph
  {Quantum simulation of dynamical maps with trapped ions},\ \href
  {https://doi.org/https://doi.org/10.1038/nphys2630} {\bibfield  {journal}
  {\bibinfo  {journal} {Nature Physics}\ }\textbf {\bibinfo {volume} {9}},\
  \bibinfo {pages} {361} (\bibinfo {year} {2013})}\BibitemShut {NoStop}%
\bibitem [{\citenamefont {Harrington}\ \emph {et~al.}(2022)\citenamefont
  {Harrington}, \citenamefont {Mueller},\ and\ \citenamefont
  {Murch}}]{harrington2022engineered}%
  \BibitemOpen
  \bibfield  {author} {\bibinfo {author} {P.~M. Harrington}, \bibinfo {author}
  {E.~J. Mueller},\ and\ \bibinfo {author} {K.~W. Murch},\ }\emph {Engineered
  dissipation for quantum information science},\ \href
  {https://doi.org/10.1038/s42254-022-00494-8} {\bibfield  {journal} {\bibinfo
  {journal} {Nature Reviews Physics}\ }\textbf {\bibinfo {volume} {4}},\
  \bibinfo {pages} {660} (\bibinfo {year} {2022})}\BibitemShut {NoStop}%
\bibitem [{\citenamefont {Terhal}\ \emph {et~al.}(2020)\citenamefont {Terhal},
  \citenamefont {Conrad},\ and\ \citenamefont {Vuillot}}]{terhal2020towards}%
  \BibitemOpen
  \bibfield  {author} {\bibinfo {author} {B.~M. Terhal}, \bibinfo {author}
  {J.~Conrad},\ and\ \bibinfo {author} {C.~Vuillot},\ }\emph {Towards scalable
  bosonic quantum error correction},\ \href
  {https://doi.org/10.1088/2058-9565/ab98a5} {\bibfield  {journal} {\bibinfo
  {journal} {Quantum Science and Technology}\ }\textbf {\bibinfo {volume}
  {5}},\ \bibinfo {pages} {043001} (\bibinfo {year} {2020})}\BibitemShut
  {NoStop}%
\bibitem [{\citenamefont {Gertler}\ \emph {et~al.}(2021)\citenamefont
  {Gertler}, \citenamefont {Baker}, \citenamefont {Li}, \citenamefont {Shirol},
  \citenamefont {Koch},\ and\ \citenamefont {Wang}}]{gertler2021protecting}%
  \BibitemOpen
  \bibfield  {author} {\bibinfo {author} {J.~M. Gertler}, et~al.,\ }\emph
  {Protecting a bosonic qubit with autonomous quantum error correction},\ \href
  {https://doi.org/10.1038/s41586-021-03257-0} {\bibfield  {journal} {\bibinfo
  {journal} {Nature}\ }\textbf {\bibinfo {volume} {590}},\ \bibinfo {pages}
  {243} (\bibinfo {year} {2021})}\BibitemShut {NoStop}%
\bibitem [{\citenamefont {Cruikshank}\ and\ \citenamefont
  {Jacobs}(2017{\natexlab{a}})}]{cruikshank2017role}%
  \BibitemOpen
  \bibfield  {author} {\bibinfo {author} {B.~Cruikshank}\ and\ \bibinfo
  {author} {K.~Jacobs},\ }\emph {The role of quantum measurements in physical
  processes and protocols},\ \href {https://doi.org/10.1088/2058-9565/aa6d3e}
  {\bibfield  {journal} {\bibinfo  {journal} {Quantum Science and Technology}\
  }\textbf {\bibinfo {volume} {2}},\ \bibinfo {pages} {033001} (\bibinfo {year}
  {2017}{\natexlab{a}})}\BibitemShut {NoStop}%
\bibitem [{\citenamefont {Cruikshank}\ and\ \citenamefont
  {Jacobs}(2017{\natexlab{b}})}]{cruikshank2017high}%
  \BibitemOpen
  \bibfield  {author} {\bibinfo {author} {B.~Cruikshank}\ and\ \bibinfo
  {author} {K.~Jacobs},\ }\emph {High-threshold low-overhead fault-tolerant
  classical computation and the replacement of measurements with unitary
  quantum gates},\ \href {https://doi.org/10.1103/PhysRevLett.119.030503}
  {\bibfield  {journal} {\bibinfo  {journal} {Physical Review Letters}\
  }\textbf {\bibinfo {volume} {119}},\ \bibinfo {pages} {030503} (\bibinfo
  {year} {2017}{\natexlab{b}})}\BibitemShut {NoStop}%
\bibitem [{\citenamefont {Ercan}\ \emph {et~al.}(2018)\citenamefont {Ercan},
  \citenamefont {Ghosh}, \citenamefont {Crow}, \citenamefont {Premakumar},
  \citenamefont {Joynt}, \citenamefont {Friesen},\ and\ \citenamefont
  {Coppersmith}}]{ercan2018measurement}%
  \BibitemOpen
  \bibfield  {author} {\bibinfo {author} {H.~E. Ercan}, et~al.,\ }\emph
  {Measurement-free implementations of small-scale surface codes for
  quantum-dot qubits},\ \href {https://doi.org/10.1103/PhysRevA.97.012318}
  {\bibfield  {journal} {\bibinfo  {journal} {Physical Review A}\ }\textbf
  {\bibinfo {volume} {97}},\ \bibinfo {pages} {012318} (\bibinfo {year}
  {2018})}\BibitemShut {NoStop}%
\bibitem [{\citenamefont {Boykin}\ \emph {et~al.}(2010)\citenamefont {Boykin},
  \citenamefont {Mor}, \citenamefont {Roychowdhury},\ and\ \citenamefont
  {Vatan}}]{boykin2010algorithms}%
  \BibitemOpen
  \bibfield  {author} {\bibinfo {author} {P.~O. Boykin}, \bibinfo {author}
  {T.~Mor}, \bibinfo {author} {V.~Roychowdhury},\ and\ \bibinfo {author}
  {F.~Vatan},\ }\emph {Algorithms on ensemble quantum computers},\ \href
  {https://doi.org/https://doi.org/10.1007/s11047-009-9133-0} {\bibfield
  {journal} {\bibinfo  {journal} {Natural computing}\ }\textbf {\bibinfo
  {volume} {9}},\ \bibinfo {pages} {329} (\bibinfo {year} {2010})}\BibitemShut
  {NoStop}%
\bibitem [{\citenamefont {Paz-Silva}\ \emph {et~al.}(2010)\citenamefont
  {Paz-Silva}, \citenamefont {Brennen},\ and\ \citenamefont
  {Twamley}}]{paz2010fault}%
  \BibitemOpen
  \bibfield  {author} {\bibinfo {author} {G.~A. Paz-Silva}, \bibinfo {author}
  {G.~K. Brennen},\ and\ \bibinfo {author} {J.~Twamley},\ }\emph {Fault
  tolerance with noisy and slow measurements and preparation},\ \href
  {https://doi.org/10.1103/PhysRevLett.105.100501} {\bibfield  {journal}
  {\bibinfo  {journal} {Physical Review Letters}\ }\textbf {\bibinfo {volume}
  {105}},\ \bibinfo {pages} {100501} (\bibinfo {year} {2010})}\BibitemShut
  {NoStop}%
\bibitem [{\citenamefont {Crow}\ \emph {et~al.}(2016)\citenamefont {Crow},
  \citenamefont {Joynt},\ and\ \citenamefont {Saffman}}]{crow2016improved}%
  \BibitemOpen
  \bibfield  {author} {\bibinfo {author} {D.~Crow}, \bibinfo {author}
  {R.~Joynt},\ and\ \bibinfo {author} {M.~Saffman},\ }\emph {Improved error
  thresholds for measurement-free error correction},\ \href
  {https://doi.org/10.1103/PhysRevLett.117.130503} {\bibfield  {journal}
  {\bibinfo  {journal} {Physical Review Letters}\ }\textbf {\bibinfo {volume}
  {117}},\ \bibinfo {pages} {130503} (\bibinfo {year} {2016})}\BibitemShut
  {NoStop}%
\bibitem [{\citenamefont {Premakumar}\ \emph {et~al.}(2020)\citenamefont
  {Premakumar}, \citenamefont {Saffman},\ and\ \citenamefont
  {Joynt}}]{premakumar2020measurement}%
  \BibitemOpen
  \bibfield  {author} {\bibinfo {author} {V.~N. Premakumar}, \bibinfo {author}
  {M.~Saffman},\ and\ \bibinfo {author} {R.~Joynt},\ }\href@noop {} {\emph
  {Measurement-free error correction with coherent ancillas}} (\bibinfo {year}
  {2020}),\ \Eprint {https://arxiv.org/abs/2007.09804} {arXiv:2007.09804
  [quant-ph]} \BibitemShut {NoStop}%
\bibitem [{\citenamefont {Chao}\ and\ \citenamefont
  {Reichardt}(2018)}]{Chao2018}%
  \BibitemOpen
  \bibfield  {author} {\bibinfo {author} {R.~Chao}\ and\ \bibinfo {author}
  {B.~W. Reichardt},\ }\emph {{Quantum error correction with only two extra
  qubits}},\ \href {https://doi.org/10.1103/PhysRevLett.121.050502} {\bibfield
  {journal} {\bibinfo  {journal} {Physical Review Letters}\ }\textbf {\bibinfo
  {volume} {121}},\ \bibinfo {pages} {050502} (\bibinfo {year}
  {2018})}\BibitemShut {NoStop}%
\bibitem [{\citenamefont {Chamberland}\ and\ \citenamefont
  {Beverland}(2018)}]{Chamberland2018}%
  \BibitemOpen
  \bibfield  {author} {\bibinfo {author} {C.~Chamberland}\ and\ \bibinfo
  {author} {M.~E. Beverland},\ }\emph {{Flag fault-tolerant error correction
  with arbitrary distance codes}},\ \href
  {https://doi.org/10.22331/q-2018-02-08-53} {\bibfield  {journal} {\bibinfo
  {journal} {Quantum}\ }\textbf {\bibinfo {volume} {2}},\ \bibinfo {pages} {53}
  (\bibinfo {year} {2018})}\BibitemShut {NoStop}%
\bibitem [{\citenamefont {Chao}\ and\ \citenamefont
  {Reichardt}(2020)}]{Chao2020}%
  \BibitemOpen
  \bibfield  {author} {\bibinfo {author} {R.~Chao}\ and\ \bibinfo {author}
  {B.~W. Reichardt},\ }\emph {{Flag fault-tolerant error correction for any
  stabilizer code}},\ \href {https://doi.org/10.1103/PRXQuantum.1.010302}
  {\bibfield  {journal} {\bibinfo  {journal} {PRX Quantum}\ }\textbf {\bibinfo
  {volume} {1}},\ \bibinfo {pages} {010302} (\bibinfo {year}
  {2020})}\BibitemShut {NoStop}%
\bibitem [{\citenamefont {Reichardt}(2020)}]{reichardt2020fault}%
  \BibitemOpen
  \bibfield  {author} {\bibinfo {author} {B.~W. Reichardt},\ }\emph
  {Fault-tolerant quantum error correction for Steane’s seven-qubit color
  code with few or no extra qubits},\ \href
  {https://doi.org/10.1088/2058-9565/abc6f4} {\bibfield  {journal} {\bibinfo
  {journal} {Quantum Science and Technology}\ }\textbf {\bibinfo {volume}
  {6}},\ \bibinfo {pages} {015007} (\bibinfo {year} {2020})}\BibitemShut
  {NoStop}%
\bibitem [{\citenamefont {Heu{\ss}en}\ \emph {et~al.}(2023)\citenamefont
  {Heu{\ss}en}, \citenamefont {Postler}, \citenamefont {Rispler}, \citenamefont
  {Pogorelov}, \citenamefont {Marciniak}, \citenamefont {Monz}, \citenamefont
  {Schindler},\ and\ \citenamefont {M{\"u}ller}}]{heussen2023strategies}%
  \BibitemOpen
  \bibfield  {author} {\bibinfo {author} {S.~Heu{\ss}en}, et~al.,\ }\emph
  {Strategies for a practical advantage of fault-tolerant circuit design in
  noisy trapped-ion quantum computers},\ \href
  {https://doi.org/10.1103/PhysRevA.107.042422} {\bibfield  {journal} {\bibinfo
   {journal} {Physical Review A}\ }\textbf {\bibinfo {volume} {107}},\ \bibinfo
  {pages} {042422} (\bibinfo {year} {2023})}\BibitemShut {NoStop}%
\bibitem [{\citenamefont {Goto}\ \emph {et~al.}(2023)\citenamefont {Goto},
  \citenamefont {Ho},\ and\ \citenamefont {Kanao}}]{goto2023measurementfree}%
  \BibitemOpen
  \bibfield  {author} {\bibinfo {author} {H.~Goto}, \bibinfo {author} {Y.~Ho},\
  and\ \bibinfo {author} {T.~Kanao},\ }\href@noop {} {\emph {Measurement-free
  fault-tolerant logical zero-state encoding of the distance-three nine-qubit
  surface code in a one-dimensional qubit array}} (\bibinfo {year} {2023}),\
  \Eprint {https://arxiv.org/abs/2303.17211} {arXiv:2303.17211 [quant-ph]}
  \BibitemShut {NoStop}%
\bibitem [{\citenamefont {Nielsen}\ and\ \citenamefont
  {Chuang}(2010)}]{nielsen2010quantum}%
  \BibitemOpen
  \bibfield  {author} {\bibinfo {author} {M.~A. Nielsen}\ and\ \bibinfo
  {author} {I.~L. Chuang},\ }\href {https://doi.org/10.1017/CBO9780511976667}
  {\emph {\bibinfo {title} {Quantum Computation and Quantum Information: 10th
  Anniversary Edition}}}\ (\bibinfo  {publisher} {Cambridge University Press},\
  \bibinfo {year} {2010})\BibitemShut {NoStop}%
\bibitem [{\citenamefont {Shor}(1996)}]{Shor1996}%
  \BibitemOpen
  \bibfield  {author} {\bibinfo {author} {P.~W. Shor},\ }in\ \href
  {https://doi.org/10.1109/SFCS.1996.548464} {\emph {\bibinfo {booktitle}
  {Proceedings of 37th Conference on Foundations of Computer Science}}}\
  (\bibinfo  {publisher} {IEEE},\ \bibinfo {year} {1996})\ p.~\bibinfo {pages}
  {56}\BibitemShut {NoStop}%
\bibitem [{\citenamefont {Steane}(1997)}]{steane1997active}%
  \BibitemOpen
  \bibfield  {author} {\bibinfo {author} {A.~M. Steane},\ }\emph {Active
  stabilization, quantum computation, and quantum state synthesis},\ \href
  {https://doi.org/10.1103/PhysRevLett.78.2252} {\bibfield  {journal} {\bibinfo
   {journal} {Physical Review Letters}\ }\textbf {\bibinfo {volume} {78}},\
  \bibinfo {pages} {2252} (\bibinfo {year} {1997})}\BibitemShut {NoStop}%
\bibitem [{\citenamefont {Calderbank}\ and\ \citenamefont
  {Shor}(1996)}]{calderbank1996good}%
  \BibitemOpen
  \bibfield  {author} {\bibinfo {author} {A.~R. Calderbank}\ and\ \bibinfo
  {author} {P.~W. Shor},\ }\emph {Good quantum error-correcting codes exist},\
  \href {https://link.aps.org/pdf/10.1103/PhysRevA.54.1098} {\bibfield
  {journal} {\bibinfo  {journal} {Physical Review A}\ }\textbf {\bibinfo
  {volume} {54}},\ \bibinfo {pages} {1098} (\bibinfo {year}
  {1996})}\BibitemShut {NoStop}%
\bibitem [{\citenamefont {Bravyi}\ and\ \citenamefont
  {Kitaev}(1998)}]{bravyi1998quantum}%
  \BibitemOpen
  \bibfield  {author} {\bibinfo {author} {S.~B. Bravyi}\ and\ \bibinfo {author}
  {A.~Y. Kitaev},\ }\href@noop {} {\emph {Quantum codes on a lattice with
  boundary}} (\bibinfo {year} {1998}),\ \Eprint
  {https://arxiv.org/abs/quant-ph/9811052} {arXiv:quant-ph/9811052 [quant-ph]}
  \BibitemShut {NoStop}%
\bibitem [{\citenamefont {Dennis}\ \emph {et~al.}(2002)\citenamefont {Dennis},
  \citenamefont {Kitaev}, \citenamefont {Landahl},\ and\ \citenamefont
  {Preskill}}]{dennis2002topological}%
  \BibitemOpen
  \bibfield  {author} {\bibinfo {author} {E.~Dennis}, \bibinfo {author}
  {A.~Kitaev}, \bibinfo {author} {A.~Landahl},\ and\ \bibinfo {author}
  {J.~Preskill},\ }\emph {Topological quantum memory},\ \href
  {https://doi.org/https://doi.org/10.1063/1.1499754} {\bibfield  {journal}
  {\bibinfo  {journal} {Journal of Mathematical Physics}\ }\textbf {\bibinfo
  {volume} {43}},\ \bibinfo {pages} {4452} (\bibinfo {year}
  {2002})}\BibitemShut {NoStop}%
\bibitem [{\citenamefont {Bombin}\ and\ \citenamefont
  {Martin-Delgado}(2006{\natexlab{a}})}]{bombin2006topological}%
  \BibitemOpen
  \bibfield  {author} {\bibinfo {author} {H.~Bombin}\ and\ \bibinfo {author}
  {M.~A. Martin-Delgado},\ }\emph {Topological quantum error correction with
  optimal encoding rate},\ \href {https://doi.org/10.1103/PhysRevA.73.062303}
  {\bibfield  {journal} {\bibinfo  {journal} {Physical Review A}\ }\textbf
  {\bibinfo {volume} {73}},\ \bibinfo {pages} {062303} (\bibinfo {year}
  {2006}{\natexlab{a}})}\BibitemShut {NoStop}%
\bibitem [{\citenamefont {Bombin}\ and\ \citenamefont
  {Martin-Delgado}(2006{\natexlab{b}})}]{bombin2006distillation}%
  \BibitemOpen
  \bibfield  {author} {\bibinfo {author} {H.~Bombin}\ and\ \bibinfo {author}
  {M.~A. Martin-Delgado},\ }\emph {Topological quantum distillation},\ \href
  {https://link.aps.org/pdf/10.1103/PhysRevLett.97.180501} {\bibfield
  {journal} {\bibinfo  {journal} {Physical Review Letters}\ }\textbf {\bibinfo
  {volume} {97}},\ \bibinfo {pages} {180501} (\bibinfo {year}
  {2006}{\natexlab{b}})}\BibitemShut {NoStop}%
\bibitem [{\citenamefont {Bombin}\ and\ \citenamefont
  {Martin-Delgado}(2007)}]{bombin2007topological}%
  \BibitemOpen
  \bibfield  {author} {\bibinfo {author} {H.~Bombin}\ and\ \bibinfo {author}
  {M.-A. Martin-Delgado},\ }\emph {Topological computation without braiding},\
  \href {https://doi.org/10.1103/PhysRevLett.98.160502} {\bibfield  {journal}
  {\bibinfo  {journal} {Physical Review Letters}\ }\textbf {\bibinfo {volume}
  {98}},\ \bibinfo {pages} {160502} (\bibinfo {year} {2007})}\BibitemShut
  {NoStop}%
\bibitem [{\citenamefont {Fowler}(2011)}]{fowler2011two}%
  \BibitemOpen
  \bibfield  {author} {\bibinfo {author} {A.~G. Fowler},\ }\emph
  {Two-dimensional color-code quantum computation},\ \href
  {https://doi.org/10.1103/PhysRevA.83.042310} {\bibfield  {journal} {\bibinfo
  {journal} {Physical Review A}\ }\textbf {\bibinfo {volume} {83}},\ \bibinfo
  {pages} {042310} (\bibinfo {year} {2011})}\BibitemShut {NoStop}%
\bibitem [{\citenamefont {Steane}(1996)}]{steane1996error}%
  \BibitemOpen
  \bibfield  {author} {\bibinfo {author} {A.~M. Steane},\ }\emph {Error
  correcting codes in quantum theory},\ \href
  {https://link.aps.org/pdf/10.1103/PhysRevLett.77.793} {\bibfield  {journal}
  {\bibinfo  {journal} {Physical Review Letters}\ }\textbf {\bibinfo {volume}
  {77}},\ \bibinfo {pages} {793} (\bibinfo {year} {1996})}\BibitemShut
  {NoStop}%
\bibitem [{\citenamefont {Tomita}\ and\ \citenamefont
  {Svore}(2014)}]{tomita2014low}%
  \BibitemOpen
  \bibfield  {author} {\bibinfo {author} {Y.~Tomita}\ and\ \bibinfo {author}
  {K.~M. Svore},\ }\emph {Low-distance surface codes under realistic quantum
  noise},\ \href {https://doi.org/10.1103/PhysRevA.90.062320} {\bibfield
  {journal} {\bibinfo  {journal} {Physical Review A}\ }\textbf {\bibinfo
  {volume} {90}},\ \bibinfo {pages} {062320} (\bibinfo {year}
  {2014})}\BibitemShut {NoStop}%
\bibitem [{\citenamefont {Barenco}\ \emph {et~al.}(1995)\citenamefont
  {Barenco}, \citenamefont {Bennett}, \citenamefont {Cleve}, \citenamefont
  {DiVincenzo}, \citenamefont {Margolus}, \citenamefont {Shor}, \citenamefont
  {Sleator}, \citenamefont {Smolin},\ and\ \citenamefont
  {Weinfurter}}]{barenco1995elementary}%
  \BibitemOpen
  \bibfield  {author} {\bibinfo {author} {A.~Barenco}, et~al.,\ }\emph
  {Elementary gates for quantum computation},\ \href
  {https://doi.org/10.1103/PhysRevA.52.3457} {\bibfield  {journal} {\bibinfo
  {journal} {Physical Review A}\ }\textbf {\bibinfo {volume} {52}},\ \bibinfo
  {pages} {3457} (\bibinfo {year} {1995})}\BibitemShut {NoStop}%
\bibitem [{\citenamefont {Goto}(2016)}]{goto2016minimizing}%
  \BibitemOpen
  \bibfield  {author} {\bibinfo {author} {H.~Goto},\ }\emph {Minimizing
  resource overheads for fault-tolerant preparation of encoded states of the
  Steane code},\ \href {https://doi.org/10.1038/srep19578} {\bibfield
  {journal} {\bibinfo  {journal} {Scientific Reports}\ }\textbf {\bibinfo
  {volume} {6}},\ \bibinfo {pages} {1} (\bibinfo {year} {2016})}\BibitemShut
  {NoStop}%
\bibitem [{\citenamefont {Maslov}(2016)}]{maslov2016advantages}%
  \BibitemOpen
  \bibfield  {author} {\bibinfo {author} {D.~Maslov},\ }\emph {Advantages of
  using relative-phase Toffoli gates with an application to multiple control
  Toffoli optimization},\ \href {https://doi.org/10.1103/PhysRevA.93.022311}
  {\bibfield  {journal} {\bibinfo  {journal} {Physical Review A}\ }\textbf
  {\bibinfo {volume} {93}},\ \bibinfo {pages} {022311} (\bibinfo {year}
  {2016})}\BibitemShut {NoStop}%
\bibitem [{\citenamefont {M{\o}lmer}\ and\ \citenamefont
  {S{\o}rensen}(1999)}]{molmer1999multiparticle}%
  \BibitemOpen
  \bibfield  {author} {\bibinfo {author} {K.~M{\o}lmer}\ and\ \bibinfo {author}
  {A.~S{\o}rensen},\ }\emph {Multiparticle entanglement of hot trapped ions},\
  \href {https://doi.org/10.1103/PhysRevLett.82.1835} {\bibfield  {journal}
  {\bibinfo  {journal} {Physical Review Letters}\ }\textbf {\bibinfo {volume}
  {82}},\ \bibinfo {pages} {1835} (\bibinfo {year} {1999})}\BibitemShut
  {NoStop}%
\bibitem [{\citenamefont {Pino}\ \emph {et~al.}(2021)\citenamefont {Pino},
  \citenamefont {Dreiling}, \citenamefont {Figgatt}, \citenamefont {Gaebler},
  \citenamefont {Moses}, \citenamefont {Allman}, \citenamefont {Baldwin},
  \citenamefont {Foss-Feig}, \citenamefont {Hayes}, \citenamefont {Mayer} \emph
  {et~al.}}]{pino2021demonstration}%
  \BibitemOpen
  \bibfield  {author} {\bibinfo {author} {J.~M. Pino}, et~al.,\ }\emph
  {Demonstration of the trapped-ion quantum CCD computer architecture},\ \href
  {https://doi.org/https://doi.org/10.1038/s41586-021-03318-4} {\bibfield
  {journal} {\bibinfo  {journal} {Nature}\ }\textbf {\bibinfo {volume} {592}},\
  \bibinfo {pages} {209} (\bibinfo {year} {2021})}\BibitemShut {NoStop}%
\bibitem [{\citenamefont {Chamberland}\ and\ \citenamefont
  {Cross}(2019)}]{chamberland2019fault}%
  \BibitemOpen
  \bibfield  {author} {\bibinfo {author} {C.~Chamberland}\ and\ \bibinfo
  {author} {A.~W. Cross},\ }\emph {Fault-tolerant magic state preparation with
  flag qubits},\ \href
  {https://doi.org/https://doi.org/10.22331/q-2019-05-20-143} {\bibfield
  {journal} {\bibinfo  {journal} {Quantum}\ }\textbf {\bibinfo {volume} {3}},\
  \bibinfo {pages} {143} (\bibinfo {year} {2019})}\BibitemShut {NoStop}%
\bibitem [{\citenamefont {Bermudez}\ \emph {et~al.}(2017)\citenamefont
  {Bermudez}, \citenamefont {Xu}, \citenamefont {Nigmatullin}, \citenamefont
  {O’Gorman}, \citenamefont {Negnevitsky}, \citenamefont {Schindler},
  \citenamefont {Monz}, \citenamefont {Poschinger}, \citenamefont {Hempel},
  \citenamefont {Home} \emph {et~al.}}]{bermudez2017assessing}%
  \BibitemOpen
  \bibfield  {author} {\bibinfo {author} {A.~Bermudez}, et~al.,\ }\emph
  {Assessing the progress of trapped-ion processors towards fault-tolerant
  quantum computation},\ \href {https://doi.org/10.1103/PhysRevX.7.041061}
  {\bibfield  {journal} {\bibinfo  {journal} {Physical Review X}\ }\textbf
  {\bibinfo {volume} {7}},\ \bibinfo {pages} {041061} (\bibinfo {year}
  {2017})}\BibitemShut {NoStop}%
\bibitem [{\citenamefont {Rasmussen}\ \emph {et~al.}(2020)\citenamefont
  {Rasmussen}, \citenamefont {Groenland}, \citenamefont {Gerritsma},
  \citenamefont {Schoutens},\ and\ \citenamefont
  {Zinner}}]{rasmussen2020single}%
  \BibitemOpen
  \bibfield  {author} {\bibinfo {author} {S.~Rasmussen}, \bibinfo {author}
  {K.~Groenland}, \bibinfo {author} {R.~Gerritsma}, \bibinfo {author}
  {K.~Schoutens},\ and\ \bibinfo {author} {N.~Zinner},\ }\emph {Single-step
  implementation of high-fidelity $n$-bit Toffoli gates},\ \href
  {https://doi.org/10.1103/PhysRevA.101.022308} {\bibfield  {journal} {\bibinfo
   {journal} {Physical Review A}\ }\textbf {\bibinfo {volume} {101}},\ \bibinfo
  {pages} {022308} (\bibinfo {year} {2020})}\BibitemShut {NoStop}%
\bibitem [{\citenamefont {Espinoza}\ \emph {et~al.}(2021)\citenamefont
  {Espinoza}, \citenamefont {Groenland}, \citenamefont {Mazzanti},
  \citenamefont {Schoutens},\ and\ \citenamefont
  {Gerritsma}}]{espinoza2021high}%
  \BibitemOpen
  \bibfield  {author} {\bibinfo {author} {J.~D.~A. Espinoza}, \bibinfo {author}
  {K.~Groenland}, \bibinfo {author} {M.~Mazzanti}, \bibinfo {author}
  {K.~Schoutens},\ and\ \bibinfo {author} {R.~Gerritsma},\ }\emph
  {High-fidelity method for a single-step $N$-bit Toffoli gate in trapped
  ions},\ \href {https://doi.org/10.1103/PhysRevA.103.052437} {\bibfield
  {journal} {\bibinfo  {journal} {Physical Review A}\ }\textbf {\bibinfo
  {volume} {103}},\ \bibinfo {pages} {052437} (\bibinfo {year}
  {2021})}\BibitemShut {NoStop}%
\bibitem [{\citenamefont {Pelegrí}\ \emph {et~al.}(2022)\citenamefont
  {Pelegrí}, \citenamefont {Daley},\ and\ \citenamefont
  {Pritchard}}]{pelegri_2022_highfidelity}%
  \BibitemOpen
  \bibfield  {author} {\bibinfo {author} {G.~Pelegrí}, \bibinfo {author}
  {A.~J. Daley},\ and\ \bibinfo {author} {J.~D. Pritchard},\ }\emph
  {High-fidelity multiqubit Rydberg gates via two-photon adiabatic rapid
  passage},\ \href {https://doi.org/10.1088/2058-9565/ac823a} {\bibfield
  {journal} {\bibinfo  {journal} {Quantum Science and Technology}\ }\textbf
  {\bibinfo {volume} {7}},\ \bibinfo {pages} {045020} (\bibinfo {year}
  {2022})}\BibitemShut {NoStop}%
\bibitem [{\citenamefont {Evered}\ \emph {et~al.}(2023)\citenamefont {Evered},
  \citenamefont {Bluvstein}, \citenamefont {Kalinowski}, \citenamefont {Ebadi},
  \citenamefont {Manovitz}, \citenamefont {Zhou}, \citenamefont {Li},
  \citenamefont {Geim}, \citenamefont {Wang}, \citenamefont {Maskara},
  \citenamefont {Levine}, \citenamefont {Semeghini}, \citenamefont {Greiner},
  \citenamefont {Vuletic},\ and\ \citenamefont {Lukin}}]{evered2023high}%
  \BibitemOpen
  \bibfield  {author} {\bibinfo {author} {S.~J. Evered}, et~al.,\ }\href@noop
  {} {\emph {High-fidelity parallel entangling gates on a neutral atom quantum
  computer}} (\bibinfo {year} {2023}),\ \Eprint
  {https://arxiv.org/abs/2304.05420} {arXiv:2304.05420 [quant-ph]} \BibitemShut
  {NoStop}%
\bibitem [{\citenamefont {Graham}\ \emph {et~al.}(2022)\citenamefont {Graham},
  \citenamefont {Song}, \citenamefont {Scott}, \citenamefont {Poole},
  \citenamefont {Phuttitarn}, \citenamefont {Jooya}, \citenamefont {Eichler},
  \citenamefont {Jiang}, \citenamefont {Marra}, \citenamefont {Grinkemeyer},
  \citenamefont {Kwon}, \citenamefont {Ebert}, \citenamefont {Cherek},
  \citenamefont {Lichtman}, \citenamefont {Gillette}, \citenamefont {Gilbert},
  \citenamefont {Bowman}, \citenamefont {Ballance}, \citenamefont {Campbell},
  \citenamefont {Dahl}, \citenamefont {Crawford}, \citenamefont {Blunt},
  \citenamefont {Rogers}, \citenamefont {Noel},\ and\ \citenamefont
  {Saffman}}]{graham2022multiqubit}%
  \BibitemOpen
  \bibfield  {author} {\bibinfo {author} {T.~M. Graham}, et~al.,\ }\emph
  {Multi-qubit entanglement and algorithms on a neutral-atom quantum
  computer},\ \href {https://doi.org/10.1038/s41586-022-04603-6} {\bibfield
  {journal} {\bibinfo  {journal} {Nature}\ }\textbf {\bibinfo {volume} {604}},\
  \bibinfo {pages} {457} (\bibinfo {year} {2022})}\BibitemShut {NoStop}%
\bibitem [{\citenamefont {Ryan-Anderson}(2018)}]{ryan2018quantum}%
  \BibitemOpen
  \bibfield  {author} {\bibinfo {author} {C.~Ryan-Anderson},\ }\emph {\bibinfo
  {title} {Quantum algorithms, architecture, and error correction}},\ \href
  {https://digitalrepository.unm.edu/phyc_etds/203/} {Ph.D. thesis},\ \bibinfo
  {school} {The University of New Mexico} (\bibinfo {year} {2018})\BibitemShut
  {NoStop}%
\bibitem [{\citenamefont {Ryan-Anderson}(2019)}]{pecos}%
  \BibitemOpen
  \bibfield  {author} {\bibinfo {author} {C.~Ryan-Anderson},\ }\href@noop {}
  {\emph {PECOS: Performance estimator of codes on surfaces}},\ \bibinfo
  {howpublished} {\url{https://github.com/PECOS-packages/PECOS}} (\bibinfo
  {year} {2019})\BibitemShut {NoStop}%
\bibitem [{\citenamefont {Levine}\ \emph {et~al.}(2019)\citenamefont {Levine},
  \citenamefont {Keesling}, \citenamefont {Semeghini}, \citenamefont {Omran},
  \citenamefont {Wang}, \citenamefont {Ebadi}, \citenamefont {Bernien},
  \citenamefont {Greiner}, \citenamefont {Vuleti\ifmmode~\acute{c}\else
  \'{c}\fi{}}, \citenamefont {Pichler},\ and\ \citenamefont
  {Lukin}}]{levine2019parallel}%
  \BibitemOpen
  \bibfield  {author} {\bibinfo {author} {H.~Levine}, et~al.,\ }\emph {Parallel
  implementation of high-fidelity multiqubit gates with neutral atoms},\ \href
  {https://doi.org/10.1103/PhysRevLett.123.170503} {\bibfield  {journal}
  {\bibinfo  {journal} {Physical Review Letters}\ }\textbf {\bibinfo {volume}
  {123}},\ \bibinfo {pages} {170503} (\bibinfo {year} {2019})}\BibitemShut
  {NoStop}%
\bibitem [{\citenamefont {Isenhower}\ \emph {et~al.}(2011)\citenamefont
  {Isenhower}, \citenamefont {Saffman},\ and\ \citenamefont
  {M{\o}lmer}}]{isenhower_2011_multibit}%
  \BibitemOpen
  \bibfield  {author} {\bibinfo {author} {L.~Isenhower}, \bibinfo {author}
  {M.~Saffman},\ and\ \bibinfo {author} {K.~M{\o}lmer},\ }\emph {Multibit
  C$_k$NOT quantum gates via Rydberg blockade},\ \href
  {https://doi.org/10.1007/s11128-011-0292-4} {\bibfield  {journal} {\bibinfo
  {journal} {Quantum Information Processing}\ }\textbf {\bibinfo {volume}
  {10}},\ \bibinfo {pages} {755} (\bibinfo {year} {2011})}\BibitemShut
  {NoStop}%
\bibitem [{\citenamefont {Khazali}\ and\ \citenamefont
  {M\o{}lmer}(2020)}]{khazali_2020_fast}%
  \BibitemOpen
  \bibfield  {author} {\bibinfo {author} {M.~Khazali}\ and\ \bibinfo {author}
  {K.~M\o{}lmer},\ }\emph {Fast multiqubit gates by adiabatic evolution in
  interacting excited-state manifolds of Rydberg atoms and superconducting
  circuits},\ \href {https://doi.org/10.1103/PhysRevX.10.021054} {\bibfield
  {journal} {\bibinfo  {journal} {Physical Review X}\ }\textbf {\bibinfo
  {volume} {10}},\ \bibinfo {pages} {021054} (\bibinfo {year}
  {2020})}\BibitemShut {NoStop}%
\bibitem [{\citenamefont {M\"uller}\ \emph {et~al.}(2009)\citenamefont
  {M\"uller}, \citenamefont {Lesanovsky}, \citenamefont {Weimer}, \citenamefont
  {B\"uchler},\ and\ \citenamefont {Zoller}}]{mueller_2009_mesoscopic}%
  \BibitemOpen
  \bibfield  {author} {\bibinfo {author} {M.~M\"uller}, \bibinfo {author}
  {I.~Lesanovsky}, \bibinfo {author} {H.~Weimer}, \bibinfo {author} {H.~P.
  B\"uchler},\ and\ \bibinfo {author} {P.~Zoller},\ }\emph {Mesoscopic Rydberg
  gate based on electromagnetically induced transparency},\ \href
  {https://doi.org/10.1103/PhysRevLett.102.170502} {\bibfield  {journal}
  {\bibinfo  {journal} {Physical Review Letters}\ }\textbf {\bibinfo {volume}
  {102}},\ \bibinfo {pages} {170502} (\bibinfo {year} {2009})}\BibitemShut
  {NoStop}%
\bibitem [{\citenamefont {Young}\ \emph {et~al.}(2021)\citenamefont {Young},
  \citenamefont {Bienias}, \citenamefont {Belyansky}, \citenamefont {Kaufman},\
  and\ \citenamefont {Gorshkov}}]{young_2021_asymmetric}%
  \BibitemOpen
  \bibfield  {author} {\bibinfo {author} {J.~T. Young}, \bibinfo {author}
  {P.~Bienias}, \bibinfo {author} {R.~Belyansky}, \bibinfo {author} {A.~M.
  Kaufman},\ and\ \bibinfo {author} {A.~V. Gorshkov},\ }\emph {Asymmetric
  blockade and multiqubit gates via dipole-dipole interactions},\ \href
  {https://doi.org/10.1103/PhysRevLett.127.120501} {\bibfield  {journal}
  {\bibinfo  {journal} {Physical Review Letters}\ }\textbf {\bibinfo {volume}
  {127}},\ \bibinfo {pages} {120501} (\bibinfo {year} {2021})}\BibitemShut
  {NoStop}%
\bibitem [{\citenamefont {Monz}\ \emph {et~al.}(2009)\citenamefont {Monz},
  \citenamefont {Kim}, \citenamefont {H\"ansel}, \citenamefont {Riebe},
  \citenamefont {Villar}, \citenamefont {Schindler}, \citenamefont {Chwalla},
  \citenamefont {Hennrich},\ and\ \citenamefont {Blatt}}]{monz2009realization}%
  \BibitemOpen
  \bibfield  {author} {\bibinfo {author} {T.~Monz}, et~al.,\ }\emph
  {Realization of the quantum Toffoli gate with trapped ions},\ \href
  {https://doi.org/10.1103/PhysRevLett.102.040501} {\bibfield  {journal}
  {\bibinfo  {journal} {Physical Review Letters}\ }\textbf {\bibinfo {volume}
  {102}},\ \bibinfo {pages} {040501} (\bibinfo {year} {2009})}\BibitemShut
  {NoStop}%
\bibitem [{\citenamefont {Kim}\ \emph {et~al.}(2022)\citenamefont {Kim},
  \citenamefont {Morvan}, \citenamefont {Nguyen}, \citenamefont {Naik},
  \citenamefont {J{\"u}nger}, \citenamefont {Chen}, \citenamefont {Kreikebaum},
  \citenamefont {Santiago},\ and\ \citenamefont {Siddiqi}}]{kim2022high}%
  \BibitemOpen
  \bibfield  {author} {\bibinfo {author} {Y.~Kim}, et~al.,\ }\emph
  {High-fidelity three-qubit iToffoli gate for fixed-frequency superconducting
  qubits},\ \href {https://doi.org/10.1038/s41567-022-01590-3} {\bibfield
  {journal} {\bibinfo  {journal} {Nature Physics}\ }\textbf {\bibinfo {volume}
  {18}},\ \bibinfo {pages} {783} (\bibinfo {year} {2022})}\BibitemShut
  {NoStop}%
\bibitem [{\citenamefont {Baker}\ \emph {et~al.}(2022)\citenamefont {Baker},
  \citenamefont {Huber}, \citenamefont {Glaser}, \citenamefont {Roy},
  \citenamefont {Tsitsilin}, \citenamefont {Filipp},\ and\ \citenamefont
  {Hartmann}}]{baker2022single}%
  \BibitemOpen
  \bibfield  {author} {\bibinfo {author} {A.~J. Baker}, et~al.,\ }\emph {Single
  shot i-Toffoli gate in dispersively coupled superconducting qubits},\ \href
  {https://doi.org/10.1063/5.0077443} {\bibfield  {journal} {\bibinfo
  {journal} {Applied Physics Letters}\ }\textbf {\bibinfo {volume} {120}},\
  \bibinfo {pages} {054002} (\bibinfo {year} {2022})}\BibitemShut {NoStop}%
\bibitem [{\citenamefont {Glaser}\ \emph {et~al.}(2023)\citenamefont {Glaser},
  \citenamefont {Roy},\ and\ \citenamefont {Filipp}}]{glaser2023controlled}%
  \BibitemOpen
  \bibfield  {author} {\bibinfo {author} {N.~J. Glaser}, \bibinfo {author}
  {F.~Roy},\ and\ \bibinfo {author} {S.~Filipp},\ }\emph
  {Controlled-controlled-phase gates for superconducting qubits mediated by a
  shared tunable coupler},\ \href
  {https://doi.org/10.1103/PhysRevApplied.19.044001} {\bibfield  {journal}
  {\bibinfo  {journal} {Physical Review Applied}\ }\textbf {\bibinfo {volume}
  {19}},\ \bibinfo {pages} {044001} (\bibinfo {year} {2023})}\BibitemShut
  {NoStop}%
\bibitem [{\citenamefont {S{\o}rensen}\ and\ \citenamefont
  {M{\o}lmer}(2000)}]{sorensen2000entanglement}%
  \BibitemOpen
  \bibfield  {author} {\bibinfo {author} {A.~S{\o}rensen}\ and\ \bibinfo
  {author} {K.~M{\o}lmer},\ }\emph {Entanglement and quantum computation with
  ions in thermal motion},\ \href {https://doi.org/10.1103/PhysRevA.62.022311}
  {\bibfield  {journal} {\bibinfo  {journal} {Physical Review A}\ }\textbf
  {\bibinfo {volume} {62}},\ \bibinfo {pages} {022311} (\bibinfo {year}
  {2000})}\BibitemShut {NoStop}%
\bibitem [{\citenamefont {M{\"u}ller}\ \emph {et~al.}(2011)\citenamefont
  {M{\"u}ller}, \citenamefont {Hammerer}, \citenamefont {Zhou}, \citenamefont
  {Roos},\ and\ \citenamefont {Zoller}}]{muller2011simulating}%
  \BibitemOpen
  \bibfield  {author} {\bibinfo {author} {M.~M{\"u}ller}, \bibinfo {author}
  {K.~Hammerer}, \bibinfo {author} {Y.~Zhou}, \bibinfo {author} {C.~F. Roos},\
  and\ \bibinfo {author} {P.~Zoller},\ }\emph {Simulating open quantum systems:
  from many-body interactions to stabilizer pumping},\ \href
  {https://doi.org/10.1088/1367-2630/13/8/085007} {\bibfield  {journal}
  {\bibinfo  {journal} {New Journal of Physics}\ }\textbf {\bibinfo {volume}
  {13}},\ \bibinfo {pages} {085007} (\bibinfo {year} {2011})}\BibitemShut
  {NoStop}%
\bibitem [{\citenamefont {Martinez}\ \emph {et~al.}(2016)\citenamefont
  {Martinez}, \citenamefont {Monz}, \citenamefont {Nigg}, \citenamefont
  {Schindler},\ and\ \citenamefont {Blatt}}]{martinez2016compiling}%
  \BibitemOpen
  \bibfield  {author} {\bibinfo {author} {E.~A. Martinez}, \bibinfo {author}
  {T.~Monz}, \bibinfo {author} {D.~Nigg}, \bibinfo {author} {P.~Schindler},\
  and\ \bibinfo {author} {R.~Blatt},\ }\emph {Compiling quantum algorithms for
  architectures with multi-qubit gates},\ \href
  {https://doi.org/10.1088/1367-2630/18/6/063029} {\bibfield  {journal}
  {\bibinfo  {journal} {New Journal of Physics}\ }\textbf {\bibinfo {volume}
  {18}},\ \bibinfo {pages} {063029} (\bibinfo {year} {2016})}\BibitemShut
  {NoStop}%
\bibitem [{\citenamefont {Bergholm}\ \emph {et~al.}(2022)\citenamefont
  {Bergholm}, \citenamefont {Izaac}, \citenamefont {Schuld}, \citenamefont
  {Gogolin}, \citenamefont {Ahmed}, \citenamefont {Ajith}, \citenamefont
  {Alam}, \citenamefont {Alonso-Linaje}, \citenamefont {AkashNarayanan},
  \citenamefont {Asadi}, \citenamefont {Arrazola}, \citenamefont {Azad},
  \citenamefont {Banning}, \citenamefont {Blank}, \citenamefont {Bromley},
  \citenamefont {Cordier}, \citenamefont {Ceroni}, \citenamefont {Delgado},
  \citenamefont {Matteo}, \citenamefont {Dusko}, \citenamefont {Garg},
  \citenamefont {Guala}, \citenamefont {Hayes}, \citenamefont {Hill},
  \citenamefont {Ijaz}, \citenamefont {Isacsson}, \citenamefont {Ittah},
  \citenamefont {Jahangiri}, \citenamefont {Jain}, \citenamefont {Jiang},
  \citenamefont {Khandelwal}, \citenamefont {Kottmann}, \citenamefont {Lang},
  \citenamefont {Lee}, \citenamefont {Loke}, \citenamefont {Lowe},
  \citenamefont {McKiernan}, \citenamefont {Meyer}, \citenamefont
  {Montañez-Barrera}, \citenamefont {Moyard}, \citenamefont {Niu},
  \citenamefont {O'Riordan}, \citenamefont {Oud}, \citenamefont {Panigrahi},
  \citenamefont {Park}, \citenamefont {Polatajko}, \citenamefont {Quesada},
  \citenamefont {Roberts}, \citenamefont {Sá}, \citenamefont {Schoch},
  \citenamefont {Shi}, \citenamefont {Shu}, \citenamefont {Sim}, \citenamefont
  {Singh}, \citenamefont {Strandberg}, \citenamefont {Soni}, \citenamefont
  {Száva}, \citenamefont {Thabet}, \citenamefont {Vargas-Hernández},
  \citenamefont {Vincent}, \citenamefont {Vitucci}, \citenamefont {Weber},
  \citenamefont {Wierichs}, \citenamefont {Wiersema}, \citenamefont {Willmann},
  \citenamefont {Wong}, \citenamefont {Zhang},\ and\ \citenamefont
  {Killoran}}]{bergholm2018pennylane}%
  \BibitemOpen
  \bibfield  {author} {\bibinfo {author} {V.~Bergholm}, et~al.,\ }\href@noop {}
  {\emph {PennyLane: Automatic differentiation of hybrid quantum-classical
  computations}} (\bibinfo {year} {2022}),\ \Eprint
  {https://arxiv.org/abs/1811.04968} {arXiv:1811.04968 [quant-ph]} \BibitemShut
  {NoStop}%
\bibitem [{\citenamefont {Pagano}\ \emph {et~al.}(2022)\citenamefont {Pagano},
  \citenamefont {Weber}, \citenamefont {Jaschke}, \citenamefont {Pfau},
  \citenamefont {Meinert}, \citenamefont {Montangero},\ and\ \citenamefont
  {B\"uchler}}]{pagano_2022_error}%
  \BibitemOpen
  \bibfield  {author} {\bibinfo {author} {A.~Pagano}, et~al.,\ }\emph {Error
  budgeting for a controlled-phase gate with strontium-88 Rydberg atoms},\
  \href {https://doi.org/10.1103/PhysRevResearch.4.033019} {\bibfield
  {journal} {\bibinfo  {journal} {Physical Review Research}\ }\textbf {\bibinfo
  {volume} {4}},\ \bibinfo {pages} {033019} (\bibinfo {year}
  {2022})}\BibitemShut {NoStop}%
\bibitem [{\citenamefont {Jandura}\ and\ \citenamefont
  {Pupillo}(2022)}]{jandura_2022_timeoptimal}%
  \BibitemOpen
  \bibfield  {author} {\bibinfo {author} {S.~Jandura}\ and\ \bibinfo {author}
  {G.~Pupillo},\ }\emph {Time-optimal two- and three-qubit gates for Rydberg
  atoms},\ \href {https://doi.org/10.22331/q-2022-05-13-712} {\bibfield
  {journal} {\bibinfo  {journal} {{Quantum}}\ }\textbf {\bibinfo {volume}
  {6}},\ \bibinfo {pages} {712} (\bibinfo {year} {2022})}\BibitemShut {NoStop}%
\bibitem [{\citenamefont {Barredo}\ \emph {et~al.}(2016)\citenamefont
  {Barredo}, \citenamefont {de~Léséleuc}, \citenamefont {Lienhard},
  \citenamefont {Lahaye},\ and\ \citenamefont
  {Browaeys}}]{barredo_2016_an_atom}%
  \BibitemOpen
  \bibfield  {author} {\bibinfo {author} {D.~Barredo}, \bibinfo {author}
  {S.~de~Léséleuc}, \bibinfo {author} {V.~Lienhard}, \bibinfo {author}
  {T.~Lahaye},\ and\ \bibinfo {author} {A.~Browaeys},\ }\emph {An atom-by-atom
  assembler of defect-free arbitrary two-dimensional atomic arrays},\ \href
  {https://doi.org/10.1126/science.aah3778} {\bibfield  {journal} {\bibinfo
  {journal} {Science}\ }\textbf {\bibinfo {volume} {354}},\ \bibinfo {pages}
  {1021} (\bibinfo {year} {2016})}\BibitemShut {NoStop}%
\bibitem [{\citenamefont {Endres}\ \emph {et~al.}(2016)\citenamefont {Endres},
  \citenamefont {Bernien}, \citenamefont {Keesling}, \citenamefont {Levine},
  \citenamefont {Anschuetz}, \citenamefont {Krajenbrink}, \citenamefont
  {Senko}, \citenamefont {Vuletic}, \citenamefont {Greiner},\ and\
  \citenamefont {Lukin}}]{endres_2016_atom}%
  \BibitemOpen
  \bibfield  {author} {\bibinfo {author} {M.~Endres}, et~al.,\ }\emph
  {Atom-by-atom assembly of defect-free one-dimensional cold atom arrays},\
  \href {https://doi.org/10.1126/science.aah3752} {\bibfield  {journal}
  {\bibinfo  {journal} {Science}\ }\textbf {\bibinfo {volume} {354}},\ \bibinfo
  {pages} {1024} (\bibinfo {year} {2016})}\BibitemShut {NoStop}%
\bibitem [{\citenamefont {Henriet}\ \emph {et~al.}(2020)\citenamefont
  {Henriet}, \citenamefont {Beguin}, \citenamefont {Signoles}, \citenamefont
  {Lahaye}, \citenamefont {Browaeys}, \citenamefont {Reymond},\ and\
  \citenamefont {Jurczak}}]{henriet_2020_quantum}%
  \BibitemOpen
  \bibfield  {author} {\bibinfo {author} {L.~Henriet}, et~al.,\ }\emph {Quantum
  computing with neutral atoms},\ \href
  {https://doi.org/10.22331/q-2020-09-21-327} {\bibfield  {journal} {\bibinfo
  {journal} {{Quantum}}\ }\textbf {\bibinfo {volume} {4}},\ \bibinfo {pages}
  {327} (\bibinfo {year} {2020})}\BibitemShut {NoStop}%
\bibitem [{\citenamefont {Kaufman}\ and\ \citenamefont
  {Ni}(2021)}]{kaufman2021quantum}%
  \BibitemOpen
  \bibfield  {author} {\bibinfo {author} {A.~M. Kaufman}\ and\ \bibinfo
  {author} {K.-K. Ni},\ }\emph {Quantum science with optical tweezer arrays of
  ultracold atoms and molecules},\ \href
  {https://doi.org/10.1038/s41567-021-01357-2} {\bibfield  {journal} {\bibinfo
  {journal} {Nature Physics}\ }\textbf {\bibinfo {volume} {17}},\ \bibinfo
  {pages} {1324} (\bibinfo {year} {2021})}\BibitemShut {NoStop}%
\bibitem [{\citenamefont {Lao}\ and\ \citenamefont
  {Almudever}(2020)}]{lao2020fault}%
  \BibitemOpen
  \bibfield  {author} {\bibinfo {author} {L.~Lao}\ and\ \bibinfo {author}
  {C.~G. Almudever},\ }\emph {Fault-tolerant quantum error correction on
  near-term quantum processors using flag and bridge qubits},\ \href
  {https://doi.org/10.1103/PhysRevA.101.032333} {\bibfield  {journal} {\bibinfo
   {journal} {Physical Review A}\ }\textbf {\bibinfo {volume} {101}},\ \bibinfo
  {pages} {032333} (\bibinfo {year} {2020})}\BibitemShut {NoStop}%
\bibitem [{\citenamefont {Chamberland}\ and\ \citenamefont
  {Noh}(2020)}]{chamberland2020very}%
  \BibitemOpen
  \bibfield  {author} {\bibinfo {author} {C.~Chamberland}\ and\ \bibinfo
  {author} {K.~Noh},\ }\emph {Very low overhead fault-tolerant magic state
  preparation using redundant ancilla encoding and flag qubits},\ \href
  {https://doi.org/https://doi.org/10.1038/s41534-020-00319-5} {\bibfield
  {journal} {\bibinfo  {journal} {npj Quantum Information}\ }\textbf {\bibinfo
  {volume} {6}},\ \bibinfo {pages} {91} (\bibinfo {year} {2020})}\BibitemShut
  {NoStop}%
\bibitem [{\citenamefont {Lang}\ and\ \citenamefont
  {B{\"u}chler}(2012)}]{lang2012minimal}%
  \BibitemOpen
  \bibfield  {author} {\bibinfo {author} {N.~Lang}\ and\ \bibinfo {author}
  {H.~P. B{\"u}chler},\ }\emph {Minimal instances for toric code ground
  states},\ \href {https://doi.org/10.1103/PhysRevA.86.022336} {\bibfield
  {journal} {\bibinfo  {journal} {Physical Review A}\ }\textbf {\bibinfo
  {volume} {86}},\ \bibinfo {pages} {022336} (\bibinfo {year}
  {2012})}\BibitemShut {NoStop}%
\bibitem [{\citenamefont {Amaro}\ \emph {et~al.}(2020)\citenamefont {Amaro},
  \citenamefont {M{\"u}ller},\ and\ \citenamefont {Pal}}]{amaro2020scalable}%
  \BibitemOpen
  \bibfield  {author} {\bibinfo {author} {D.~Amaro}, \bibinfo {author}
  {M.~M{\"u}ller},\ and\ \bibinfo {author} {A.~K. Pal},\ }\emph {Scalable
  characterization of localizable entanglement in noisy topological quantum
  codes},\ \href {https://doi.org/10.1088/1367-2630/ab84b3} {\bibfield
  {journal} {\bibinfo  {journal} {New Journal of Physics}\ }\textbf {\bibinfo
  {volume} {22}},\ \bibinfo {pages} {053038} (\bibinfo {year}
  {2020})}\BibitemShut {NoStop}%
\bibitem [{\citenamefont {Nigg}\ \emph {et~al.}(2014)\citenamefont {Nigg},
  \citenamefont {M{\ifmmode\ddot{u}\else\"{u}\fi}ller}, \citenamefont
  {Martinez}, \citenamefont {Schindler}, \citenamefont {Hennrich},
  \citenamefont {Monz}, \citenamefont {Martin-Delgado},\ and\ \citenamefont
  {Blatt}}]{nigg2014quantum}%
  \BibitemOpen
  \bibfield  {author} {\bibinfo {author} {D.~Nigg}, et~al.,\ }\emph {{Quantum
  computations on a topologically encoded qubit}},\ \href
  {https://doi.org/10.1126/science.1253742} {\bibfield  {journal} {\bibinfo
  {journal} {Science}\ }\textbf {\bibinfo {volume} {345}},\ \bibinfo {pages}
  {302} (\bibinfo {year} {2014})}\BibitemShut {NoStop}%
\bibitem [{\citenamefont {Maslov}(2017)}]{maslov2017basic}%
  \BibitemOpen
  \bibfield  {author} {\bibinfo {author} {D.~Maslov},\ }\emph {Basic circuit
  compilation techniques for an ion-trap quantum machine},\ \href
  {https://doi.org/10.1088/1367-2630/aa5e47} {\bibfield  {journal} {\bibinfo
  {journal} {New Journal of Physics}\ }\textbf {\bibinfo {volume} {19}},\
  \bibinfo {pages} {023035} (\bibinfo {year} {2017})}\BibitemShut {NoStop}%
\end{thebibliography}%

\end{document}